\def\lsim{\raise0.3ex\hbox{$<$\kern-0.75em\raise-1.1ex\hbox{$\sim$}}}
\def\gsim{\raise0.3ex\hbox{$>$\kern-0.75em\raise-1.1ex\hbox{$\sim$}}}

\def\simle{\stackrel{<}{\sim} }
\def\mean#1{\left<#1\right>}
\def\Journal#1#2#3#4#5{ #5 {#1} {\it  {#2}} {\bf #3} #4}

\def\IJMPE{{Int. J. Mod. Phys. E}}
\def\EPJC{{Eur. Phys. J. C}}
\def\JETPL{{JETP Lett.\ }}
\def\NPA{{Nucl. Phys. A}}
\def\NPB{{Nucl. Phys. B}}
\def\PLB{{Phys. Lett. B}}
\def\PLC{Phys. Repts.\ }
\def\PRD{{Phys. Rev. D}}
\def\PRC{{Phys. Rev.  C}}
\def\PPNP{{Prog. Part. Nucl. Phys.}}
\def\ZPC{{Z. Phys. C}}
\def\ARNS{{Ann. Rev. Nucl. Part. Sci.\ }} 

\documentclass[12pt]{iopart}
\usepackage{epsfig}
\begin{document}

\review[Recent results in relativistic heavy ion collisions]{Recent results in relativistic heavy ion collisions: from ``a new state of matter" to ``the perfect fluid".}
\author{M J Tannenbaum\footnote{Research supported by U. S. Department of Energy, DE-AC02-98CH10886.}}

\address{Physics Dept. 510c
Brookhaven National Laboratory
Upton, NY 11973-5000 USA}
\ead{mjt@bnl.gov}
\begin{abstract}
 Experimental Physics with Relativistic Heavy Ions dates from 1992 when a beam of $^{197}$Au of energy greater than $10 A$ GeV/$c$ first became avalilable at the Alternating Gradient Synchrotron (AGS) at Brookhaven National Laboratory (BNL) soon followed in 1994 by a $^{208}$Pb beam of $158A$ GeV/$c$ at the Super Proton Synchrotron (SPS) at CERN (European Center for Nuclear Research). Previous pioneering measurements at the Berkeley Bevalac~\cite{PlasticBall} in the late 1970's and early 1980's were at much lower bombarding energies ($\simle 1A$ GeV/$c$) where nuclear breakup rather than particle production is the dominant inelastic process in A+A collisions. More recently, starting in 2000, the Relativistic Heavy Ion Collider (RHIC) at BNL has produced head-on collisions of two $100 A$ GeV beams of fully stripped Au ions, corresponding to nucleon-nucleon center-of-mass energy, $\sqrt{s_{NN}}=200$ GeV, total c.m. energy $200 A$ GeV. The objective of this research program is to produce nuclear matter with extreme density and  temperature, possibly resulting in a state of matter where the quarks and gluons normally confined inside individual nucleons ($r < 1$fm) are free to act over distances an order of magnitude larger. Progress from the period 1992 to the present will be reviewed, with reference to previous results from light ion and proton-proton collisions where appropriate. Emphasis will be placed on the measurements which formed the basis for the announcements by the two  major laboratories:``A new state of matter", by CERN on Feb 10, 2000 and ``The perfect fluid", by BNL on April 19, 2005.

\end{abstract}

Published 23 June 2006:  \RPP {\bf 69} (2006) 2005--2059\\[0.25in]
\maketitle

\section {Introduction} \label{sec:Intro}
    In the early 1970's, it became clear that the nucleon was 
not an elementary particle but was composed of a substructure of 3 valence quarks confined into a bound state by a strong interaction, Quantum Chromo Dynamics (QCD), which is mediated by the exchange of color-charged vector gluons.~\cite{Fritzsch73} In sharp distinction to the behavior of the uncharged quanta of quantum electrodynamics (QED), the color-charged gluons of QCD interact with each other. This leads to the property of asymptotic freedom,~\cite{Politzer, GrossWilczek} the reduction of the effective coupling constant at short distances, and is believed to provide the confinement property at long distances where the quarks and gluons behave as if attached to each other by a color string. It is worth reviewing some developments leading up to and immediately following the discovery of QCD which bear on the importance of Relativistic Heavy Ion collisions as a probe of this fundamental interaction. 
    
    In the early 1960's, with the construction of proton accelerators with energies well above the threshold for anti-proton production, a veritable `zoo' of new particles and resonances was discovered~\cite{Cahn}. Gell-Mann~\cite{MGM62} and Ne'eman~\cite{YN61}  noticed that particles sharing the same quantum numbers (spin, parity) follow the symmetry of the mathematical group SU(3) which is based on 3 elementary generators, up, down, strange, or $u$, $d$, $s$, with spin 1/2 and fractional electrical charge,~\cite{MGM64,Zweig} which Gell-Mann called quarks. Mesons are described as states made of a quark-anti-quark ($q\bar{q}$) pair and baryons as states of 3 quarks $(qqq)$.  This led to the prediction of a new baryon, the $\Omega^{-}$ $(sss)$ with strangeness -3, which was observed shortly thereafter.~\cite{NPS} However, the $\Omega^{-}$ had a problem: 3 identical $s$ quarks in the same state, apparently violating the Pauli Exclusion Principle. To avoid this problem, it was proposed~\cite{Greenberg} that quarks come in 3 `colors', i.e. distinguishing characteristics which would allow 3 otherwise identical quarks to occupy the same state (formally, para-Fermi statistics of rank 3). A major breakthrough was the realization that the real SU(3) symmetry was not the original 3 quarks $uds$ (now called `flavor'), but the 3 colors; and that color-charged gluons are the quanta of the `asymtotically-free' strong interaction which binds hadrons.~\cite{Fritzsch73} The fourth ``charm" or $c$ quark, proposed to explain the absence of certain channels in weak decays of strange-particles,~\cite{GIM} thus had no problem fitting into this scheme--the quark symmetry became SU(2), groups of doublets. Elegant as these theories were, the experimental results were what made them believable. 
    
    There were 3 key experimental observations that made the composite theory of hadrons believable: 1) the discovery of pointlike constituents (`partons') inside the proton, in deeply inelastic (large energy loss, $\nu$, large 4-momentum transfer, $Q$) electron-proton ($ep$) scattering (DIS) at the Stanford Linear Accelerator Center (SLAC)~\cite{DIS68}; 2) the observation of particle production at large transverse momenta ($p_T$) in p-p collisions at the CERN-Intersecting Storage Rings (ISR)~\cite{CCR,SS,BS}, which proved that the partons of DIS interacted much more strongly with each other than the electromagnetic scattering observed at SLAC; 3) the observation of the J/$\Psi$, a narrow bound state of $c\bar{c}$, in both p+Be collisions at the BNL-AGS~\cite{Ting}, and in $e^{+}e^{-}$ annihilations at SLAC~\cite{Richter}, shortly followed by observation of the $\Psi^{'}$, a similar state with higher mass~\cite{Richter2} which corresponded to $c\bar{c}$ bound states in a simple couloumb-like potential with a string-like linear confining potential.~\cite{AppelquistPolitzer,DeRujulaGlashow,Cornell} These discoveries turned 
Gell-Mann and Zweig's quarks from mere mathematical concepts to the fundamental constituents of matter, the components of the nucleon.~\cite{KutiWeisskopf,Close,Owens78,FFF78} 
   
    \subsection{From Bjorken Scaling to QCD to the QGP}  
    
     The fundamental idea to emerge from DIS, which was the basis of much of the subsequent theoretical developments leading to QCD, was the concept of Bjorken scaling~\cite{Bjorken69} which indicated that protons consist of point-like objects (partons). The structure function $F_2(Q^2,\nu)$ which describes the inelastic $ep$ scattering cross section was predicted to~\cite{Bjorken69} and observed to~\cite{DIS68} ``scale'' i.e. to be a  function only of the ratio of the variables, $Q^2/\nu$. The deeply inelastic scattering of an electron from a proton is simply incoherent quasi-elastic scattering of the electron from point-like partons of effective mass $Mx$, where $\nu=Q^2/2Mx$, where $M$ is the rest mass of the proton. Thus Bjorken `$x$' is the fraction of the nucleon momentum (or mass) carried by the parton. Similar ideas for the scaling of longitudinal momentum distributions in p-p collisions were also given~\cite{FeynmanScaling,LimFrag}. However these ideas related to the ``soft" (low $p_T$) particle production rather than the large $p_T$ or ``hard scattering" processes described by Bjorken~\cite{egseeDis71}.   
     
     Bjorken scaling was the basis of QCD~\cite{Fritzsch73}, the MIT Bag model~\cite{Chodos74} and led to the conclusion~\cite{CollinsPerry} that ``superdense matter (found in neutron-star cores, exploding black holes, and the early big-bang universe) consists of quarks rather than of hadrons", because the hadrons overlap and their individuality is confused. Collins and Perry~\cite{CollinsPerry} called this state ``quark soup" but used the equation of state of a gas of free massless quarks from which the interacting gluons acquire an effective mass, which provides long-range screening. They anticipated superfluidity and superconductivity in  nuclear matter at high densities and low temperatures. They also pointed  out that for the theory of strong interactions (QCD), ``high density matter is the second situation where one expects to be able to make reliable calculations---the first is Bjorken scaling". In the Bjorken scaling region, the theory is asymptotically free at large momentum transfers while in high-density nuclear matter long range interactions are screened by many-body effects, so they can be ignored and short distance behavior can be calculated with the asymptotically-free QCD and relativistic many-body theory. Shuryak~\cite{Shuryak80} codified and elaborated on these ideas and provided the name ``QCD (or quark-gluon) plasma" for ``this phase of matter", a plasma being an ionized gas.   
     \subsection{Relativistic Heavy Ions Collisions}
     It was soon realized that the collisions of relativistic heavy ions could provide the means of obtaining superdense nuclear matter in the laboratory.~\cite{WillisCocconi,Teller,LeeWick,BearMountain} 
 The kinetic energy of the incident projectiles would be dissipated in the large 
volume of nuclear matter involved in the reaction.  The system is expected 
to come to equilibrium, thus heating and compressing the nuclear matter so that it undergoes a phase transition from a state of nucleons containing bound quarks and gluons to a state of deconfined quarks and gluons, the Quark Gluon Plasma, in chemical and thermal equilibrium, covering the entire
volume of the colliding nuclei or a volume that corresponds to many units of the characteristic length scale. In the terminology of high energy physics, 
this is called a ``soft'' process, related to the QCD confinement scale
~\cite{PDG}
\begin{equation}
\Lambda^{-1}_{\rm QCD} \simeq {\rm (0.2\ GeV)}^{-1} \simeq 1 \, 
\mbox{fm}\qquad .
\label{eq:LambdaQCD}
\end{equation}
Two energy regimes are discussed for the QGP~\cite{Anishetty80}. 
At lower energies, typical of the AGS fixed target program, the colliding nuclei are 
expected to stop each other, leading to a baryon-rich system. This 
will be the region of maximum baryon density. At very high energy,
100 to 200 GeV per nucleon pair in the center of mass, nuclei become transparent and the nuclear
fragments will be well separated from a central region of particle
production. This is the region of the baryon-free or gluon plasma. In the nuclear fragmentation regions a baryon-rich plasma may also be formed~\cite{Anishetty80,Stocker05}. 

\begin{figure}[!thb]
\begin{center}
\begin{tabular}{cc}
\includegraphics[scale=0.65,angle=0]{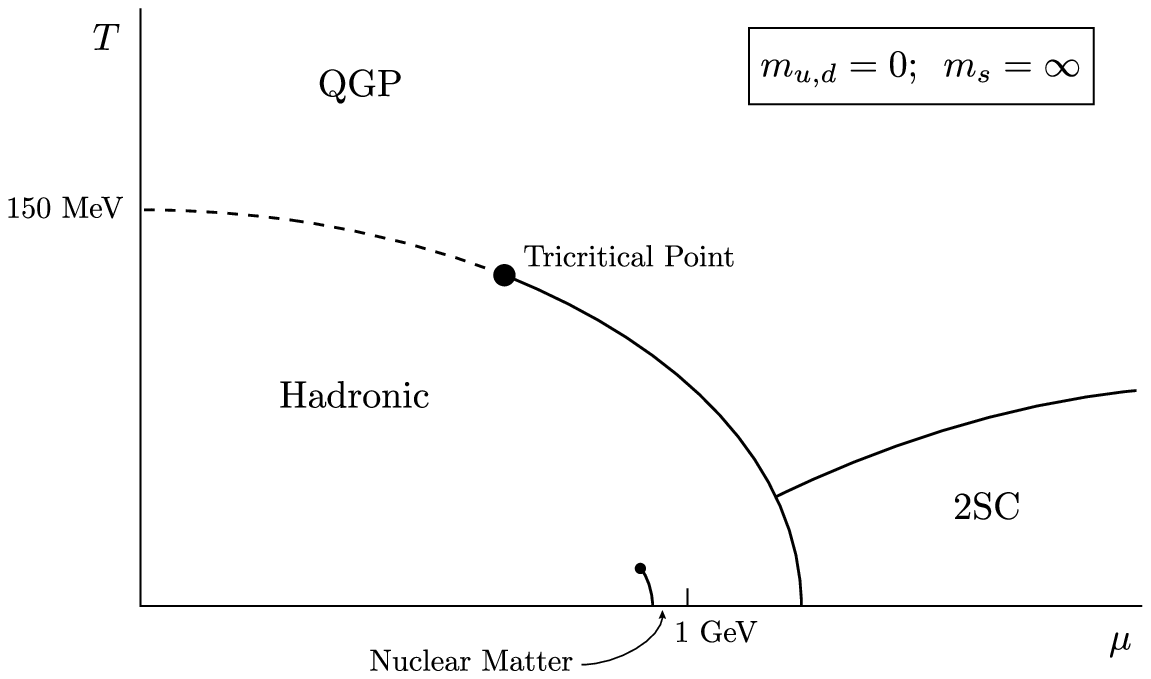}&
\includegraphics[scale=0.60,angle=0]{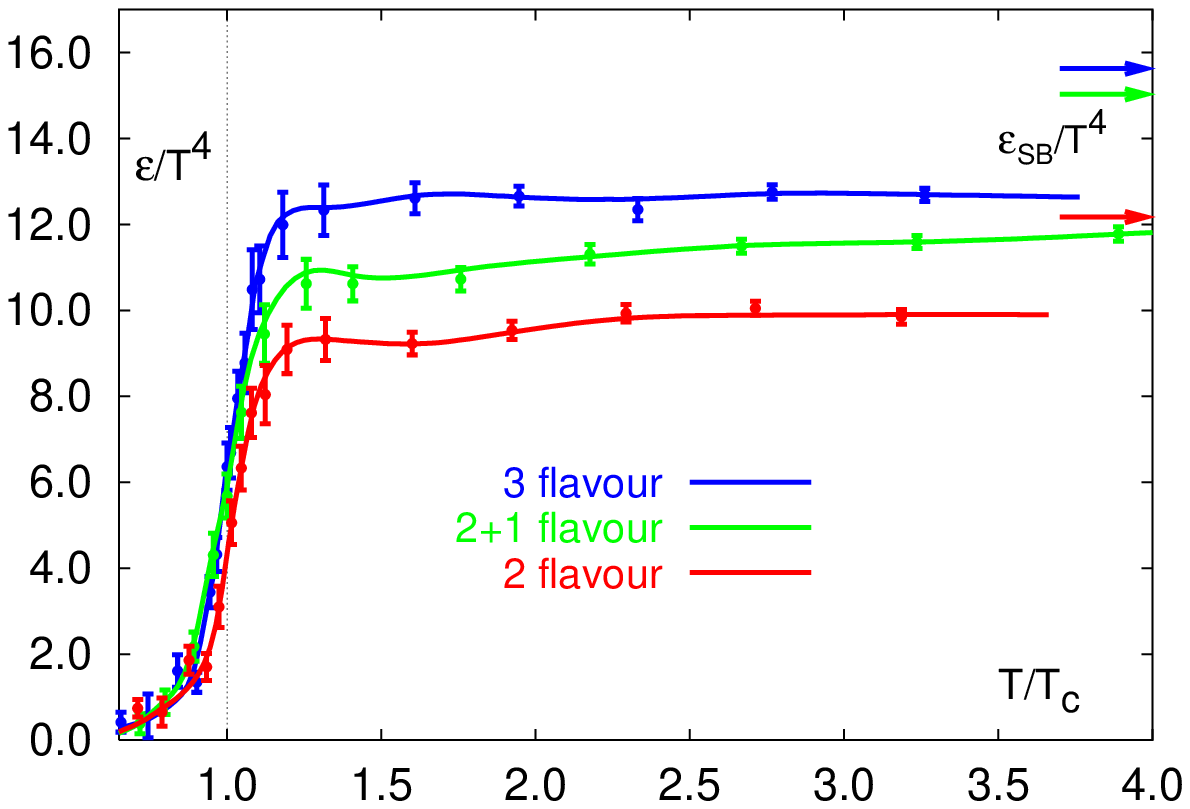} 

\end{tabular}
\end{center}\vspace*{-0.25in}
\caption{a)(left) Proposed phase diagram for nuclear matter~\cite{Krishna99}: Temperature,  $T$,  vs Baryon Chemical Potential, $\mu$. b) (right) Lattice calculation~\cite{Karsch02}  of energy density, $\epsilon$/$T^4$ as a function of the number of active flavors: 2 flavor $(u,d)$, 3 flavor $(u,d,s)$. \label{fig:phaselat}}

\end{figure}

   There has been considerable work over the past 
three decades in 
making quantitative predictions for the QGP~\cite{CollinsPerry,CabibboParisi75,BaymOthers,SatzRPP63,egQM}. The predicted transition temperature from a state of hadrons to the QGP varies, from $T_c\sim 150$ MeV at zero baryon density, to zero temperature at a critical baryon density
~\cite{BaymOthers,HeinzLeeMRB87} roughly 1 GeV/fm$^3$, 
$\sim$ 6.5 times the normal density of cold nuclear matter,  
$\rho_0 = 0.14\,  {\rm nucleons}/ {\rm fm}^3$, $\mu_B\simeq 930$ MeV, 
where $\mu_B$ is the Baryon chemical potential. A typical expected phase diagram of nuclear matter~\cite{Krishna99} is shown in Fig.~\ref{fig:phaselat}a. Not distinguished on Fig.~\ref{fig:phaselat}a in the hadronic phase are the liquid self-bound ground state of nuclear matter and the gas of free nucleons~\cite{DAgostino05}. 

Predictions for the transition temperature for $\mu_B\sim 0$ are constrained to a relatively narrow range $140 < T_c < 250\,  $MeV, while the critical energy density is predicted to be 5 to 20 times the
normal nuclear energy density, $\epsilon_{0}=0.14$ GeV/fm$^3$. Presumably, the most accurate predictions of the phase transition are given by numerical solutions of the QCD Lagrangian on a lattice~\cite{Creutz}, see Fig.~\ref{fig:phaselat}b~\cite{Karsch02}. Here, the critical energy density at the transition temperature, $T_c\sim 150-170$~MeV, is stated to be presently known only with large errors~\cite{KarschHP05}, $\epsilon_c$=(0.3--1.3) GeV/fm$^3$.

   One of the nice features of the search for the QGP is that it 
requires the integrated use of many disciplines in Physics: High Energy Particle Physics, Nuclear Physics, Relativistic Mechanics, Quantum Statistical Mechanics, etc.~\cite{Recently,Policastro,Nastase-BlackHole} From the point of view of an experimentalist there are two major questions in this field. The first is how to relate the thermodynamical properties (temperature, energy density, entropy ...) of the QGP or hot nuclear matter to properties that can be measured in the laboratory. The second question is how the QGP can be detected. 

    One of the
major challenges in this field is to find signatures that are unique to the QGP so that this new state of matter can be distinguished from the ``ordinary physics" of relativistic nuclear collisions.  Another more general challenge is to find effects which are specific to A+A collisions, such as collective or coherent phenomena, in distinction to cases for which  A+A collisions can be considered as merely an incoherent superposition of nucleon-nucleon collisions~\cite{specificity,Weiner05,Alexopoulos02}.

    Many signatures of the Quark Gluon Plasma~\cite{MJTIND,BMuellerRPP,RLTAPP}  have been proposed over the past two decades, which cover the experimental RHI programs at the AGS, the SPS, RHIC and soon the Large Hadron Collider (LHC) at CERN. In recent peer reviewed articles summarizing the first 3 years of RHIC operation~\cite{BRWP,PHWP,STWP,PXWP,THWPs}, the four experiments and many distinguished theorists presented their results and opinions on QGP signatures and whether the QGP had been detected at RHIC. These were summarized in a BNL press release on April 18, 2005: ``instead of behaving like a gas of free quarks or and gluons, as was expected, the matter created in RHIC's heavy ion collisions appears to be more like a {\em liquid}." This matter interacted much more strongly than expected, causing the theorists~\cite{THWPs} to give it the new name ``sQGP" (strongly interacting QGP). Weighing heavily on this process was the CERN press release~\cite{CERNbaloney} and unpublished preprint~\cite{HeinzJacob} on February 10, 2000, just at the start of RHIC operations, which discussed properties of the QGP and announced that ``The collected data from the [SPS] experiments gives compelling evidence that a new state of matter has been created. This state of matter found in heavy ion collisions at the SPS features many of the characteristics of the theoretically predicted quark-gluon plasma..." ``The data from any one experiment is not enough to give the full picture but the combined results from all experiments agree and fit. Whereas all attempts to  explain them using established particle interactions have failed, many of the observations are consistent with the predicted signatures of a quark-gluon plasma." In other words, several features of the CERN measurements were consistent with the expected properties of the QGP at that time,  i.e. a gas of quarks and gluons. (Although not mentioned in any press release, the same could be said for several features of measurements at the AGS fixed target program.) In light of this apparent contradiction, the data from the AGS, SPS and RHIC experiments will be examined with emphasis on the QGP signatures outlined in these review articles and press releases. 
\section{Observables in Relativistic Heavy Ion Collisions}
\begin{figure}[!thb]
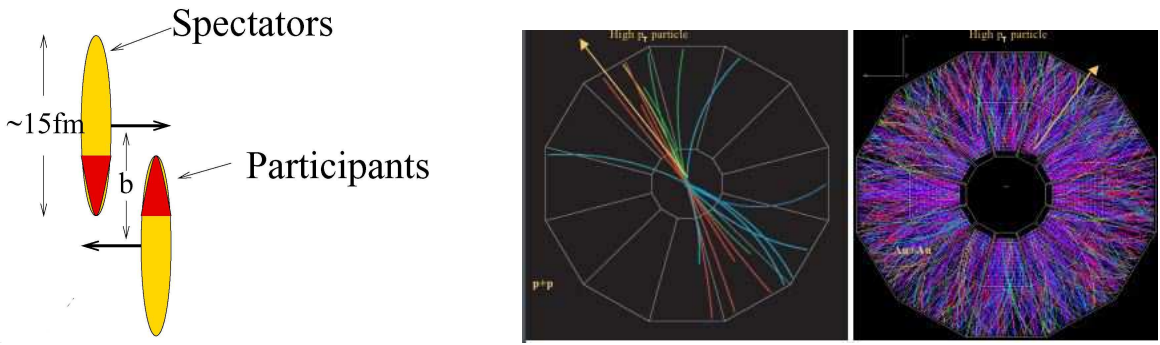

\begin{center}
\begin{tabular}{cc}

\includegraphics[scale=1,angle=0]{figs/Lars-ps.epsf}\hspace*{0.25in}&
\includegraphics[scale=0.5,angle=0]{figs/STARJet+AuAu-g.epsf}
\end{tabular}
\end{center}\vspace*{-0.25in}
\caption[]{a) (left) Schematic of collision of two nuclei with radius $R$ and impact parameter $b$. b) (center) A p-p collision in the STAR detector viewed along the collision axis; c) (right) Au+Au central collision at $\sqrt{s_{NN}}=200$ GeV in the STAR detector.  
\label{fig:collstar}}

\end{figure}

   A schematic drawing of a Relativistic Heavy Ion collision is shown in Fig.~\ref{fig:collstar}a. In the center of mass system of the nucleus-nucleus collision, the two Lorentz-contracted nuclei of radius $R$ approach each other with impact parameter $b$. In the region of overlap, the ``participating" nucleons interact with each other, while in the non-overlap region, the ``spectator" nucleons simply continue on their original trajectories and can be measured in Zero Degree Calorimeters (ZDC), so that the number of participants can be determined. The degree of overlap is called the centrality of the collision, with $b\sim 0$, being the most central and $b\sim 2R$, the most peripheral. The maximum time of overlap is $\tau_O=2R/\gamma\,c$ where $\gamma$ is the Lorentz factor and $c$ is the velocity of light. 
The energy of the inelastic collision is predominantly dissipated by multiple particle production. 

     For any observed particle of momentum $p$, energy $E$, the momentum can be 
resolved into transverse ($p_T$) and longitudinal ($p_L$) components; and in
many cases the mass ($m$) of the particle can be determined. The longitudinal
momentum is conveniently expressed in terms of the rapidity ($y$):
\begin{equation}
y=\ln\left({E+p_L\over m_T}\right)
\label{eq:1} 
\end{equation}
\begin{equation}
\cosh y=E/m_T \qquad \sinh y=p_L/m_T \qquad dy=dp_L/E
\label{eq:2} 
\end{equation}
where
\begin{equation}
m_T=\sqrt{m^2+p_T^2} \quad {\rm and}\quad 
      E=\sqrt{p_L^2+m_T^2}=\sqrt{p^2+m^2}
\label{eq:3}
\end{equation}
In the limit when ($m\ll E$) the rapidity reduces to the 
pseudorapidity ($\eta$)
\begin{equation}
\eta=-\ln\tan\theta/2
\label{eq:4} \end{equation}
\begin{equation}
\cosh\eta=\csc\theta \qquad \sinh\eta=\cot\theta
\label{eq:5}
\end{equation}
where $\theta$ is the polar angle of emission.
The rapidity variable has the useful property that it 
is additive under a Lorentz transformation. 
\subsection{Kinematics of the collision}

    For any collision, the center-of-mass (c.m.) system---in which the momenta 
of the incident projectile and target are equal and opposite---is at rapidity 
$y^{\rm cm}$. The total energy in the c.m. system is denoted $\sqrt{s}$, 
which, evidently, is also the ``invariant mass'' of the c.m. system. For  
a collision of an incident projectile of energy $E_1$, 
mass $m_1$, in the ``Laboratory System'', 
where the target, of mass $m_2$, is at rest (appropriate for fixed-target experiments): 
\begin{equation}
s=m_1^2+m_2^2+2 E_1 m_2  \quad .
\label{eq:syk1} 
\end{equation}
The c.m. rest frame moves in the laboratory system (along the 
collision axis) with a velocity $\beta^{\rm cm} c$ corresponding to:
\begin{equation}
\gamma^{\rm cm}={{E_1+m_2} \over \sqrt{s}}\qquad {\rm and} \qquad 
y^{\rm cm}=\cosh^{-1}\,\gamma^{\rm cm} \quad , 
\label{eq:syk2}
\end{equation}
where $\gamma^2=1/(1-\beta^2)$. 
Another useful quantity is $y^{\rm beam}$, the rapidity of the incident 
particle in the laboratory system
\begin{equation}
y^{\rm beam}=\cosh^{-1}\,{E_1 \over m_1} \quad ,
\label{eq:syk3}
\end{equation}
and note that for equal mass projectile and target:
\begin{equation}
y^{\rm cm}=y^{\rm beam}/2 \quad .
\label{eq:syk4}
\end{equation}
In the region near the projectile or 
target rapidity, the Feynman $x$ fragmentation variable is also used:
\begin{equation}
x_F=2p^*_L/\sqrt{s} \quad ,
\label{eq:syk8}\end{equation}
where $p^*_L$ is the longitudinal momentum of a particle in the c.m frame. 

    The kinematics are considerably simpler in the nucleon-nucleon c.m. frame in which the two nuclei approach each other with the same $\gamma$. (This is the reference frame of the detectors at RHIC). The nucleon-nucleon c.m. energy is denoted $\sqrt{s_{NN}}$, and the total c.m. energy is  $\sqrt{s}=A\cdot\sqrt{s_{NN}}$ for symmetric A+A collisions. The colliding nucleons approach each other with energy $\sqrt{s_{NN}}/2$ and equal and opposite momenta. The rapidity of the nucleon-nucleon center of mass is $y_{NN}=0$, and, taking $m_1=m_2=m_N=$931~MeV, the projectile and target nucleons are at equal and opposite rapidities~\cite{Kinematicnote}: 
    \begin{equation}
y^{\rm proj}=-y^{\rm target}=\cosh^{-1}\,{\sqrt{s_{NN}} \over {2 m_N}}=y^{\rm beam}/2\qquad . 
\label{eq:syk9}
\end{equation} 

\subsection{A brief overview of relativistic collisions of nucleons and nuclei}

	The challenge of RHI collisions can be understood from  Fig.~\ref{fig:collstar}b, which compares a p-p collision to an Au+Au central collision in the STAR detector~\cite{STWP}. It would appear to be a daunting task to reconstruct all the particles in such events. Consequently, it is more common to use single-particle or multi-particle inclusive variables to analyze these reactions.

A single particle ``inclusive'' reaction involves the measurement of just one
particle coming out of a reaction,  
\[ a + b \rightarrow c +\mbox{\rm anything} \;\;\; .\]
The terminology~\cite{FeynmanScaling} comes from the fact that all final states with the particle 
$c$ are summed over, or {\em included}. A ``semi-inclusive'' 
reaction\cite{KNO} refers to the measurement of all events of a given  
topology or class, e.g. 
\[ a + b \rightarrow n_1 \mbox{\rm \ particles in class 1}  
+\mbox{\rm anything} \;\;\; ,\]
where ``centrality'' is the most common class in relativistic heavy ion collisions.

\begin{figure}[htb]
\begin{center}
\begin{tabular}{cc}
\includegraphics[scale=0.525,angle=0]{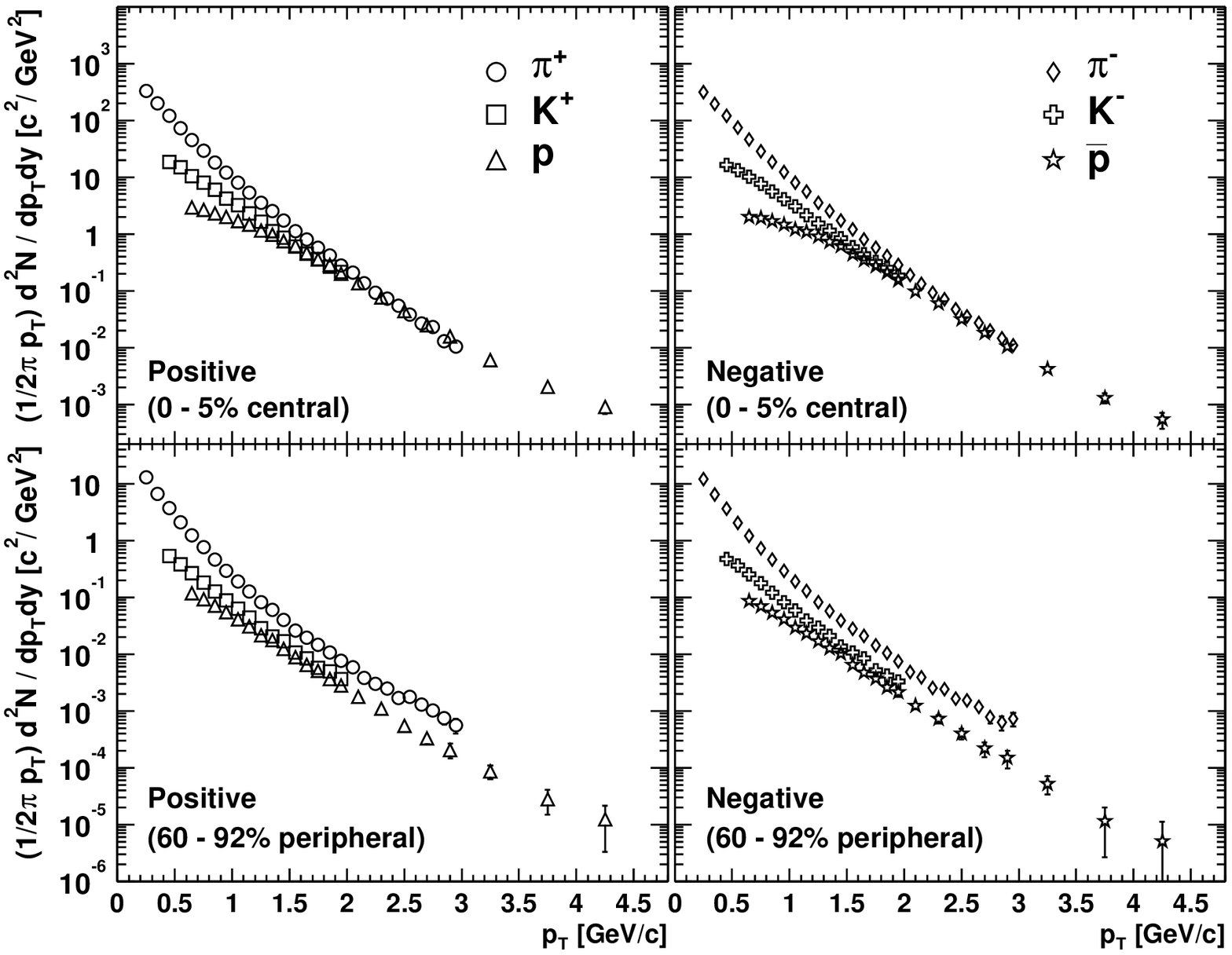}
&\hspace*{-0.35in}
\includegraphics[scale=0.525,angle=0]{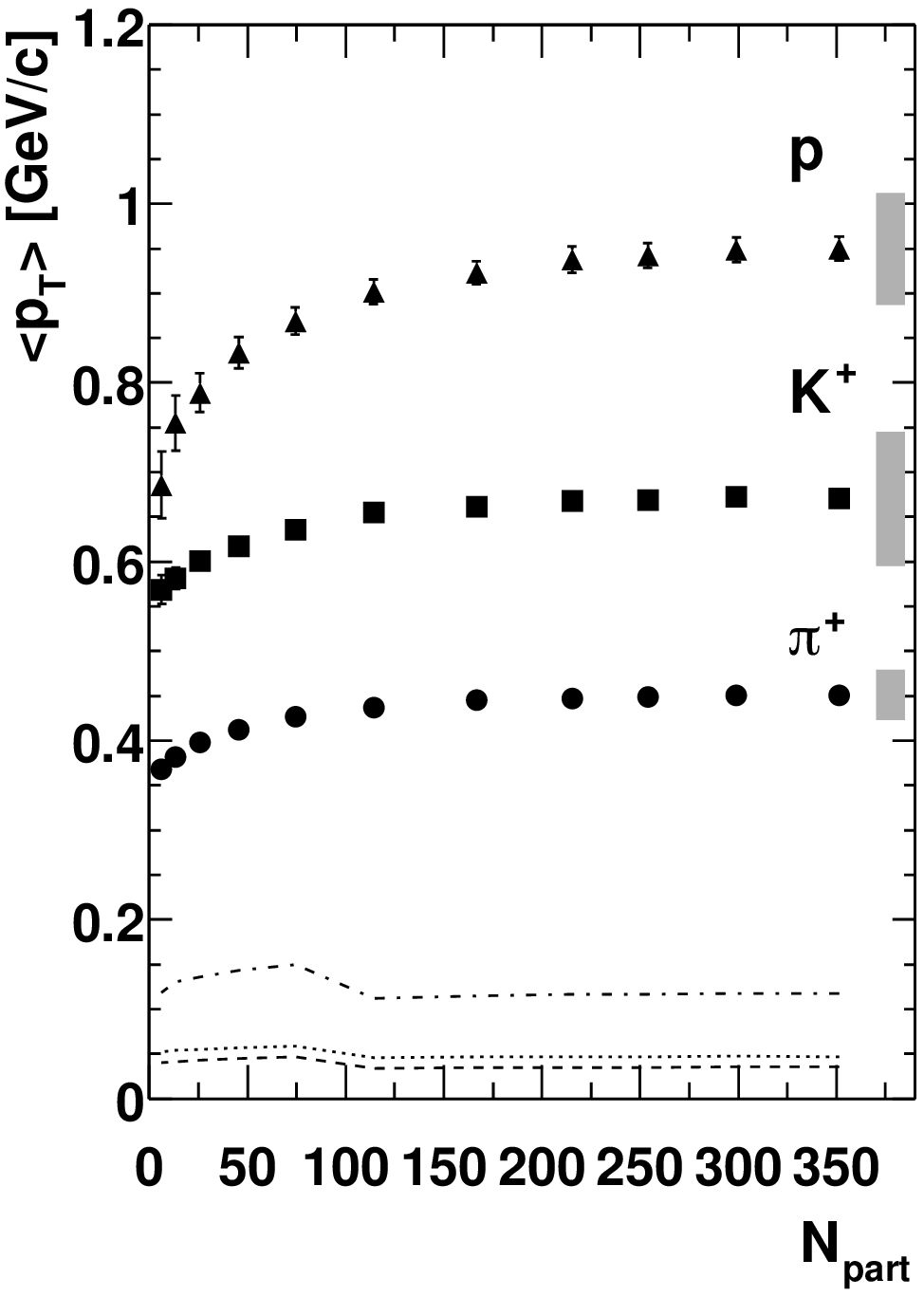}
\end{tabular}
\end{center}\vspace*{-0.25in}
\caption[]{a) (left) Semi-inclusive invariant $p_T$ spectra for $\pi^{\pm}$, $K^{\pm}$, $p^{\pm}$ in Au+Au collisions at $\sqrt{s_{NN}}=200$ GeV~\cite{PXWP}.  
b) (right) $\mean{p_T}$ of positive particles as a function of centrality ($N_{\rm part}$) from the same data.
\label{fig:PXpTspectra} }
\end{figure}

	 Measurements are presented in terms of 
the (Lorentz) invariant single particle inclusive differential cross section (or Yield per event in the class if semi-inclusive):
\begin{equation}
{Ed^3\sigma\over dp^3}={d^3\sigma\over p_T dp_T dy d\phi}=
{1\over 2\pi}\, {\bf f}(p_T,y) \quad ,
\label{eq:siginv} \end{equation} 
where $y$ is the rapidity, $p_T$ is the transverse momentum, and $\phi$ is the azimuth of the particle (see Fig.~\ref{fig:PXpTspectra}). 
The average transverse momentum, $\mean{p_T}$, or the mean transverse kinetic 
energy, $\mean{m_T}- m$, or the asymptotic slope are taken as measures of the 
temperature, $T$, of the reaction.  

	It is important to be aware that the integral of the single particle inclusive cross section over all the variables is not equal to $\sigma_I$ the interaction cross section, but 
rather is equal to the mean multiplicity times the interaction cross section: 
$\mean{n} \times\, \sigma_I$. Hence the mean multiplicity per interaction is 
\begin{equation}
\mean{n} ={1\over \sigma_I}\, \displaystyle\int {d\phi \over {2\pi}} dy\,dp_T\,\, p_T\,\, {\bf f}(p_T,y)= {1\over \sigma_I}\int dy\,{{d\sigma}\over {dy}} 
=\int dy\, \rho(y) \;\;\; , 
\label{eq:P4} \end{equation}
where the terminology for the multiplicity density in rapidity is $(1/\sigma_I)\,d\sigma/dy= \rho(y)=dn/dy$ for identified particles ($m$ known), $dn/d\eta$ for non-identified particles ($m$ unknown, assumed massless). The total charged particle multiplicity is taken as a measure of the total entropy, $S$ and $dn/d\eta$ is taken as a measure of the entropy density in restricted intervals of rapidity.  

\subsubsection{ The rapidity density and the rapidity plateau.} 
\begin{figure}[!hbt]
\begin{center}
\includegraphics[scale=1.0,angle=0]{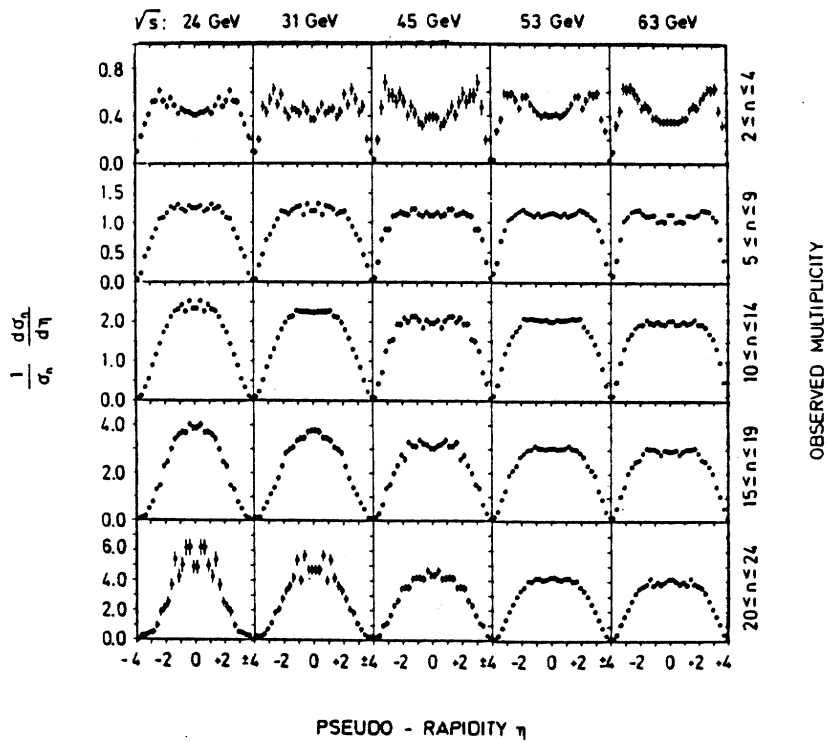}
\end{center}\vspace*{-0.25in}
\caption[]{Measurements in a streamer chamber at the CERN ISR~\cite{Thome77} of the normalized charged particle densities ${1\over \sigma_n} {{d\sigma_n}\over d\eta}$ (correctedfor acceptance up to $|\eta|\simeq 4$) in various intervals of the total observed multiplicity as a function of the c.m. energy $\sqrt{s}$ of the p-p collision.  \label{fig:in4}}

\end{figure}

\begin{figure}[!hbt]
\begin{center}
\hspace*{1.4in}\includegraphics[scale=0.60,angle=0]{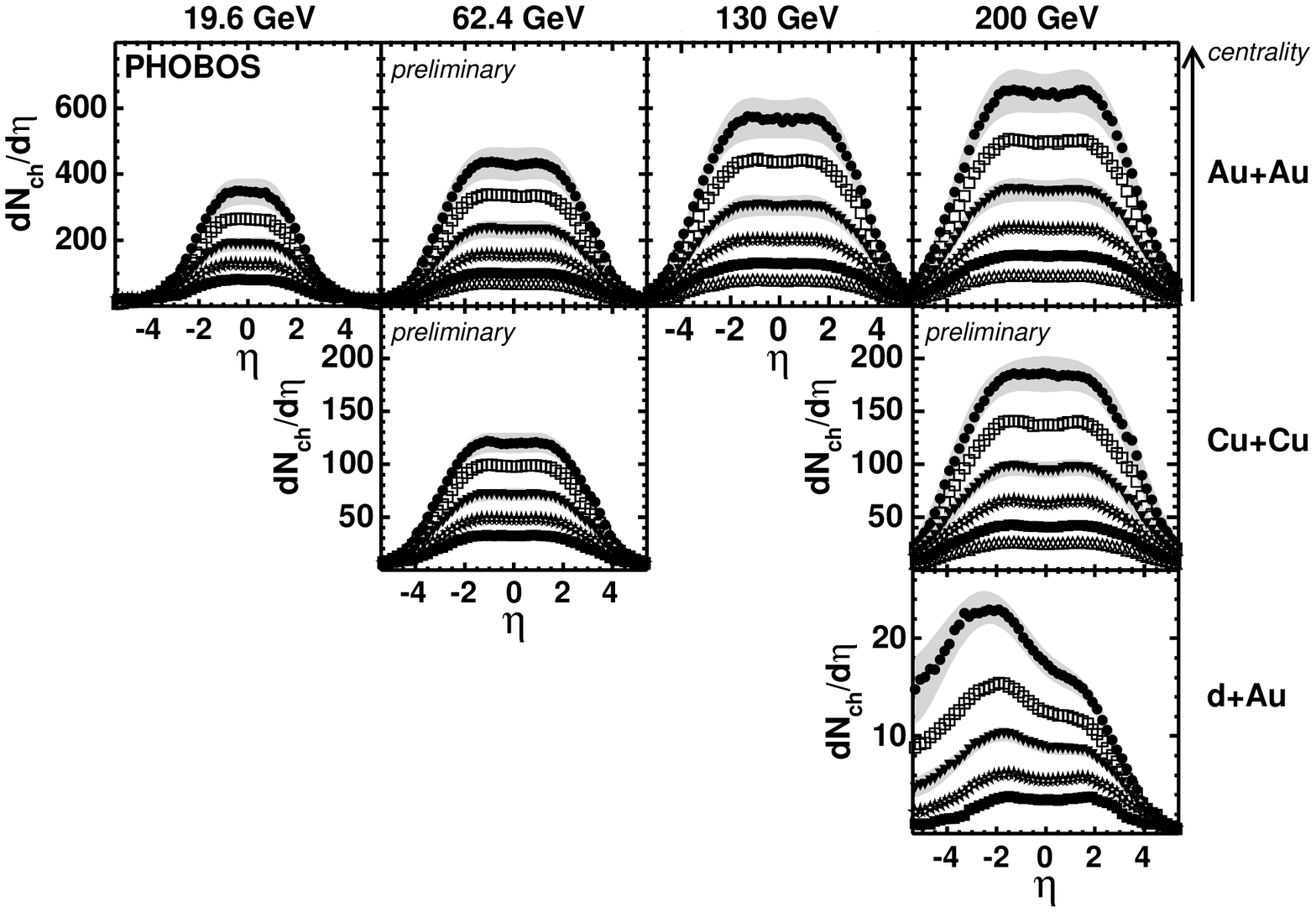}
\end{center}
\vspace*{-1.05in}
\includegraphics[scale=0.42,angle=0]{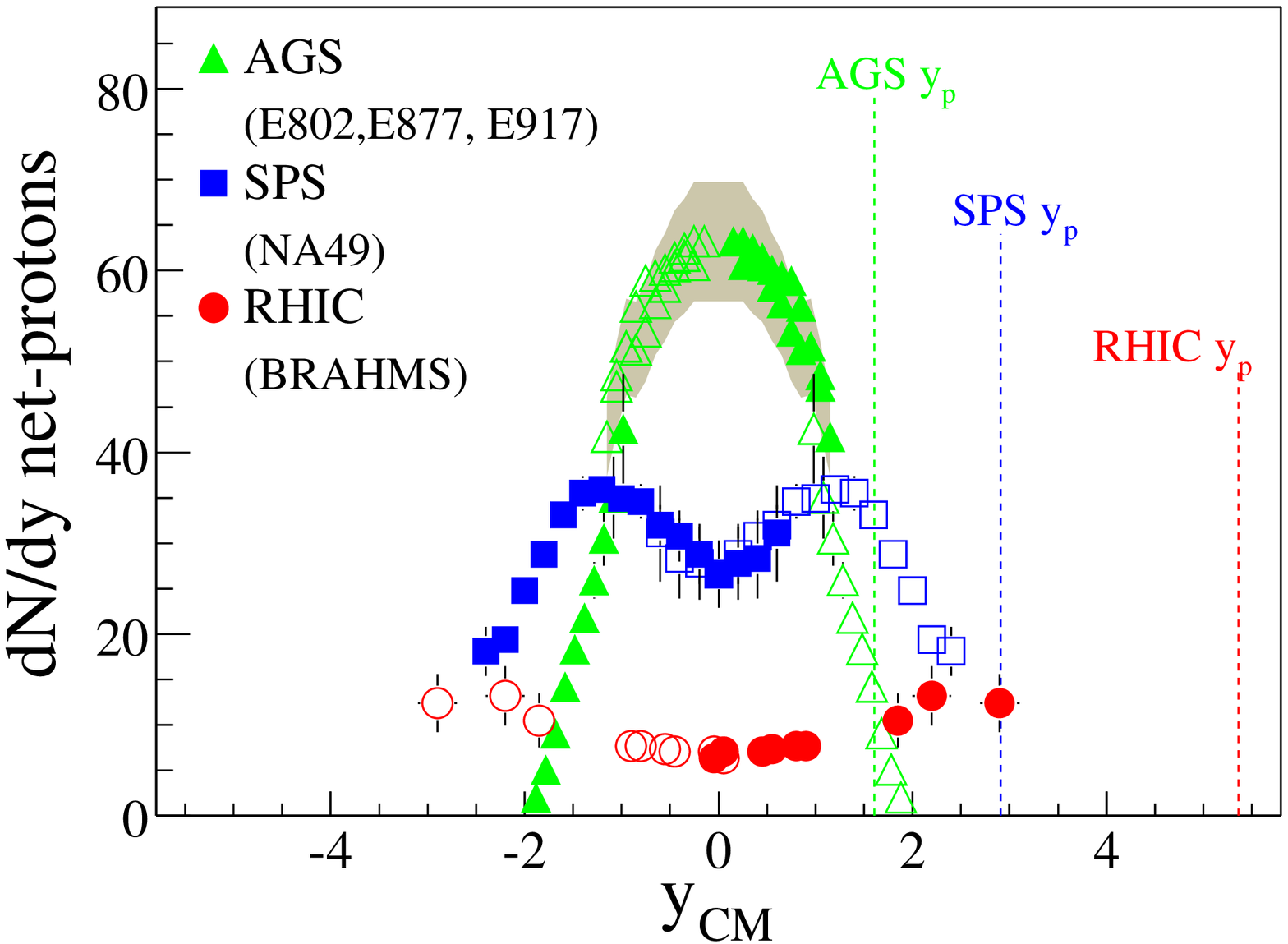}
\caption[]{a) (top right) $dn/d\eta$ for A+A and d+Au collisions at RHIC as a function of $\sqrt{s_{NN}}$~\cite{RolandQM05}. b) (bottom left) $dn/dy|_{p} - dn/dy|_{\bar{p}}$ at AGS, SPS and RHIC, $y^{\rm proj}=1.6, 2.9, 5.4$~\cite{BrahmsPRL93}. 
\label{fig:PHOBOSQM05}}

\end{figure}

	The shape and evolution with $\sqrt{s}$ of the charged particle 
density in rapidity, $dn/dy$, provide a graphic description of high energy 
collisions. Data from a classical measurement in a streamer chamber from p-p 
collisions at the CERN ISR~\cite{Thome77} are shown in Fig.~\ref{fig:in4}.  
Regions of nuclear fragmentation take up the first 1-2 units around the projectile and target rapidity and if the center-of-mass energy is sufficiently high, a
central plateau is exhibited. The data in p+A and A+A collisions follow a similar trend 
(Fig~\ref{fig:PHOBOSQM05}a)~\cite{RolandQM05}.  
The distributions increase in width with increasing $\sqrt{s_{NN}}$ but by a smaller amount than the increase in $y^{\rm beam}$ and show a small decrease in width with increasing centrality. $dn/d\eta$ increases with increasing centrality, $\sqrt{s_{NN}}$ and A in A+A collisions. In the asymmetric d+Au collision, $dn/d\eta$ in the target rapidity region is larger than in the projectile region, but not by much, only about ~50\%. Also the nuclear transparency is evident, there is no reduction of particles at the projectile rapidity with increasing centrality. 

	Subtleties of the distributions in A+A collisions become apparent when identified particles are used~\cite{BrahmsPRL93}. In Fig.~\ref{fig:PHOBOSQM05}b, the difference of $dn/dy$ for protons and anti-protons, i.e net-protons is shown as a function of c.m. energy, 
$\sqrt{s_{NN}}=$ 5 (AGS, Au+Au), 17 (SPS, Pb+Pb), 200 (RHIC, Au+Au), $y^{\rm proj}=$1.6, 2.9, 5.4. As $\sqrt{s_{NN}}$ is reduced, 
stopping 
of the participating nucleons 
is indicated by the nuclear fragmentation peak moving from the fragmentation region (not visible for RHIC) to mid-rapidity.

\subsubsection{The Bjorken energy density}
       Another variable, closely related to multiplicity,  is the
transverse energy density in rapidity or $dE_T/dy \sim \mean{p_T}\times dn/dy$, usually measured in calorimeters by summing over all particles on an event in a fixed but relatively large solid angle~\cite{MJTErice03}: $E_T=\sum_i E_i \sin\theta_i$.  
$dE_T/dy$ is thought to be related to the co-moving energy density in
a longitudinal 
expansion~\cite{BjorkenPRD27,PXWP}, and taken by all experiments as a measure of 
the energy density in space $\epsilon$:
\begin{equation}
\epsilon_{Bj}={d\mean{E_T}\over dy} {1\over \tau_F\pi R^2}
 \label{eq:eBj}
 \end{equation}
where $\tau_F$, the formation time, is usually taken as 1 fm/c,
$\pi R^2$ is the effective area of the collision, and $d\mean{E_T}/dy$ is the 
co-moving energy density.

\begin{figure}[!thb]
\begin{center}
\begin{tabular}{cc}

\includegraphics[scale=0.3,angle=-90]{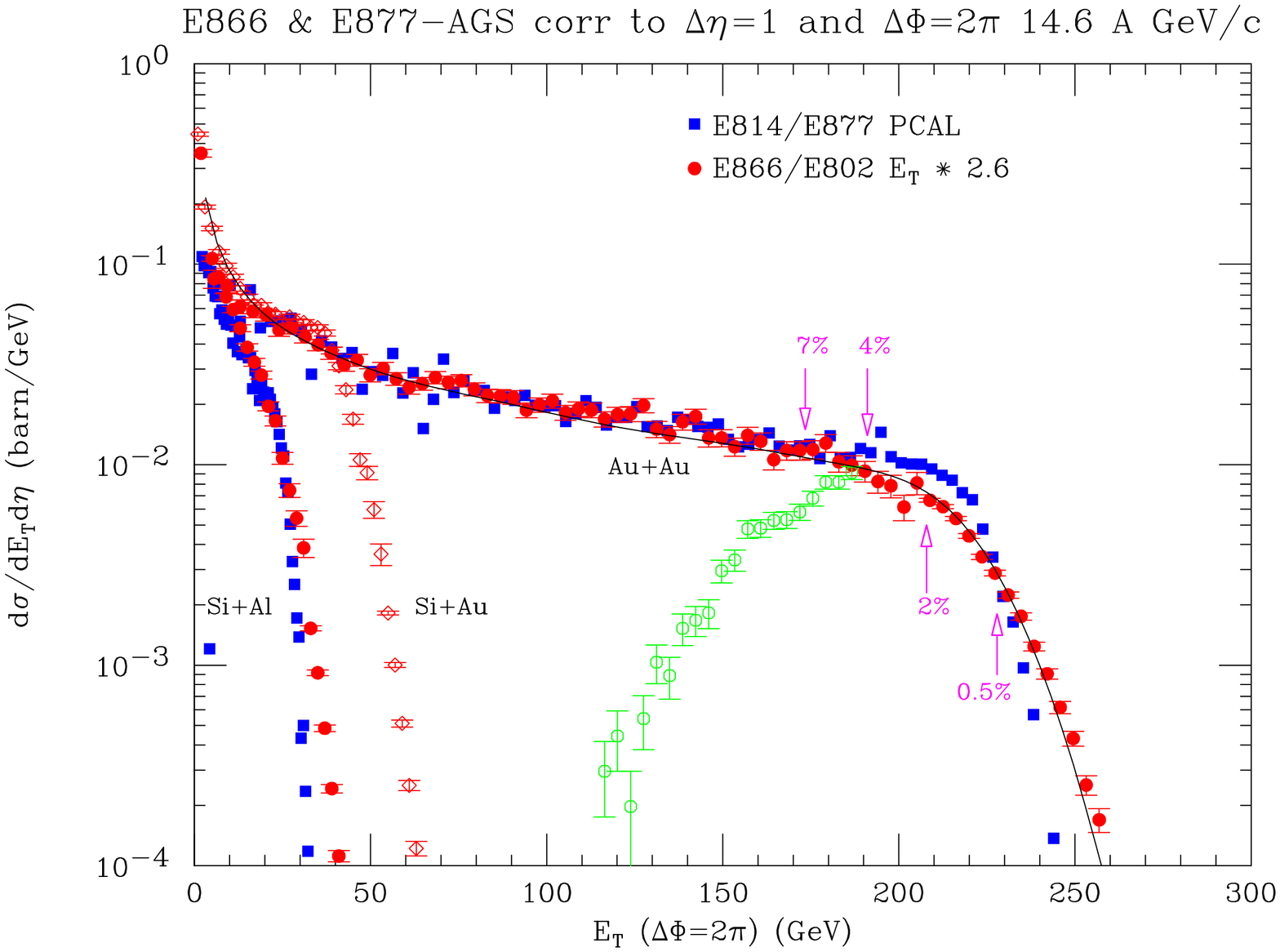}\hspace*{0.25in}&
\includegraphics[scale=0.3,angle=-90]{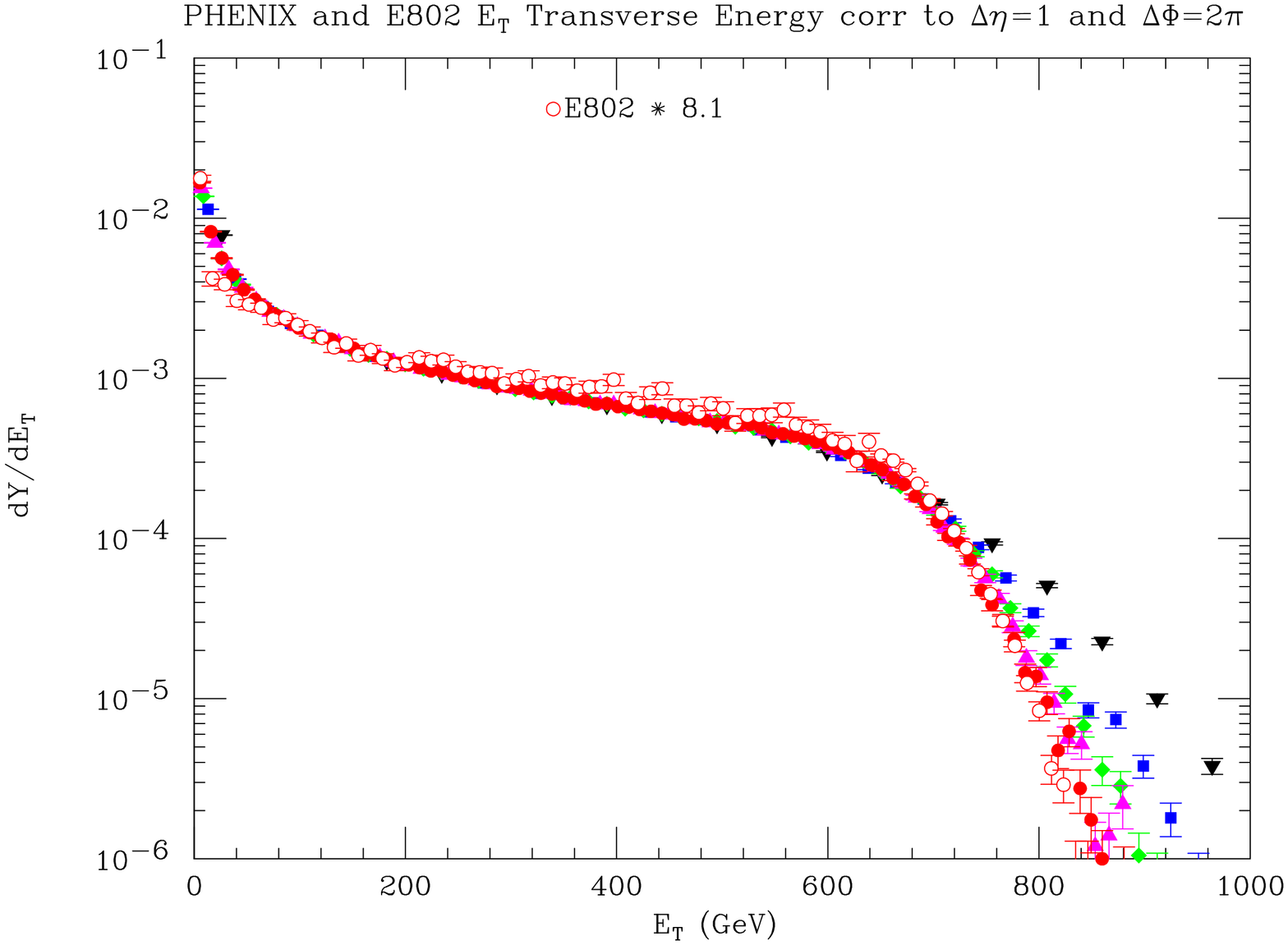}
\end{tabular}
\end{center}\vspace*{-0.25in}
\caption[]{a) (left) Mid-rapidity $E_T$ spectra (corrected to $\Delta\phi=2\pi$, $\Delta\eta=1$) from AGS measurements by E814/E877 \fullsquare~\cite{Barrette93} and E802/E866~\fullcircle\opencircle~\cite{AbbottPRC63} for 14.6$\cdot A$ GeV/c Si+Al, Au and 11.6$\cdot A$ GeV/c Au+Au (corrected to 14.6$\cdot A$ GeV/c). The E802 data have been scaled by the factor indicated to match the E877 measurement. Upper percentiles of the distribution are indicated. The light \opencircle points on the Au+Au data are from a ZDC trigger of (0-7\%) centrality. b) (right) $E_T$ distribution for $\sqrt{s_{NN}}=200$ GeV Au+Au at RHIC~\fullcircle~\cite{PXppg019} together with the E802 measurement~\opencircle from the left panel scaled as indicated~\cite{CalScales}. The solid circles represent a measurement in an azimuthal aperture of $\Delta\phi=5\times 22.4^\circ$, corrected to full azimuth. 
The other solid symbols represent measurements in $\Delta\phi=4,3,2,1 \times 22.4^\circ$, corrected to full azimuth, where the smaller the aperture, the flatter the slope of the data above the knee.       
\label{fig:ETdists}}

\end{figure}
Besides the Bjorken energy density, the importance of $E_T$ distributions in RHI collisions (see Fig.~\ref{fig:ETdists}) is that they are sensitive primarily to the nuclear geometry of the reaction, and hence can be used to measure the centrality of individual interactions on an event-by-event basis~\cite{RHICcent}. Fig.~\ref{fig:ETdists}a shows mid-rapidity $E_T$ distributions for Si+Al, Si+Au and Au+Au at the AGS~\cite{Barrette93,AbbottPRC63}. The increase of $dE_T/d\eta$ with increasing atomic mass of projectile and target is evident. Fig.~\ref{fig:ETdists}b shows that $E_T$ distributions at RHIC and the AGS are the same shape, when scaled to match at the knee, which shows that the shape of the distribution is essentially independent of $\sqrt{s_{NN}}$ and thus dominated by the nuclear geometry of the Au+Au reaction. Above the knee, the slope of the fall-off is sensitive to fluctuations and depends somewhat on the solid angle~\cite{AbbottPRC63,MJTErice03,PXWP}. A consistent evaluation~\cite{PXppg019,CalScales} of the Bjorken energy density for 0--5\% centrality gives $\tau_F \epsilon_{Bj}=$ 1.0, 2.0, 5.4 GeV/fm$^2$ at $\sqrt{s_{NN}}=$ 5, 17, 200 GeV.     
\subsection{Space-time and quantum mechanical issues} 
	 An interaction of two relativistic heavy ions can be viewed initially as the superposition of successive collisions of the participating nucleons in the overlap region (recall Fig.~\ref{fig:collstar})~\cite{Glauber}. Collective effects may subsequently develop due to rescattering of the participants with each other or with produced particles. Conceivably, a cascade of interacting particles could develop. However, the actual situation is considerably more fascinating. 
	 
     Space-time and quantum mechanical issues play an important role in the physics of RHI collisions and the considerations are different in the longitudinal and transverse directions. In QED and QCD, which have the same $1/r$ form of the potential at small distances, two particles of charge $Z_1$ and $Z_2$, velocity $v$, which pass each other at impact parameter $b$ each acquire a momentum $p_T= 2 Z_1 Z_2/b v$ transverse to the direction of motion. Thus large transverse momenta correspond to small impact parameters in both QED and QCD.  
     
     In the longitudinal direction because of the large $\gamma$ factors involved, small excitations/deexcitations of the colliding nucleons, take place over long distances. When a nucleon with momentum, $p_L$, mass, $M$, makes a collision, the only thing it can do consistent with relativity and quantum mechanics is to get excited to a state with invariant mass $M^* \geq M$, with roughly the same energy and reduced $p'_L=p_L -\Delta p_L$, where $\Delta p_L^2 = -\Delta m_T^2\simeq -\Delta m^2$ from Eq.~\ref{eq:3}. By the uncertainty, principle, a distance $\delta z=\hbar/\Delta p_L=\gamma\beta/\Delta m$(=14 fm for $\gamma=10$ and $m=m_{\pi}$)  is required to resolve $\Delta m$. The large $\gamma$ factor in relativistic collisions ensures that the excited nucleons pass through the entire target nucleus before de-exciting into e.g. a nucleon + a pion. Thus, nuclei are transparent to relativistic nucleons; and pions are produced outside the target nucleus, thus avoiding a nuclear cascade. 
     
     For instance, in the collision of a relativistic proton with a 15 interaction-length-thick lead brick, a cascade develops and all particles are absorbed. Nothing comes out the back. By contrast, in the collision of a relativistic proton with a lead nucleus, which is roughly 15 interaction mean-free-paths thick through the center, the (excited) proton comes out the back! This is relativity and quantum mechanics in action.

\subsection{Participant ($N_{\rm part}$) scaling-the Wounded Nucleon Model}
\begin{figure}[htb]
\begin{center}
\begin{tabular}{ccc}
\hspace*{-0.2in}\includegraphics[scale=0.52,angle=-1]{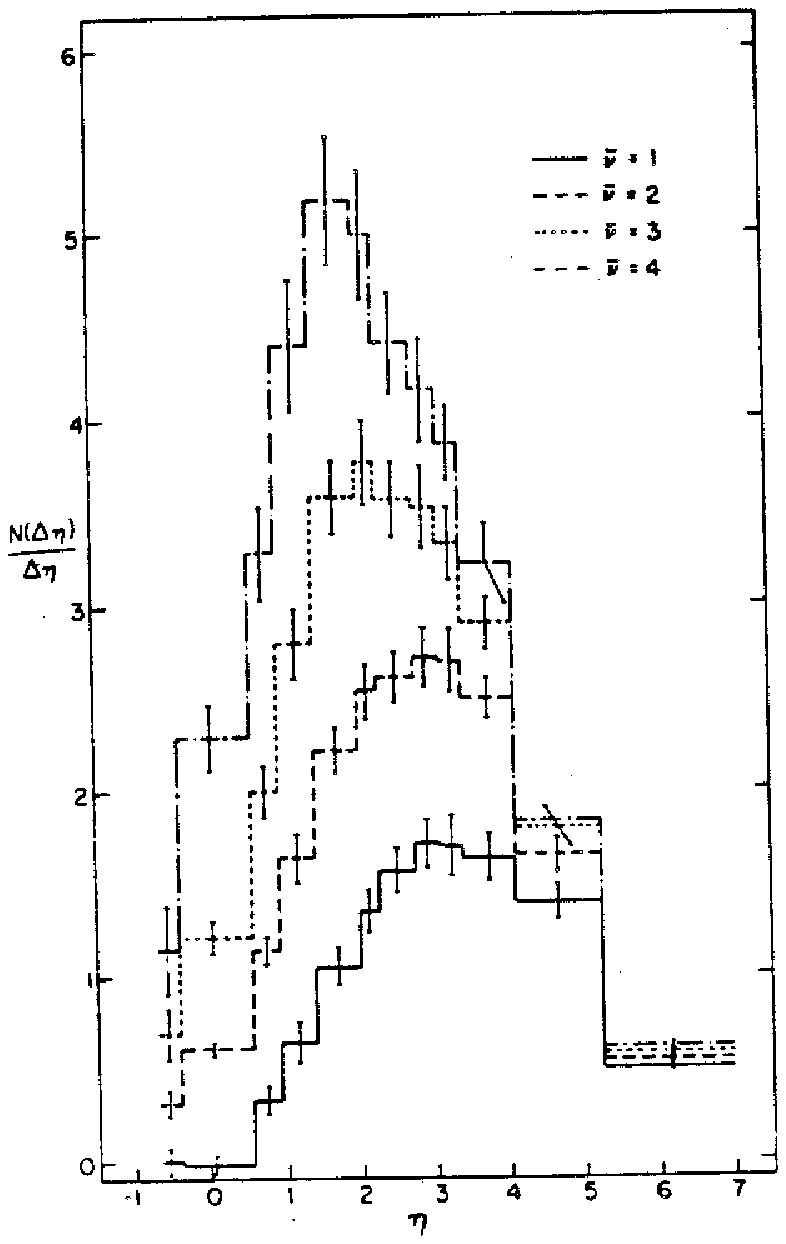}&\hspace*{-0.35in}
\includegraphics[scale=0.90,angle=+0.5]{figs/Busza-200cGC.epsf}&\hspace*{-0.35in}
\includegraphics[scale=0.33,angle=0]{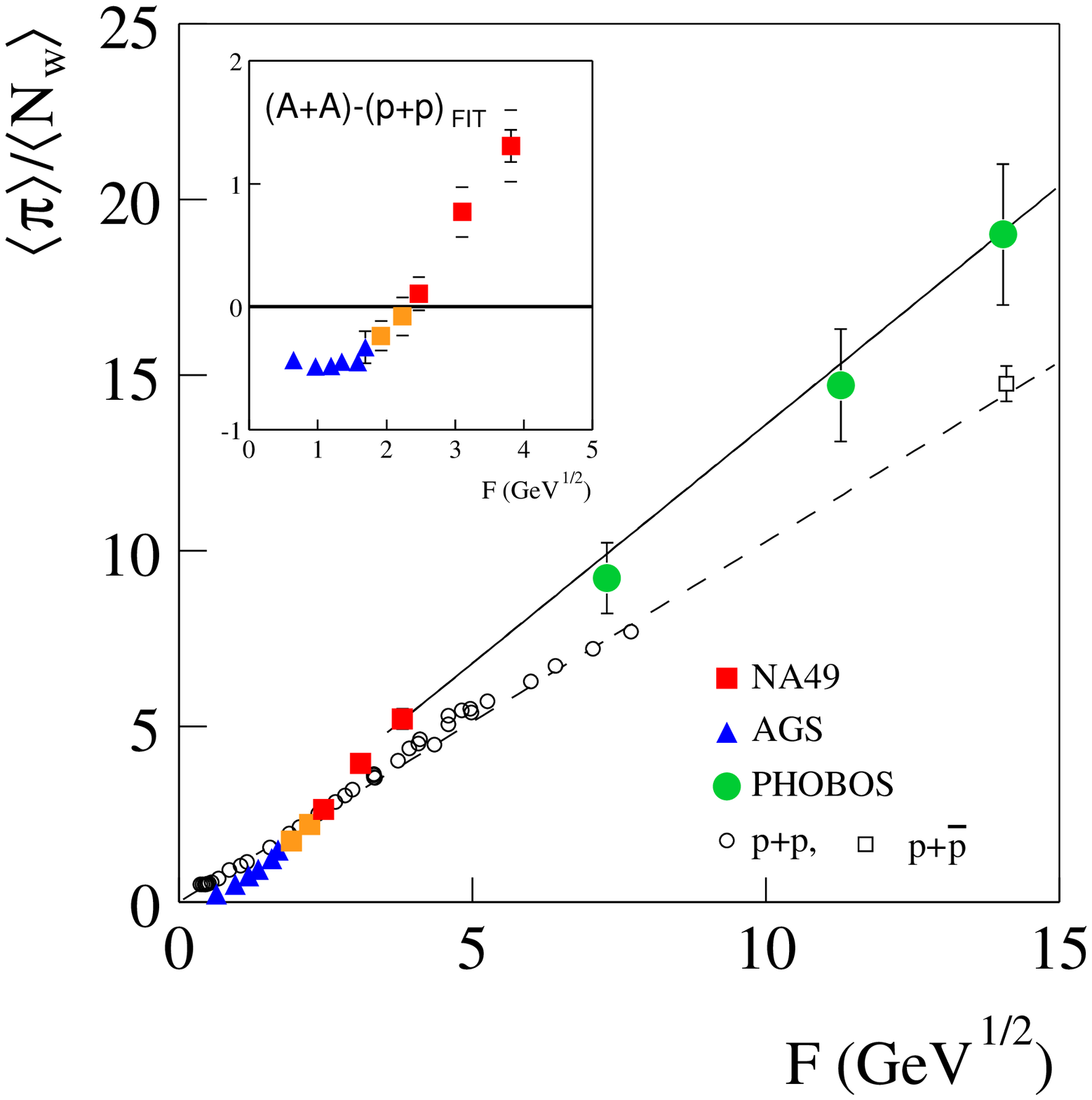}
\end{tabular}
\end{center}
\caption[]{a) (left) Pseudorapidity distributions $dn/d\eta$ of relativistic charged particles for various values of $\bar\nu$ in 200 GeV/c proton-nucleus 
interactions \cite{Busza75}. b) (center) $R_A$ as a function of $\bar{\nu}$ for collisions of 200 GeV $p$, $\pi^{\pm}$ on various nuclei. The solid line shows  the linear trend of the data $\sim$ Eq.~\ref{eq:pa2} \cite{Busza75}. c) (right) Mean number of pions per participant for A+A central collisions and $p(\bar{p})-p$ collisions as a function of $\sqrt{s_{NN}}$ represented by $F=(\sqrt{s_{NN}} -2m_N)^{3/4}/\sqrt{s_{NN}}^{1/4}$ ~\cite{GazdzickiQM04} \label{fig:IJ31}}
\end{figure}

   These concepts of relativity and quantum mechanics are dramatically illustrated by particle production in proton nucleus (p+A) interactions measured at bombarding energy of 200 GeV $\sqrt{s_{NN}}=19.4$ GeV.  When a high energy proton (or any hadron) passes through a nucleus, it can make several successive collisions. However, the charged particle multiplicity density, $dn/dy$, observed in p+A interactions is not simply 
proportional to the number of collisions, but increases much more slowly. 
These features are strikingly illustrated~\cite{Busza75} in 
Fig.~\ref{fig:IJ31}, which is the pseudorapidity distribution of relativistic charged particles ($v/c >  0.85$) produced by 200 GeV proton interactions in 
various nuclear targets. The sizes of the nuclei are discussed in terms of $\bar\nu$, the average number of collisions 
encountered by an incident hadron passing through a nucleus of atomic mass $A$ in which an interaction occurred:
\begin{equation}
   \bar\nu ={A\sigma_{hp}\over \sigma_{hA}},
\label{eq:pa1}\end{equation}
where $\sigma_{hp}$ and $\sigma_{hA}$ are the absorption cross sections for the incident hadron on a nucleon and a nucleus, respectively \cite{Fishbane71}.
 
The most dramatic feature of Fig.~\ref{fig:IJ31}  
is that there is virtually no change in the forward fragmentation region ($\eta > 5.0$) with increasing $A$. By contrast, there is tremendous activity in the target region ($\eta\leq 0.5$). In the mid-rapidity region, $dn/d\eta$ increases with $A$ and the peak of the distribution shifts backwards. The integral of the distribution (the average multiplicity) and $dn/d\eta$ at mid-rapidity show a linear increase with $\bar\nu$ for all targets given by the relation 
\begin{equation}
R_A={\mean{n}_{hA}\over\mean{n}_{hp}}={1+\bar\nu\over 2} \qquad ,
\label{eq:pa2}\end{equation}
which is simply the ratio of the average number of participating nucleons (participants or $N_{\rm part}$) for the 2 cases: $N_{\rm part}=1+\bar\nu$ for the p+A collision and $N_{\rm part}=2$ for a p+p collision. In addition to the quantum and relativistic arguments given above, this result can be explained by the further assumptions that the excited nucleon interacts with the same cross section as an unexcited nucleon and that the successive collisions of the excited nucleon do not greatly affect the excited state or its eventual fragmentation products. \cite{midFrankel} 
This leads to the conclusion that the multiplicity in nuclear interactions should be proportional to the total number of projectile and target participants, rather than to the total number of collisions. This is called the Wounded Nucleon Model (WNM) \cite{WNM}. 

	Interestingly, the WNM works well only at roughly $\sqrt{s_{NN}}\sim20$ GeV where it was discovered. At lower $\sqrt{s_{NN}}$, at the AGS, particle production in p+A and A+A collisions is smaller than the WNM and for $\sqrt{s_{NN}}\geq 31$ GeV is larger than the WNM~\cite{BCMOR,Ochiai,PhobosdA} as emphasized by Gazdzicki~\cite{GazdzickiQM04} ({\em ``the kink"}). However as the effect is seen in p+A collisions, it is not likely to be related to the QGP~\cite{butAQM,Voloshin,De} although the physics is surely very interesting and needs further investigation.

\section{ Signatures of the Quark-Gluon Plasma }

   One of the more interesting signatures proposed for the QGP
is that it could trigger a catastrophic transition from the 
metastable vacuum of the present universe to a lower energy state~\cite{LeeWick,Hut84}, ``a possibility naturally occurring in many spontaneously broken quantum field theories". This type signature gives laboratory directors nightmares, and such a possibility must be ruled out to a high degree of certainty~\cite{Jaffe00}. However this is actually one example of a class of signatures of the QGP called ``Chiral Symmetry Restoration". It is convenient to group QGP signatures into classes according to their sensitivity to expected properties of the QGP~\cite{BMuellerRPP}: (i) signals sensitive to the equation of state; (ii) signals of chiral symmetry restoration; (iii) probes of the response function of the medium (including deconfinement). 

	All authors imagine that the QGP is in a state of chemical and thermal equilibrium at temperature $T_F$ which is established over a short formation time, $\tau_F$, following the collision and that hydrodynamics describes the evolution of the system of quarks and gluons as it expands and cools and eventually freezes out at transition temperature, $T_c$, at time, $\tau_c$, to a system of hadrons which are observable. Another important time is the overlap time of the colliding Lorentz contracted nuclear pancakes as seen in the nucleus-nucleus c.m. system, $\tau_O=2R/\gamma$, where $R$ is the radius of the nucleus in a symmetric A+A collision. For Au+Au or Pb+Pb collisions,  $R\sim 7.5$ fm , $\tau_O=5.6$, 1.7, 0.14 fm/c at $\sqrt{s_{NN}}=$5, 17, 200 GeV.   

\subsection{Degrees of freedom, equation of state}
	It is important to emphasize that the operative degrees of freedom in the QGP are taken to be quarks and gluons and the operative ``charge" is taken be color, unscreened over many units of the characteristic length scale. Since the spin-1 gluons have 8 colors and the spin-1/2 quarks have 3 colors, with  presumably 3 active flavors in the QGP ($u,d,s$), there are many more degrees of freedom in the QGP than in the final state `gas' of hadrons (HG); and this is one of the signatures of the transition from one phase to the other. The QGP is described by the thermodynamic quantities, entropy density, $s$, energy density, $\epsilon$, temperature, $T$, pressure $P$, volume, $V$, and it is important to understand how to relate these quantities to properties of the observable final state particles. As in classical thermodynamics, the equation of state (EOS) is the relationship among the three quantities, P,V,T, which are not independent for a fixed amount of matter. A relation between any two quantities is sufficient to define the EOS. For instance, the large difference in degrees of freedom~\cite{PXWP} between a QGP (taken as a Stefan-Boltzmann gas) and a hadron gas dominated by spin 0 pions (3 d.o.f: $\pi^+, \pi^-,\pi^0$) is shown in Lattice calculations (Fig.~\ref{fig:phaselat}b) :
	\begin{equation}
	\epsilon_{ SB}^{ QGP}=47.5 {\pi^2\over 30} T^4 \qquad
	\epsilon_{ HG}^{ QGP}=3 {\pi^2\over 30} T^4 \qquad .
	\label{eq:epsSB} 
	\end{equation}
However, it is not obvious how this effect could be measured---possibly by evidence of discontinuities or rapid changes as a function of an observable~\cite{VanHove8283,RajaWilcz-crit} or by fluctuations due to passing through the phase transition~\cite{KrishnaShuryak,egPXncfluct}. (Recall that $T\leftrightarrow \mean{p_T}$, $s \leftrightarrow dn/d\eta$ and $\epsilon\leftrightarrow dE_T/dy$). 

	In general, fluctuations are a standard probe of thermodynamic systems and phase transitions~\cite{MGSM92,EVS89,Krishna99,HH01} and have been extensively studied at the SPS and RHIC.  However, no strong signals have yet been observed in relativistic A+A collisions.~\cite{egCFRNC05}. 

	The volume of a thermalized source is thought to be measurable by identical
particle interferometry using the GGLP effect~\cite{GGLP60}, commonly called Bose-Einstein correlations since the measurement is predominantly performed with identical bosons ($\pi^{\pm},K^{\pm}$)~\cite{WeinerHBTreview,Femtoscopy}. Also, collective effects, where all partons (or particles) moving in the same direction have a common (flow) velocity, are sensitive to the EOS.~\cite{DLLsci,Bhalerao05}  

\subsubsection{Thermal Equilibrium}
One of the best probes of thermal equilibrium is thermal lepton pair production~\cite{KKMM86} or thermal photon production~\cite{ShuryakPLB78}. This is effectively the `black-body radiation' of the QGP and should follow a Boltzmann distribution in the local rest frame~\cite{CooperFrye}:
\begin{equation}
{{d^2\sigma} \over {dp_L p_T dp_T}}={1\over {e^{E/T} \pm 1}}\sim e^{-E/T} \qquad .
\label{eq:boltz}
\end{equation}
Since $p_T dp_T=m_T dm_T$ and $E=m_T \cosh y$, a signal of thermal production is that the $p_T$ and mass dependence of the cross section are not independent
but depend only on the transverse mass $m_T$. This means that at any
fixed value of $m_{ee}$ and rapidity, the $\mean{p_T(m_{ee})}$ is linearly proportional to 
$m_{ee}$. As neither the photons nor $e^+ e^-$ pairs are strongly interacting, they emerge from a QGP or hadronic system without interacting and thus are sensitive to the entire thermal history of the system, especially the early stage where the QGP should be dominant. In principle, the initial temperature of the system, $T_i$ can be determined from the rate of thermal photon or $e^+ e^-$ pair production~\cite{ShuryakPLB78,KKMM86}. However, even though radiation from the QGP comes from the interaction of `massless' quarks and gluons: 
$$ q\,\bar{q}\rightarrow e^+ e^- \qquad g\,q\rightarrow \gamma\, q \qquad q\,q\rightarrow q\,q\,\gamma \qquad {\rm etc,} $$ 
the latest `state-of-the-art' calculations~\cite{Turbide,AMY04} indicate that for the same temperature, the thermal photon production rates in a QGP or a hadron gas are very similar~\cite{ReygersHP04}.  
 It is important to distinguish thermal photons (lepton pairs), which dominate at low $p_T$ (mass), near $T_c$,  and are exponential, from the high $p_T$ (mass) direct photons (Drell-Yan pairs)  produced by hard-scattering in some of the same subprocesses,  $q\,g\rightarrow\gamma\,q$ ($q\,\bar{q}\rightarrow e^+ e^-)$, which follow a power-law. Hard-scattering (see below) is not relevant for thermal equilibrium.

\subsection{Chiral Symmetry Restoration}
    In the QGP~\cite{PisarskiWilczek,BMuellerRPP} 
    the light quarks are expected to have the same mass or be massless, thus exhibiting exact ``chiral symmetry". These are the so-called ``current quarks'' of DIS. This is quite different from the normal vacuum state of QCD~\cite{LeeWick} in which chiral symmetry is ``spontaneously broken" and the quarks are massive, called ``constituent quarks". This leads to several interesting effects which might be measurable. 
\subsubsection{Strangeness Enhancement} 
\label{sec:strange}
  In the QGP, the gluons, quarks and anti-quarks 
continuously react with each other via the QCD subprocesses:
$$g\,g\rightarrow q\,\bar{q} \qquad q\,\bar{q}\rightarrow g\, g \qquad 
 q\, \bar{q}\rightarrow q'\, \bar{q}' $$
where $q'$ represents a different flavor quark $(u,d\, {\rm or}\, s)$. After 
several interactions have taken place, the reaction rates and the abundances
of the gluons and the different flavor quarks (and anti-quarks) will become
equilibrated, so that they no longer change with time. This is called
chemical equilibrium. Since the masses of $u,d,s$ quarks should be 
nearly 
degenerate if chiral symmetry is restored, they should reach 
nearly 
the same equilibrium value so that the strange quarks $s,\bar s$ should have 
nearly 
the same abundance as the $u,\bar u$ and $d, \bar d$ in the gluon plasma. Thus strangeness should be enhanced in the QGP~\cite{MuellerRafelski,KMR86,LRIJMPE} compared to p-p collisions where the abundance of strange particles, e.g. $K^{\pm}$, is much below their thermal equilibrium value relative to $\pi^{\pm}$. In the baryon-rich plasma, the $s, \bar s$ will be enhanced compared to $u$ and $d$, since $u, \bar u$ and $d, \bar d$ are ``Pauli'' blocked by valence $u$ and $d$ quarks.

   Since the quarks and gluons in a QGP can not be observed directly, the principal probe of chemical equilibrium and strangeness enhancement in the QGP is the particle composition of observed hadrons. For instance, the abundance of strange mesons and baryons as well as anti-baryons should be quite different in a QGP than in a hadron gas or in an ordinary nuclear collision. 
   \subsubsection{Disordered Chiral Condensate} An interesting anomaly possible when the medium returns to the normal QCD ground state from the chirally symmetric QGP state is that ``misaligned" QCD vacuum regions might occur such that instead of emitting $\pi^0$ and $\pi^+\pi^-$ with a binomial distribution with  probability of 1/3 for the fraction $R=\pi^0/(\pi^0 +\pi^+ +\pi^- )$ of $\pi^0$ emitted, individual ``misaligned" domains could emit, for instance, only $\pi^0$ or only $\pi^+ \pi^-$ pairs~\cite{RajaWilczDCC}. This would give a distribution of the form $P(R)=1/(2\sqrt{R})$, which has the same mean value, $\mean{R}=1/3$ but a totally different event-by-event distribution. Incredibly, so far there is only one (null) measurement~\cite{WA98chgneutfluct} (at the SPS)  due to the inherent difficulty of reconstructing $\pi^0\rightarrow \gamma+\gamma$ at low $p_T$. 

   \subsubsection{mass-shifts, branching ratio changes} 
   While it is clear how chiral symmetry breaking and restoration affect the quarks, it is not so clear how they affect the particles we observe which are by definition on their mass-shell, with their well defined masses and are probably formed during the hadronization process, well after the QGP has cooled.  Pisarski~\cite{Pisarski82} has suggested that if the temperature for chiral symmetry restoration, $T_{ch}$ is greater that the QGP critical temperature, $T_c$, then all hadrons except pions would unbind at $T\geq T_c$, with pions massive quarks and gluons in coexistence until $T\geq T_{ch}$, when all quarks and gluons become massless. This might mean a state of constituent quarks in the region $T_c\leq T\leq T_{ch}$. Also, particles which are formed and which decay to leptons or photons in this phase can be detected, since the leptons and photons do not interact with the QGP and simply emerge unscathed. As the width of the $\rho$ is 150 MeV (lifetime 1.3 fm/c), this might give rise to~\cite{Pisarski82} an ``extraordinary signal" of a thermal $\rho^0\rightarrow$~dilepton peak ``quite distinct from the familiar $\rho^0$ peak". To a lesser extent, a similar effect is predicted for the $\phi$ meson~\cite{ShuryakLissauer} which has a width of 4.3 MeV (lifetime $\sim 46$~fm/c). The $\phi$ mass is just above the $K^+ K^-$ mass but its width is dominated by $K^+ K^-$ (or $K^0_L K^0_S$) decay and therefore is very sensitive to relative shifts of its mass with respect to the masses of the Kaons. The width of the $\phi$ is predicted to increase by a factor of 2-3 due to this effect, which may be detectable directly by high resolution measurements of $\phi\rightarrow$~dileptons, or by an increase of the observed $\phi\rightarrow e^+ e^-/K^+ K^-$ branching ratio from the standard (vacuum) value. 

\subsection{Probes of the response function of the medium including deconfinement}
\subsubsection{Deconfinement}
   Since 1986, the `gold-plated' signature of deconfinement was thought to be $J/\Psi$ suppression. Matsui and Satz proposed~\cite{MatsuiSatz86} that $J/\Psi$ production in A+A
collisions will be suppressed by Debye screening of the quark
color charge in the QGP. The $J/\Psi$ is produced when two gluons
interact to produce a $c, \bar c$ pair which then resonates to form the
$J/\Psi$. In the plasma the $c, \bar c$ interaction is screened so that the 
$c, \bar c$ go their separate ways and eventually pick up other quarks at
the periphery to become {\it open charm}. Due to the fact that it takes time for the initial $c, \bar c$ pair to form a $J/\Psi$, for the quarks to separate to the correct Bohr orbit, the $J/\Psi$ suppression should vanish characteristically with increasing $p_T$ as the $c,\bar{c}$ pair leaves the medium before the $J/\Psi$ is formed or screened~\cite{BlaizotOllitraultPLB199}. $J/\Psi$ suppression would be quite a spectacular 
effect since the naive expectation was that $J/\Psi$ production,
due to the relatively large $\sim 1.5$ GeV scale of the charm quark mass, should behave like a pointlike process, proportional to $A\times A$ in an A+A collision, and thus
would be enhanced relative to the total interaction cross section,
which increases only as $A^{2/3}$.  

\begin{figure}[!thb]
\begin{center}
\begin{tabular}{cc}

\includegraphics[scale=0.6,angle=0]{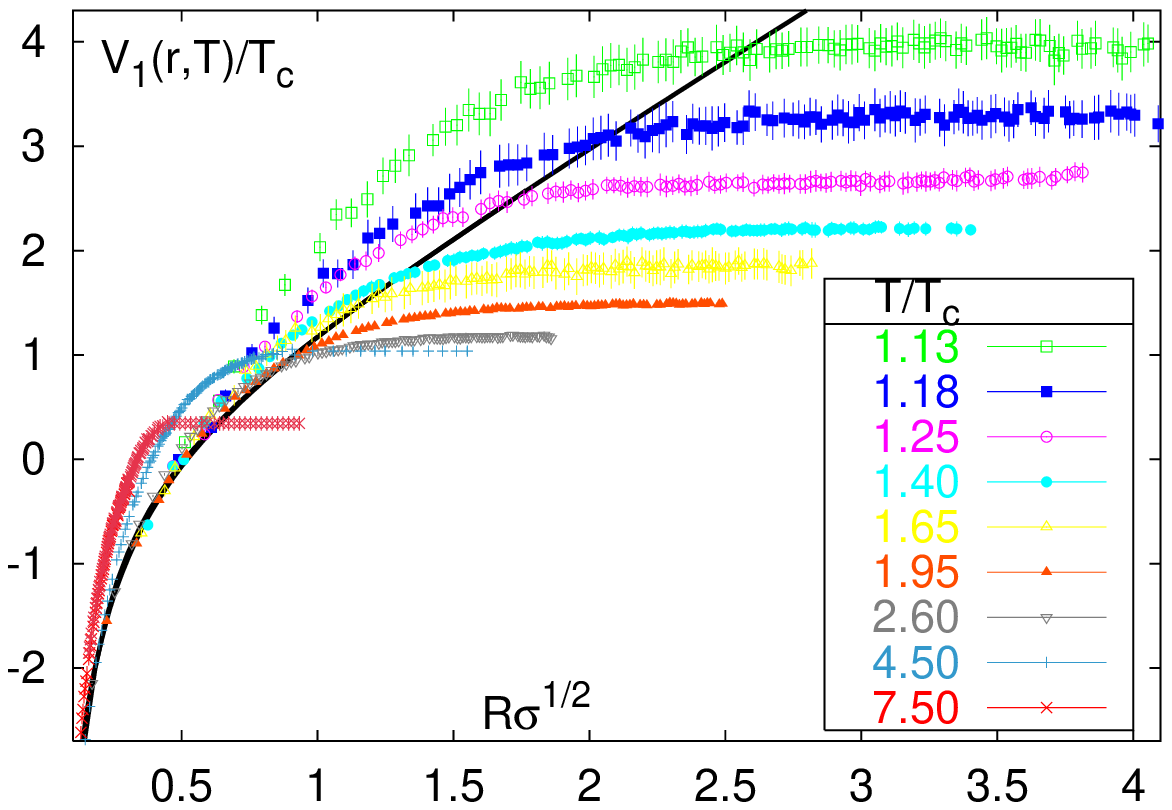}&
\includegraphics[scale=0.6,angle=0]{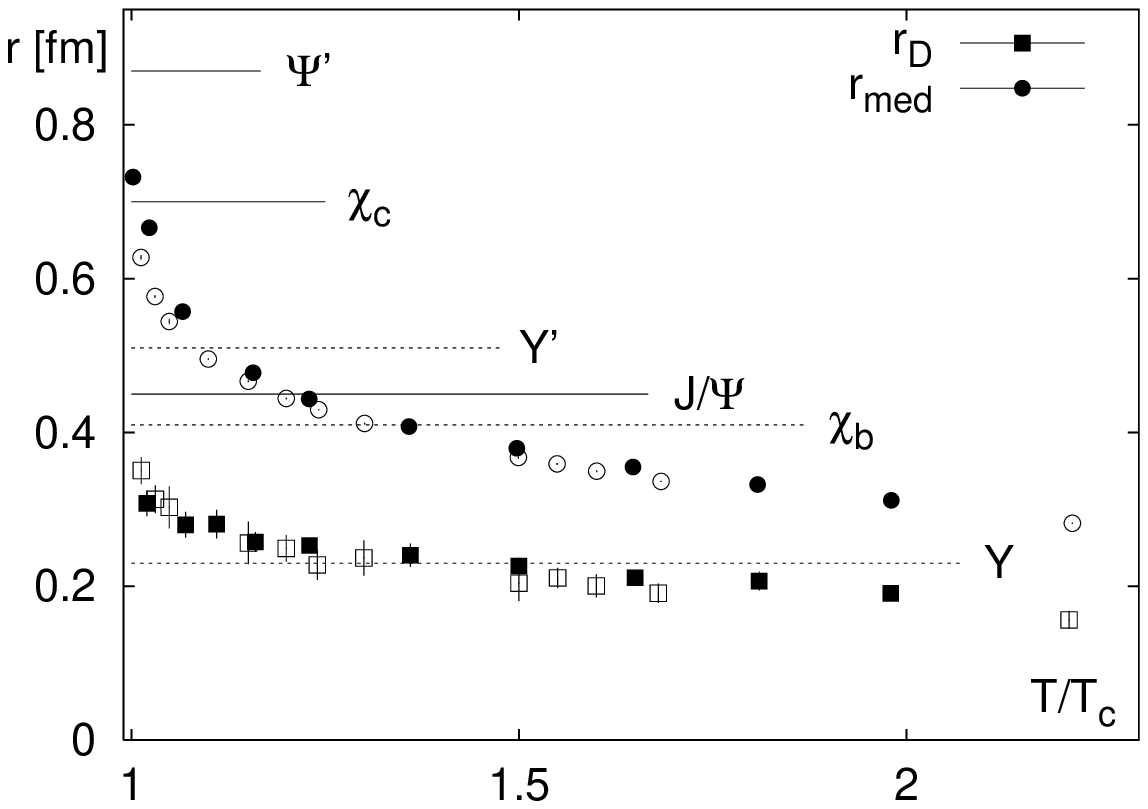}

\end{tabular}
\end{center}\vspace*{-0.25in}
\caption[]{ a) (left) Lattice gauge calculations of heavy quark potential $V(r)$ as a function of $T/T_c$~\cite{Kaczmarek04}. The solid line is the normal ($T=0$)  potential; b) (right) Estimate~\cite{KarschHP05} of $r_{med}$, the distance beyond which the force between a static quark anti-quark pair is strongly modified by temperature effects, compared to the Debye screening radius, $r_D=1/\mu$, and the mean square charge radii of $c\bar{c}$ and $b\bar{b}$ bound states (lines). Note that the actual melting points of the $\Psi^{'}$, $\chi_c$ and $J/\Psi$ are no longer expected to be $\sim$0.2,0.7 and 1.1 $T_c$ as in the calculation illustrated~\cite{KarschHP05}, but rather 1.1, 1.1 and $\sim 2\,T_c$ by more recent calculations~\cite{Wong05,Alberico05,Satz05}. \label{fig:latjpsi} }

\end{figure}

    The screening of the coulomb-like QCD potential for heavy quarks is supported by lattice gauge calculations~\cite{Kaczmarek04,KarschHP05} (Fig.~\ref{fig:latjpsi}).  
   With increasing temperature, $T$, in analogy to increasing $Q^2$, the strong coupling constant $\alpha_{s}(T)$ becomes smaller, reducing the binding energy,  and the string tension, $\sigma(T)$, becomes smaller, increasing the confining radius, effectively screening the potential\cite{SatzRPP63}: 
  \begin{equation}
  V(r)=-{4\over 3}{\alpha_{s}\over r}+\sigma\,r \rightarrow 
-{4\over 3}{\alpha_{s}\over r} e^{-\mu\,r}+\sigma\,{{(1-e^{-\mu\,r})}\over \mu}
\label{eq:VrT}
\end{equation} 
where $\mu=\mu(T)=1/r_D$ is the Debye screening mass.~\cite{SatzRPP63}. When the potential in Fig.~\ref{fig:latjpsi}a becomes constant with increasing radius, the binding force vanishes as illustrated~\cite{KarschHP05} in Fig.~\ref{fig:latjpsi}b. 

	Since the radii of higher excitations of the bound $c\bar{c}$ (charmonium) and $b\bar{b}$ states increase with their masses and the binding energies decrease, the states should melt sequentially with increasing $T$ from the highest mass to the lowest in a given family ($c\bar{c}$, $b\bar{b}$). Since 60\% of the observed $J/\Psi$ are directly produced, 30\% are from $\chi_c$ decay and 10\% from $\Psi^{'}$ decay, this should give a characteristic sequential nature to $J/\Psi$ suppression with increasing centrality (presumably increasing $T$) of an A+A collision---first the 10\% of the $J/\Psi$ from the $\Psi^{'}$ are suppressed, then the 30\% from the $\chi_c$ are suppressed and eventually the other 60\% vanish when the direct $J/\Psi$ melts.  
Such an effect was apparently observed at the SPS~\cite{NA50PLB450} and wonderful models were woven to explain the results~\cite{SatzQM99,NardiHP05}. However, recent measurements from RHIC (see below), and an increase in the expected dissociation temperatures for the charmonium states~\cite{Wong05,Alberico05,Satz05} put the whole concept of $J/\Psi$ suppression as a probe of deconfinement into question. 
\subsection{Jet Quenching}\label{sec:quenching}
\begin{figure}[!thb]
\begin{center}
\begin{tabular}{cc}

\includegraphics[scale=0.44,angle=0]{figs/dAuAuAu.epsf}&
\includegraphics[scale=0.5,angle=0]{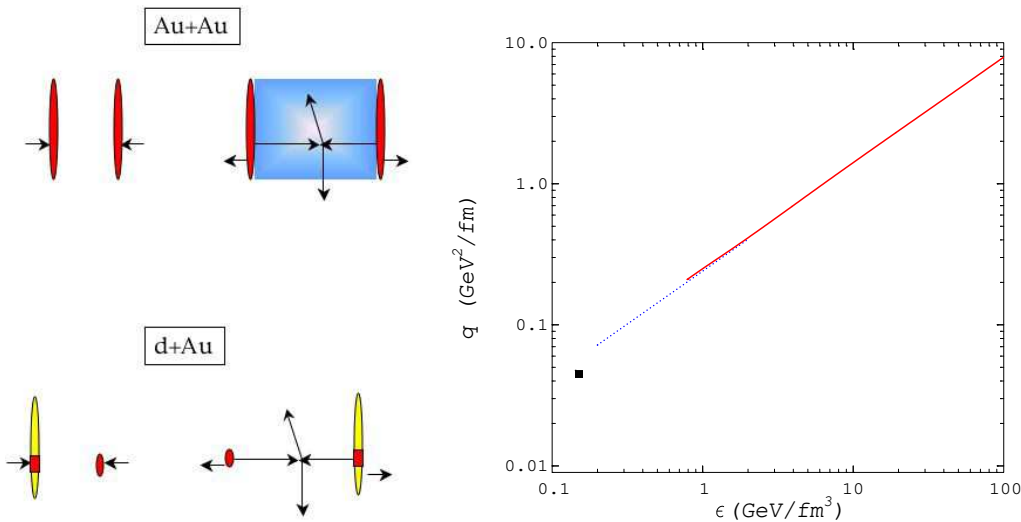}

\end{tabular}
\end{center}\vspace*{-0.25in}
\caption[]{ a) (left) Schematic diagram of hard scattering in A+A and d(p)+A collision. For Au+Au, high $p_T$ partons from scattering when the nuclei overlap, emerge sideways through the medium formed. For d+Au no medium is formed and the outgoing partons travel in the vacuum until they fragment. b) (right) Transport coefficient $\hat{q}$ as a function of energy density for different media~\cite{BaierQM02}: cold nuclear matter (\fullsquare), massless hot pion gas (dotted) and `ideal' QGP (solid curve)  \label{fig:dAuAuAu} }

\end{figure}

   A new tool for probing the color response function of the medium was developed in the early 1990's~\cite{Gy1,Gy2,Gy3} and given a firm basis in QCD~\cite{BDPS,Zakh1,BDMPS2,BSZARNS00,BaierQM02} just before RHIC turned on. In the initial collision of the two nuclei in an A+A collision when the Lorentz contracted nuclei are overlapped, hard scatterings can occur which produce pairs of high $p_T$ outgoing partons with their color charge fully exposed as they travel through the  medium before they fragment to jets of particles. If the medium also has a large density of color charges, then the partons are predicted to lose energy by `coherent' (LPM) gluon bremsstrahlung which  is sensitive to the properties of the medium. This leads to a reduction in the $p_T$ of both the partons and their fragments and hence a reduction in the number of partons or fragments at a given $p_T$, which is called jet quenching. The effect is absent in p+A or d+A collisions due to the lack of a medium produced (see Fig.~\ref{fig:dAuAuAu})

   The screened coulomb potential (Eq.~\ref{eq:VrT}) resulting from the thermal mass $\mu$ acquired by gluons in the medium modifies the elastic scattering of outgoing partons such that the average 4-momentum transfer per collision is finite: 
\begin{equation}
\mu^2=\mean{q^2}\equiv \frac { \displaystyle\int q^2 {{d\sigma_{el}}\over dq^2} dq^2}{ \displaystyle\int  {{d\sigma_{el}}\over dq^2} dq^2}=\rho\, \lambda_{\rm mfp} { \displaystyle\int q^2 {{d\sigma_{el}}\over dq^2} dq^2} \equiv\hat{q}\,\lambda_{\rm mfp} \qquad .  
\label{eq:mqsq}
\end{equation}          
The integral in the denominator is $\sigma_{el}\equiv 1/\rho\, \lambda_{\rm mfp}$, where $\rho$ is the density of scatterers and $\lambda_{\rm mfp}$ is the mean free path for elastic scattering.  The `transport coefficient' $\hat{q}$ is the mean 4-momentum transfer/collision expressed as mean 4-momentum transfer per mean free path, so that the mean 4-momentum transfer for length $L$ in the medium is 
$\mean{q(L)^2}=\hat{q}\,L=\mu^2\,L/\lambda_{\rm mfp}$. Any gluon radiation that is contained within the cone defined by the transverse momentum, $\mean{q(L)}$, is coherent over all the scatterings (LPM effect) so that the energy loss of a parton due to gluon bremsstrahlung per unit length ($x$) of the medium takes the form~\cite{BaierQM02}:
\begin{equation}
{-dE \over dx }\simeq \alpha_s \mean{q(L)^2}=\alpha_s\, \hat{q}\, L=\alpha_s\, 
\mu^2\, L/\lambda_{\rm mfp} \qquad .
\end {equation}
 Thus the total energy loss in the medium goes like $L^2$~\cite{BDMPS2}. It is also important to note that the screening mass, $\mu^2$, plays the role of $t_{\rm min}=q^2_L$ for the gluon bremsstrahlung in the medium~\cite{BDPS}. 
 
 The solution for the energy loss in pQCD for a static medium can be calculated analytically in only two limits, many soft scatterings (like multiple coulomb scattering) and 1 hard scattering, so that many treatments have been developed in varying degrees of approximation in the opacity, $L/\lambda_{\rm mfp}$  (like $\bar{\nu}$, Eq.~\ref{eq:pa1}), the average number of partonic collisions in the medium~\cite{GLV,Wang,Wiedemann}. 
Also, a formulation convenient for numerical simulations has been given~\cite{Zakharov04}.   
However, the medium is not static but is expanding, flowing, etc, which gives rise to further complications. 
It is generally agreed that the energy loss measures $\hat{q}$, which depends on the color charge density in the medium within the screening radius around the probe  (Fig.~\ref{fig:dAuAuAu}b). However it is not sensitive to the vanishing of the string tension beyond $r_D=1/\mu$, which represents true `extended' deconfinement~\cite{Weiner05}, because the formation time to resolve a pion in the fragmentation process means that quarks do not fragment to on-the-mass-shell pions until they are well outside medium where ordinary confinement applies.  
 \section{The Search for the QGP}
   In the early 1980's, before measurements with relativistic heavy ions of c.m. energy $\sqrt{s_{NN}}\gg 2$ GeV ($\sim 2$ nucleon masses) became available, it was expected that the search for the QGP would proceed by looking for its predicted properties, or by the discovery of ``anomalies" or discontinuities as a function of some experimental observable in A+A collisions~\cite{VanHove87}. However, it was also realized that ``systematic studies and comparison of pp, pA and AA data are equally important to understand basic processes hiding behind the phenomena observed in AA collisions"~\cite{NagamiyaStMalo}. Thus it was not unreasonable to expect a few surprises: 
   \begin{itemize}
\item Many of the predicted properties will be found, but will not be the QGP
\item The QGP will have a few unpredicted or unexpected properties
\item  The search will uncover many unexpected backgrounds and new properties of pp and AA collisions, some of which may be very interesting pheonmena in their own right.
\end{itemize}   
  
	Consequently, before proclaiming a perceived anomaly in A+A collisions as the QGP, it is imperative to learn and understand the 
``ordinary physics" of relativistic nuclear collisions as well as p-p and p+A collisions. 
As emphasized by Van Hove~\cite{VanHove87} much could be gained in 
this regard by studying the history of strong interactions for the past 30 
years. An understanding of the properties of high energy p-p 
and p+A interactions is vital to the ability to distinguish the 
``ordinary physics'' of relativistic nuclear interactions from the 
signatures of production of a new phenomenon like the QGP. 
It is also  possible that the ``ordinary physics'' may in itself be quite interesting. Furthermore, by 
studying the hard won knowedge of the past, one might hope to avoid some 
pitfalls in the future. 

\subsection{p-p physics}
The theory and phenomenology of secondary particle production in 
ultrarelativistic hadron-hadron collisions originated with the study 
of cosmic rays~\cite{Cocconi1958}. 
	It is hard to overstate the importance of the fundamental observation 
made by cosmic ray physicists, that the average transverse momentum of 
secondary particles is limited to $\sim$0.5 GeV/c, independent of the 
primary energy~\cite{Nishimura,Cocconi1958}. Cocconi, Koester and Perkins~\cite{Cocconi1961} proposed the prescient empirical formula for the transverse momentum spectrum of meson production: 
\begin{equation}
{d\sigma \over {p_T dp_T}}=A e^{-6p_T} \qquad , 
\label{eq:CKP}
\end{equation}
where $p_T$ is the transverse momentum in GeV/c and 
$\langle p_T\rangle=2/6=$0.333 GeV/c. 	
The observation by Orear~\cite{Orear1964} that large angle p-p elastic 
scattering at AGS energies (10 to 30 GeV incident energy) ``can be fit 
by a single exponential in transverse momentum, and that this exponential 
is the very same exponential that describes the transverse momentum 
distribution of pions produced in nucleon-nucleon collisions", led to 
the interpretation~\cite{Hagedorn} that particle production was ``statistical'', 
with Eq.~\ref{eq:CKP} as a thermal  Boltzmann spectrum, with 1/6=0.167 GeV/c 
representing the ``temperature", $T$ at which the mesons or protons 
were emitted~\cite{ErwinLanderKo}.

It was natural in a thermal scenario~\cite{Hagedorn1970, Barshay1972} 
to represent the invariant cross section, as a function of the longitudinal 
variable rapidity, $(y)$, in terms of the transverse mass, $m_T=\sqrt{p_T^2+m^2}$, 
with a universal temperature parameter $T$. This description nicely explained the 
observed successively increasing $\langle p_T\rangle$ of $\pi$, $K$, $p$, 
with increasing rest mass (e.g. see Fig.~\ref{fig:PXpTspectra})  
and had the added advantage of explaining, by the simple factor 
$e^{-6(m_K-m_{\pi})}\sim$ 11\% , the low value of $\sim$10\% observed for the 
$K/\pi$ ratio at low $p_T$ at ISR energies ($\sqrt{s}\sim 20-60$ GeV). 
     
The introduction of the constituent quark model, which used SU(3) symmetry to 
explain the hadron flavor spectrum and the static properties of hadrons~\cite{MGM64,Zweig}, 
as a dynamical model to calculate the flavor dependence of identified hadrons in
soft multiparticle production,~\cite{Anisovich74} together with the inclusive 
reaction formalism~\cite{FeynmanScaling,LimFrag,AMueller}, which showed that 
there was much to be learned by simply measuring a single particle spectrum,  
brought the study of identified inclusive single particle production into the 
mainstream of p-p physics. However, in the constituent quark model, the 
``suppression" of strange quarks, evident from the small $K/\pi$ ratio, 
was not explained, but simply quantified~\cite{Anisovich73} by a parameter 
$\lambda=2s\bar{s}/(u\bar{u}+d\bar{d})$, which represents the ratio of the numbers of $s\bar{s}$ pairs produced to the average of $u\bar{u}$ and $d\bar{d}$ pairs. 

\subsubsection{Thermodynamics and Hydrodynamics in p-p collisions}  
	One of the burning issues in the early 1950's in soft multiparticle production was whether more than one meson could be produced in a single nucleon-nucleon collision 
(``multiple production'') or whether the multiple meson production observed in nucleon-nucleus interactions was the result of several successive nucleon-nucleon collisions, with each collision producing only a single meson (``plural
production") \cite{Camerini52}. The issue was decided when multiple meson production was first observed in 1954 in collisions between neutrons with energies up to 2.2 GeV, produced at the Brookhaven Cosmotron, and protons in a hydrogen filled cloud chamber \cite{Fowler1954,HagedornComment}. 

The observation of multiparticle production occurring not only in 
nucleon-nucleus, but also in nucleon-nucleon collisions motivated Fermi and 
Landau to develop the statistical~\cite{Fermi} and hydrodynamical~\cite{Landau} 
approach to multi-particle production. Belenkij and Landau observed that although the statistical model of Fermi is sufficient to describe the particle numbers in terms of only a temperature and a chemical potential, this model has to be extended to hydrodynamics, when particle spectra are also considered. To quote Carruthers~\cite{CarruthersPRD8}, ``One envisions a thin slab of hot hadronic matter in thermal equilibrium just after the collision; strictly speaking this is a `head-on' collision picture, but one can imagine that a fraction of the collision products are described by this initial condition while leading particles carry away a sizeable fraction (of the order of 1/2) of the energy and perhaps most of the angular momentum. In Landau's model the particles do not jump right out into phase space (which leads to too many heavy particles in Fermi's picture), but undergo an expansion phase before breaking up. The force responsible for the expansion is large in the longitudinal direction (the pressure gradient is mainly in the longitudinal direction because of the Lorentz contraction) and provides a natural dynamics for the well-known transverse-longitudinal asymmetry of secondary momenta." In the light of present-day discussions, it is worthwhile continuing the direct quote: ``The detailed calculations are made on the basis of the classical relativistic hydrodynamics of a perfect fluid, whose energy-momentum tensor $T^{\mu\nu}$ is:
\begin{equation}
T^{\mu\nu}=(\epsilon +P) u^{\mu}u^{\nu}-g^{\mu\nu} P \qquad ,
\label{eq:pete11}
\end{equation}
where $u^{\mu}(x)[=(u_0, \vec{u})$ with $u_0=\sqrt{1+\vec{u}\cdot\vec{u}}\,]$ is the four-velocity field and $\epsilon$, $P$ are the scalar densities of energy and pressure. The hydrodynamical equations are simply 
$\partial^{\mu}T_{\mu\nu}=0$. In order to solve these equations one needs in addition an equation of state, taken by Landau to be $P=\epsilon/3$ which is characteristic of black body radiation. It is not surprising that this is equivalent to the vanishing of the trace of $T^{\mu}_{\nu}$, $T^{\mu}_{\mu}=\epsilon-3P$." The expansion phase is taken as scale-invariant ($\vec{u}=\vec{x}/t$)~\cite{OllitraultPRD46} and adiabatic (conserves entropy).  ``The details of the final stage, in which the fluid breaks up into asymptotic states of the system, remain quite obscure, as in the parton model." 

    The Landau hydrodynamical model was never popular with particle physicists because it seemed not relevant to small systems such as p-p collisons and because it predicted a gaussian rapidity distribution, whereas a flat rapidity distribution---the rapidity plateau---had been discovered at the CERN ISR~\cite{FlatRapidity}. 
        \begin{figure}[!hbt]
\begin{center}
\begin{tabular}{cc}

\includegraphics[scale=0.4,angle=0]{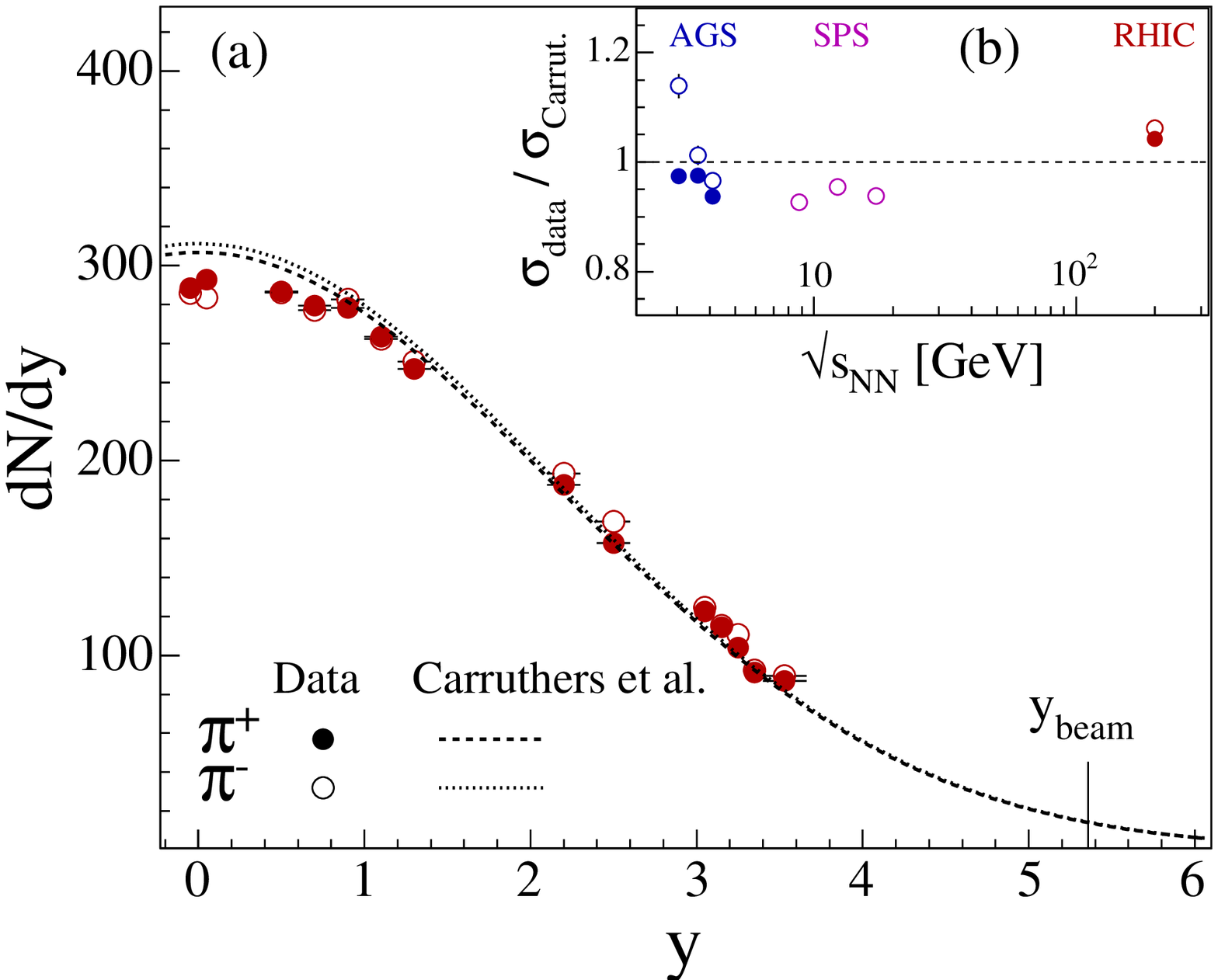}&
\includegraphics[scale=0.4,angle=0]{figs/ISRt.epsf}
\end{tabular}
\end{center}\vspace*{-0.25in}
\caption[]{a) Comparison of $dN/dy$ of $\pi^+$ and $\pi^-$ in Au+Au central collisions at $\sqrt{s_{NN}}=200$ GeV and Landau's prediction.~\cite{BrahmsPRL94} b) Ratio of data to Landau's prediction for rms width ($\sigma$) of the distribution~\cite{BrahmsPRL94}  c) (right) The invariant cross sections for $\pi^+$, $\pi^-$, $K^+$, $K^-$, $p$ and $\bar{p}$ for $p_T=0.4$ GeV/c versus $y_{\rm lab}=y-y^{\rm proj}$ at the CERN-ISR~\cite{FlatRapidity}. \label{fig:plateau?}}

\end{figure}

It is interesting to note that Bjorken used Landau hydrodynamics in his seminal paper on A+A collisions~\cite{BjorkenPRD27} with an initial condition to assure 1 dimensional longitudinal `boost invariant' expansion (resulting a flat rapidity distribution). 
    The 1d expansion lasts until the rarefaction front which starts at the outer transverse edge of the collision zone propagates at the speed of sound to the center and the whole fluid (being `aware' of the edge) starts expanding 3-dimensionally. Recent results from Au+Au collisions at RHIC indicate a relatively more Gaussian than flat rapidity distribution for identified $\pi^{\pm}$~\cite{BrahmsPRL94} (Fig.~\ref{fig:plateau?}), leading to suggestions that the p-p situation be reevaluated both with respect to the rapidity plateau and searches for evidence of other hydrodynamic phenomena~\cite{Weiner05,Alexopoulos02,Buda-Lund}.

	The applicability of the relativistic hydrodynamics of a ``perfect fluid" (zero viscosity) to both p-p and A+A collisions means that the issue for the QGP will be not whether hydrodynamics is applicable, but rather what exactly 
are the EOS and the initial conditions;	
and what exactly flows---hadrons, constituent quarks or current quarks.    
 
\subsubsection{Hard-scattering}\label{sec:hardscattering}
It is important to distinguish the soft multi-particle physics, with limited 
transverse momentum corresponding to the $\sim 1$ fm scale of the nucleon radius, which is described by constituent-quarks, thermodynamics and hydrodynamics, from the high transverse momentum phenomena ($p_T\geq 2$ GeV/c), due to the hard-scattering of point-like current-quarks, which correspond to a very short distance scale $\sim 0.1$ fm $\ll 1$ fm~\cite{BBK1971}, and contribute $\ll$ 1\% of the 
particles produced. 
These two different pictures of particle production in p-p collisions, the hard and the soft mechanisms, indicate that the ``elementary'' p-p collisions are actually rather complicated processes. 

\begin{figure}[!thb]
\begin{center}
\begin{tabular}{cc}

\includegraphics[scale=0.68,angle=0]{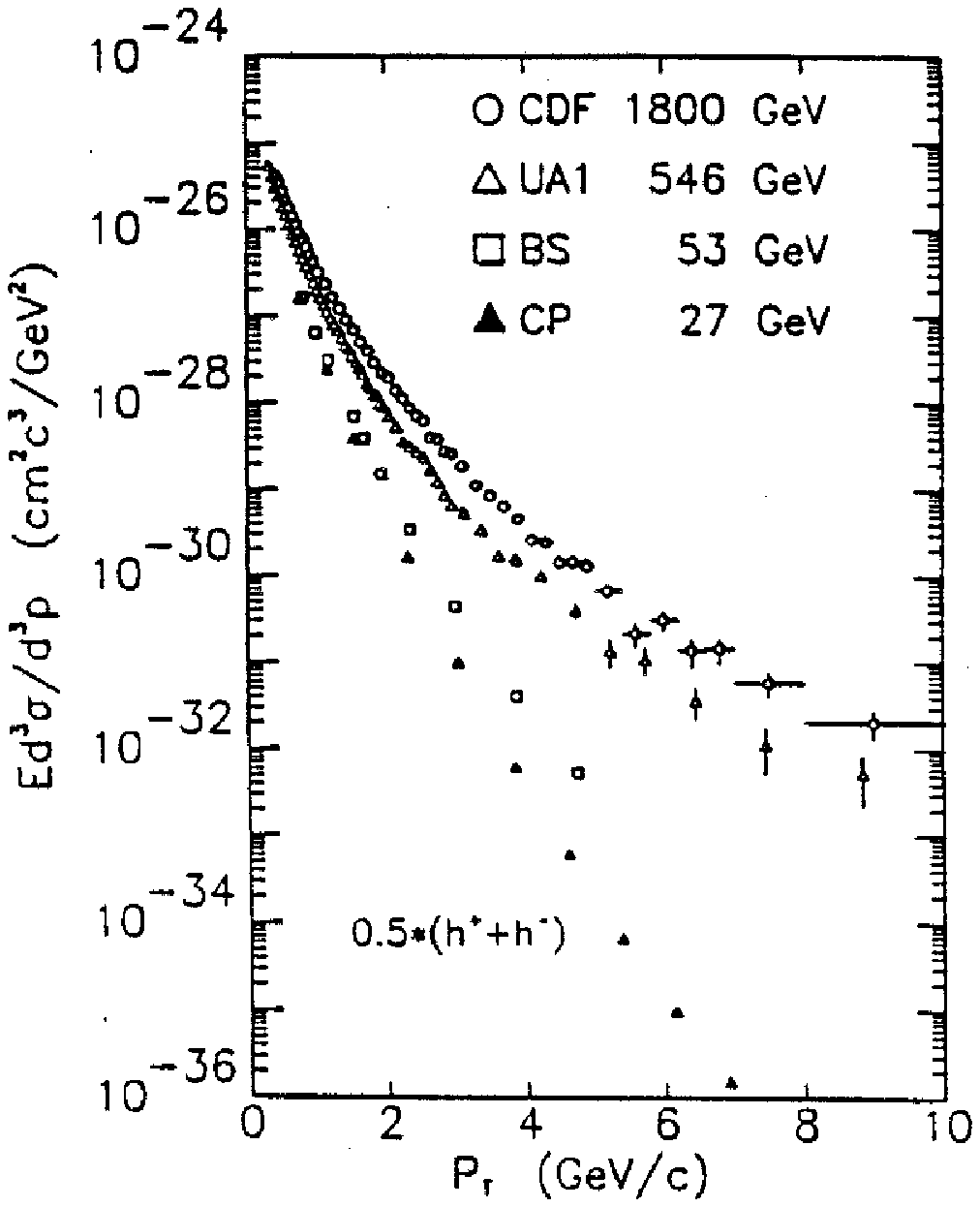}&
\includegraphics[scale=0.4,angle=0]{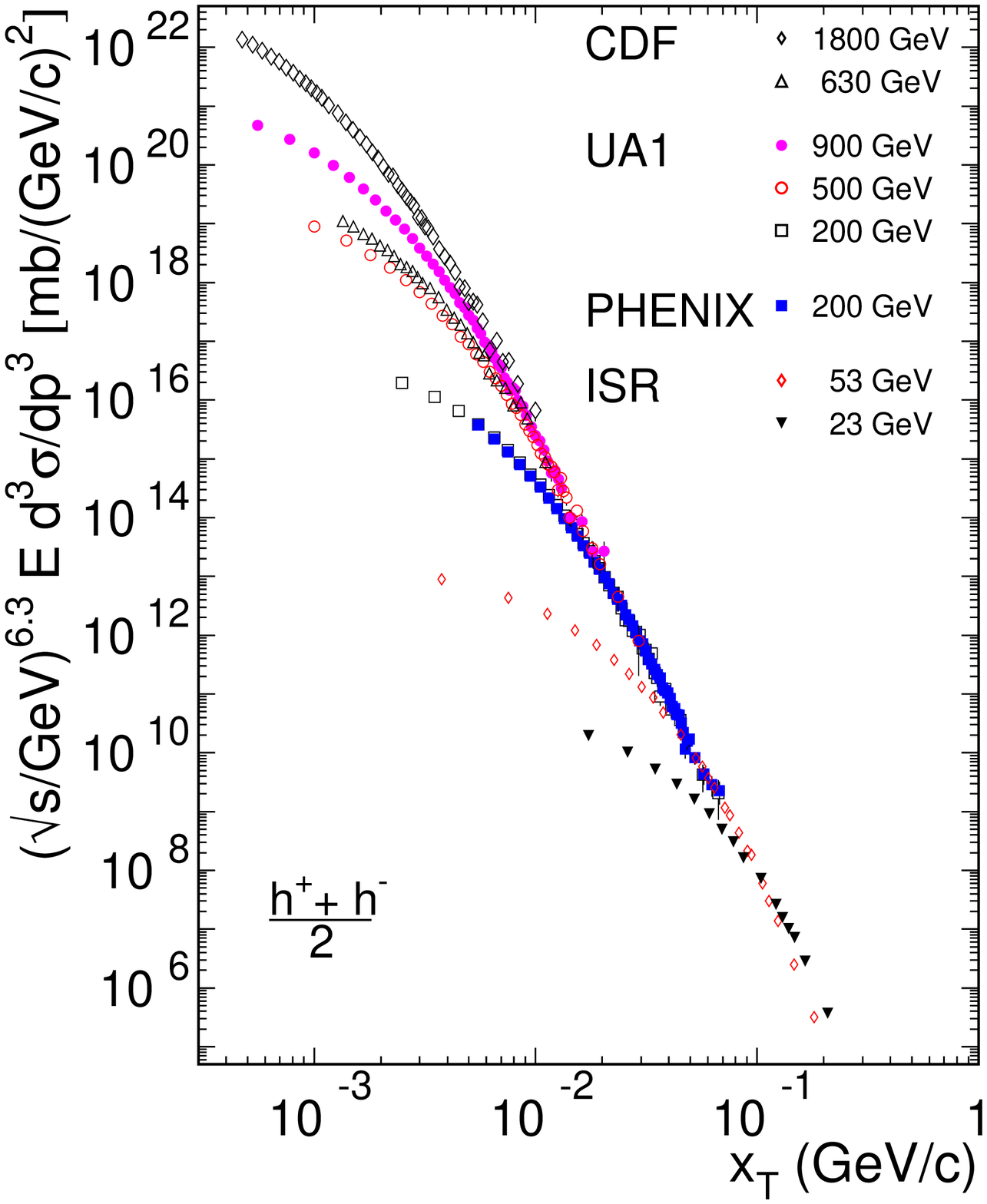}

\end{tabular}
\end{center}\vspace*{-0.25in}
\caption[]{a) (left) $E {d^3\sigma}(p_T)/{d^3p}$ at mid-rapidity as a function of $\sqrt{s}$ in $p+p$ collisions~\cite{CDF}. b) (right) $\sqrt{s}({\rm GeV})^{6.3}\times Ed^3\sigma/d^3p$ vs $x_T=2{p_T}/\sqrt{s}$~\cite{PXWP}. }
\label{fig:hpTxT}

\end{figure}

As noted above (Sec.~\ref{sec:Intro}) the observation of particle production at large transverse momentum ($p_T$), well above the extrapolation of the low $p_T$ exponential behavior (see Fig.~\ref{fig:hpTxT}a) proved that the partons of DIS strongly interacted with each other. The subject developed before the discovery of QCD, but QCD eventually provided a quantitative description of this phenomenon~\cite{OwensRMP}. 

The overall p-p hard-scattering cross section in ``leading logarithm" pQCD   
is the sum over parton reactions $a+b\rightarrow c +d$ 
(e.g. $g+q\rightarrow g+q$) at parton-parton center-of-mass (c.m.) energy $\sqrt{\hat{s}}$:   
\begin{equation}
\frac{d^3\sigma}{dx_1 dx_2 d\cos\theta^*}=
\frac{1}{s}\sum_{ab} f_a(x_1) f_b(x_2) 
\frac{\pi\alpha_s^2(Q^2)}{2x_1 x_2} \Sigma^{ab}(\cos\theta^*)
\label{eq:QCDabscat}
\end{equation} 
where $f_a(x_1)$, $f_b(x_2)$, are parton distribution functions, 
the differential probabilities for partons
$a$ and $b$ to carry momentum fractions $x_1$ and $x_2$ of their respective 
protons (e.g. $u(x_2)$), and where $\theta^*$ is the scattering angle in the parton-parton c.m. system. 
The parton-parton c.m. energy squared is $\hat{s}=x_1 x_2 s$,
where $\sqrt{s}$ is the c.m. energy of the $p+p$ collision. The parton-parton 
c.m. system moves with rapidity $\hat{y}=1/2 \ln (x_1/x_2)$ in the $p+p$ c.m. system.

Equation~\ref{eq:QCDabscat} gives the $p_T$ spectrum of outgoing parton $c$, which then
fragments into hadrons, e.g. $\pi^0$.  The fragmentation function
$D^{\pi^0}_{c}(z)$ is the probability for a $\pi^0$ to carry a fraction
$z=p^{\pi^0}/p^{c}$ of the momentum of outgoing parton $c$. Equation~\ref{eq:QCDabscat}
must be summed over all subprocesses leading to a $\pi^0$ in the final state. In this formulation, $f_a(x_1)$, $f_b(x_2)$ and $D^C_c (z)$ 
represent the ``long-distance phenomena" to be determined by experiment;
while the characteristic subprocess angular distributions,
{\bf $\Sigma^{ab}(\cos\theta^*)$},
and the coupling constant,
$\alpha_s(Q^2)=\frac{12\pi}{25\, \ln(Q^2/\Lambda^2)}$,
are fundamental predictions of QCD~\cite{Cutler:1978qm,Cutler:1977mw,Combridge:1977dm}. 

Equation~\ref{eq:QCDabscat} leads to a general `$x_T$-scaling' form for the invariant cross
section of high-$p_T$ particle production: 
\begin{equation}
E \frac{d^3\sigma}{d^3p}=\frac{1}{p_T^{n}} F({x_T}) = 
 \frac{1}{\sqrt{s}^{n}} G({x_T}) \qquad ,
 \label{eq:bbg}
 \end{equation} 
where $x_T=2p_T/\sqrt{s}$. 
The cross section has two factors, a function $F({x_T})$ ($G({x_T})$) which `scales',
i.e. depends only on the ratio of momenta, and a dimensioned factor,
${1/p_T^{n}}$ ($1/\sqrt{s}^{\, n}$),   
where $n$ equals 4  in lowest-order (LO) calculations, analogous to the $1/q^4$ form
of Rutherford Scattering in QED. The structure and fragmentation
functions   
are all in the
$F(x_T)$ ($G(x_T)$) term. Due to higher-order effects such as the running of
the coupling constant, $\alpha_s(Q^2)$, the evolution of the
structure and fragmentation functions, etc, $n$ is not a constant but is a function of $x_T$, $\sqrt{s}$.
Measured values of ${\,n(x_T,\sqrt{s})}$ in $p+p$ 
collisions are between 5 and 8~\cite{egcMJT05}.

	The scaling and power-law behavior of hard scattering are evident from the $\sqrt{s}$
dependence of the $p_T$ dependence of the $p+p$ invariant cross sections.  This is
shown for nonidentified charged hadrons,
$(h^+ + h^-)/2$, in Fig.~\ref{fig:hpTxT}a. 
At low $p_T\leq 1$ GeV/$c$ the cross sections exhibit a ``thermal" 
$\exp {(-6 p_T)}$ dependence, which is largely independent of $\sqrt{s}$, while at high $p_T$
there is a power-law tail, due to hard scattering, which depends strongly on $\sqrt{s}$. The characteristic variation with $\sqrt{s}$ at high $p_T$ is produced by the fundamental
power-law and scaling dependence of Eq.~\ref{eq:QCDabscat},~\ref{eq:bbg}. This is best
illustrated by a plot of 
\begin{equation}
\sqrt{s}^{{\,n(x_T,\sqrt{s})}} \times E \frac{d^3\sigma}{d^3p}=G(x_T) \qquad ,
\label{eq:xTscaling}
\end{equation}
as a function of $x_T$, with ${{\,n(x_T,\sqrt{s})}}=6.3$, which is valid for the $x_T$
range of the present RHIC measurements (Fig.~\ref{fig:hpTxT}b).  The data show an
asymptotic power law with increasing $x_T$. Data at a given $\sqrt{s}$ fall
below the asymptote at successively lower values of $x_T$ with
increasing $\sqrt{s}$, corresponding to the transition region from
hard to soft physics in the $p_T$ region of about 2 GeV/$c$. 
\section{``A new state of matter" vs. ``The perfect fluid".}

 With the terminology, concepts and other basics thoroughly discussed, it should be straightforward to present the relevant measurements and to evaluate whether the data support the conclusions. 
 \subsection{Particle production, thermal/chemical equilibrium, strangeness enhancement} 
\begin{figure}[!thb]
\begin{center}
\begin{tabular}{cc}
\includegraphics[height=6.0cm,width=6.0cm,angle=0]{figs/NuXuplot.epsf}&  
\includegraphics[scale=0.4,angle=0]{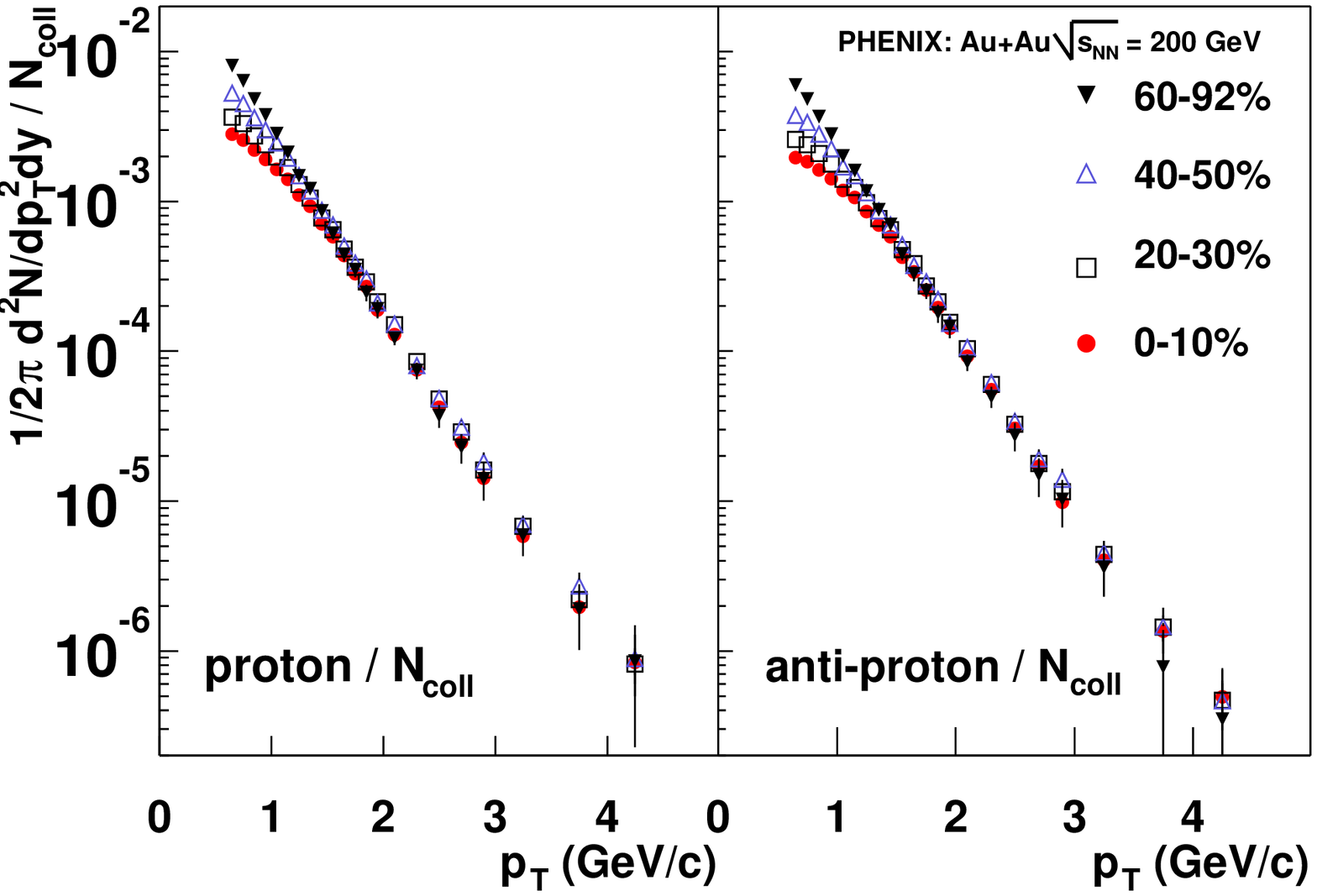}
\end{tabular}
\end{center}\vspace*{-0.25in}
\caption[]{a) (left) Inverse slope of the $m_T$ distribution (colloquially called $T$) for $\pi^{\pm}$, $K^{\pm}$, $p$, $\bar{p}$ in $158A$ GeV Pb+Pb central collisions~\cite{NA44PRL78}. b) (right) Inviarant yield of $p$ and $\bar{p}$ as a function of centrality scaled (by the number of binary collisions $N_{\rm  coll}$) to lie on top of each other for $p_T\gsim 2$ GeV/c~\cite{PXscalingPRL91}        \label{fig:radialflow}}
\end{figure}

    At RHIC energies, as shown in Fig.~\ref{fig:PXpTspectra}, the $\mean{p_T}$ of $\pi^{\pm}$, $K^{\pm}$, $p$, $\bar{p}$, increase smoothly from peripheral to  central Au+Au collisions, and as in pp collisions increase with increasing mass as would be expected for a thermal distribution (Eq.~\ref{eq:boltz}). The $m_T$-scaling property, of thermal distributions is well illustrated at SPS energies~\cite{NA44PRL78} where the effects of hard-scattering are small (see Fig.~\ref{fig:radialflow}a). 
The increase of the $\mean{p_T}$ with centrality for all particles, nearly linearly with rest mass, is evidence for collective motion (`radial flow', see below) and is seen in RHI collisions at AGS~\cite{AkibaQM96}, SPS~\cite{NA44PRL78} and RHIC~\cite{BRWP,PHWP,STWP,PXWP} energies. The effect is primarily at low $p_T$ where the slope flattens with increasing centrality as illustrated in Fig.~\ref{fig:radialflow}b. 

\begin{figure}[!thb]
\begin{center}
\begin{tabular}{cc}

\includegraphics[scale=0.4,angle=0]{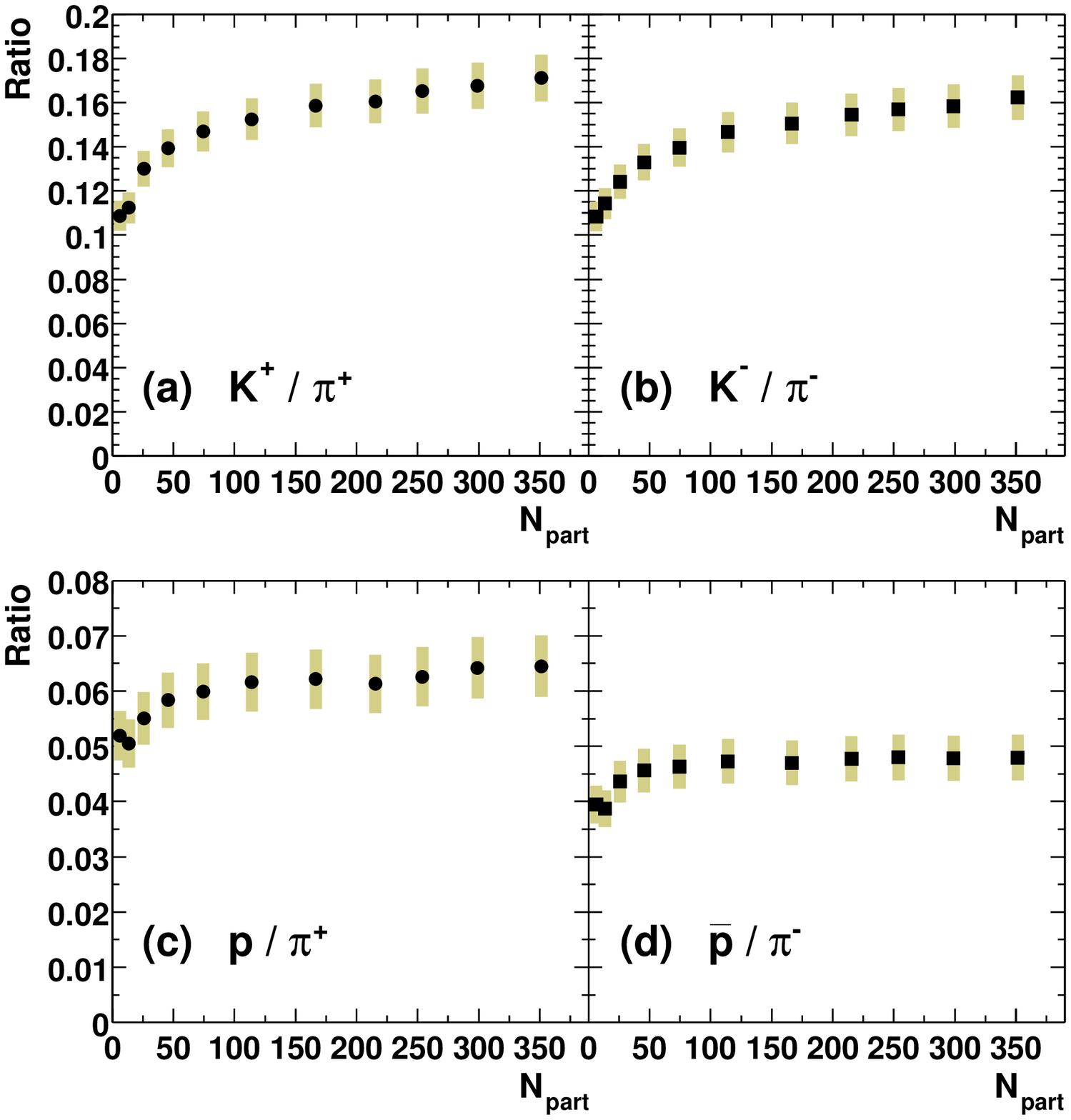}&
\includegraphics[height=7.0cm,width=7.0cm,angle=0]{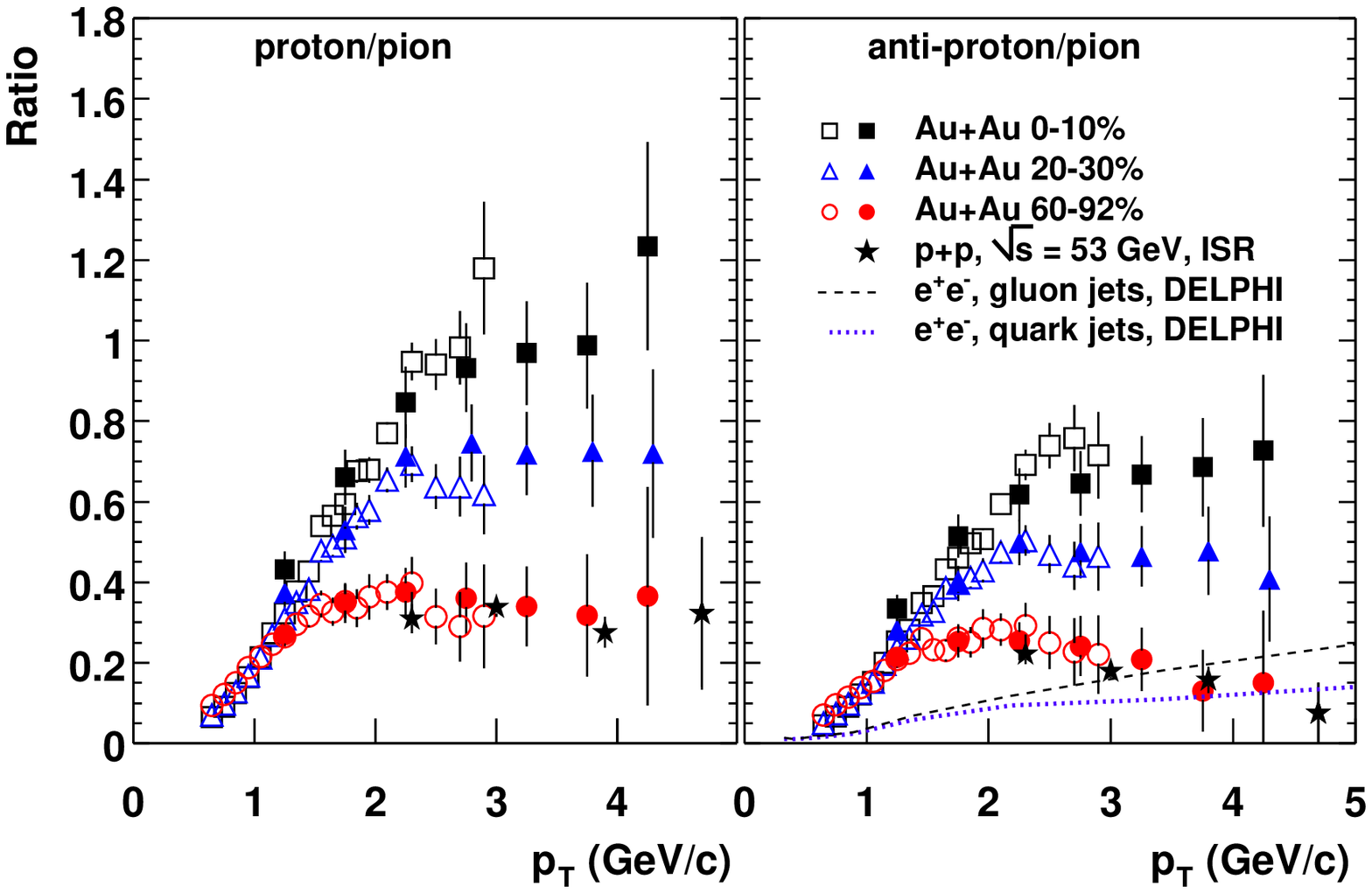}

\end{tabular}
\end{center}\vspace*{-0.25in}
\caption[]{a) (left) $K^{\pm}/\pi^{\pm}$ and $p^{\pm}/\pi^{\pm}$ ratios as a function of centrality ($N_{\rm part}$) in $\sqrt{s_{NN}}$ Au+Au collisions.~\cite{PXWP}. b) (right) $p/\pi$ and $\bar{p}/\pi$ ratio as a function of $p_T$ from the same data~\cite{PXWP,PXscalingPRL91}. Open (filled) points are for $\pi^{\pm}$ ($\pi^0$), respectively. \label{fig:ratios}}

\end{figure}

     The semi-inclusive ratios of different particle abundances also vary smoothly as a function of centrality in Au+Au collisions at RHIC (see Fig.~\ref{fig:ratios}a) with a considerably larger increase in $K^{\pm}$ production than $p$, $\bar{p}$ production relative to $\pi^{\pm}$. 
  However the $p/\pi^{+}$ and $\bar{p}/\pi^{-}$ ratios as a function of $p_T$ (Fig.~\ref{fig:ratios}b) show a dramatic increase as a function of centrality at RHIC~\cite{PXscalingPRL91} which was totally unexpected and is still not fully understood (see below). 
  
\begin{figure}[!thb]
\begin{center}
\begin{tabular}{cc}

\includegraphics[scale=0.44,angle=0]{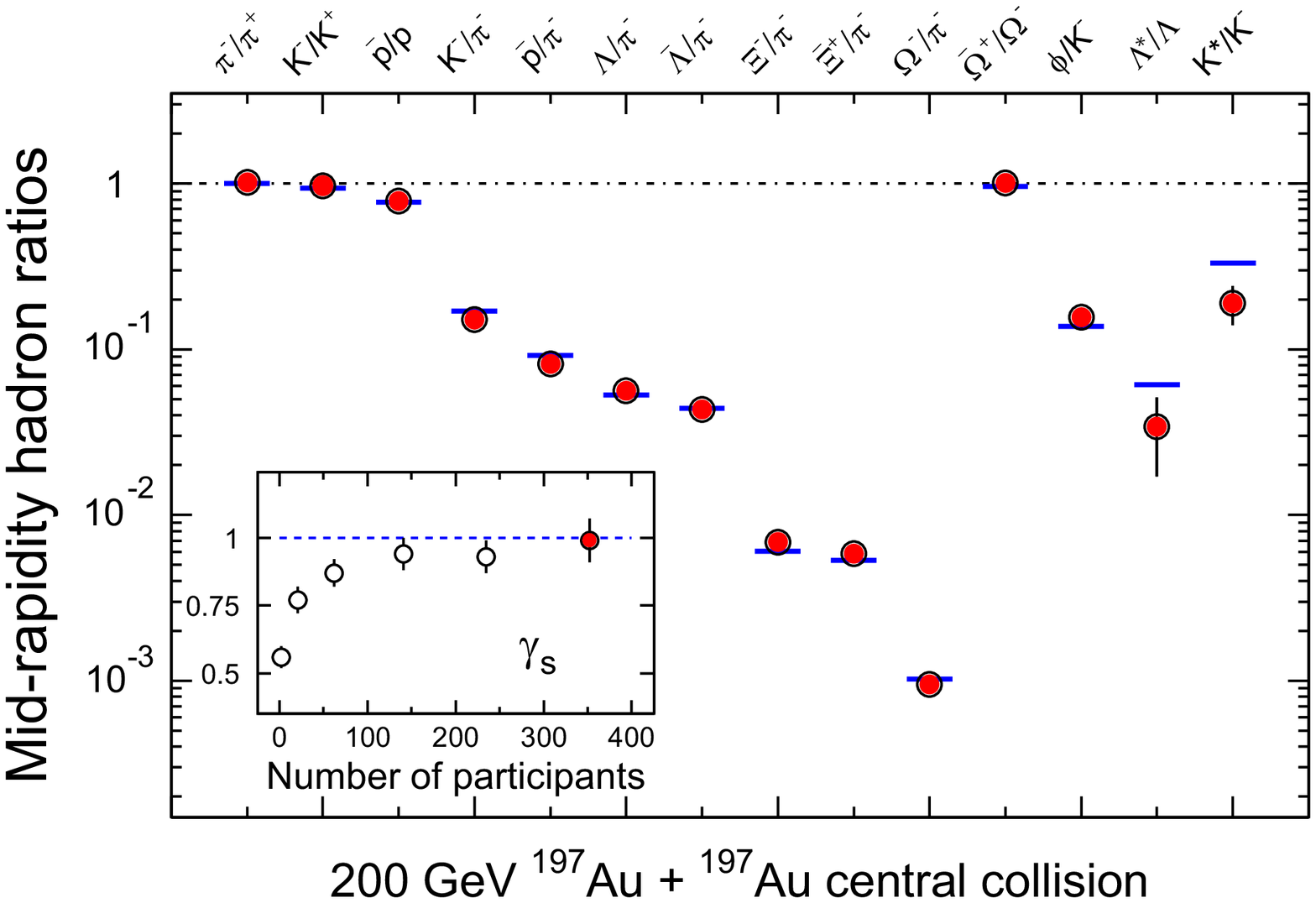}&
\includegraphics[scale=0.55,angle=0]{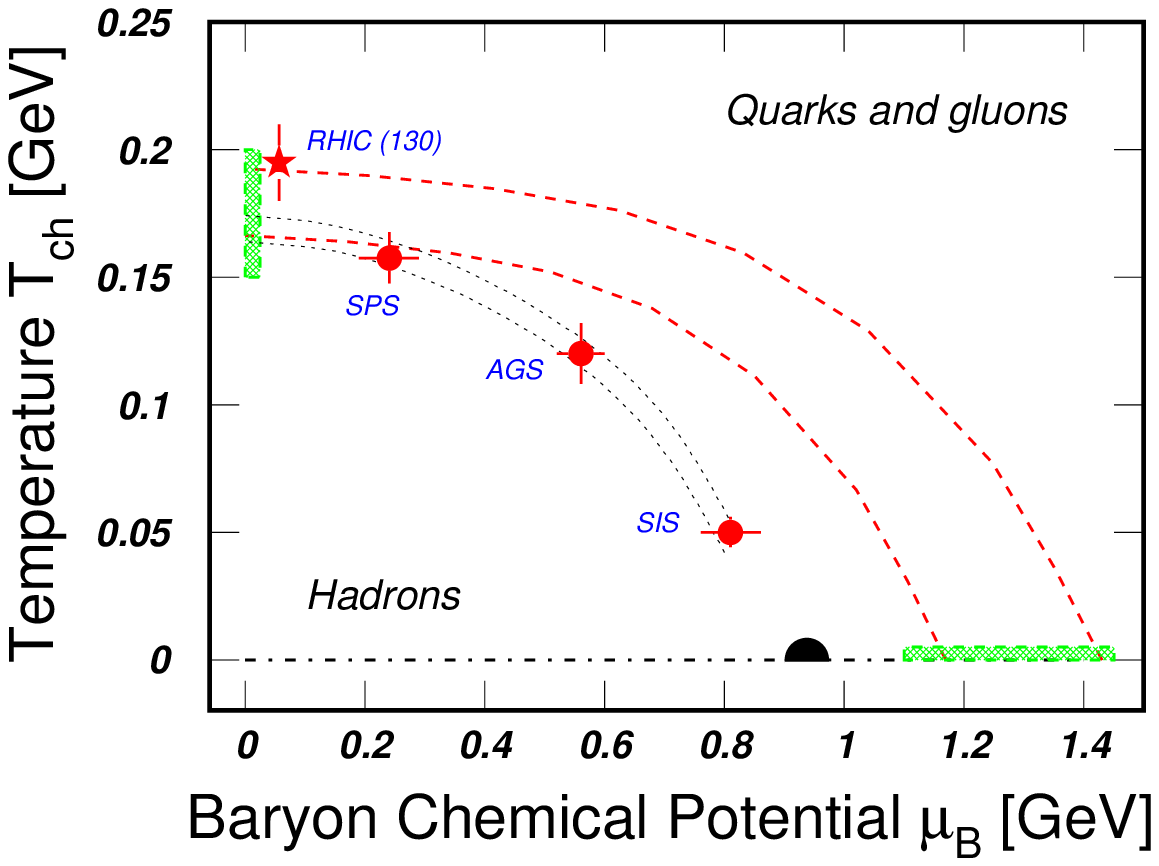}
\end{tabular}
\end{center}\vspace*{-0.25in}
\caption[]{a)(left) Ratios of $p_T$-integrated mid-rapidity yields for different hadron species measured in Au+Au central central collisions at $\sqrt{s_{NN}}=200$ GeV~\cite{STWP}. The variation of $\gamma_s$ as a function of centrality is shown as an inset including measurements for 200 GeV p-p collisions.  b) (right) $T_{ch}$ versus $\mu_B$ for thermal model fits as a function of $\sqrt{s_{NN}}$~\cite{XuKanetaQM01}, where the dashed lines represent the possible phase boundary of the QGP transition and the dotted lines represent the result of~\cite{CleymansRedlichPRL81}\label{fig:thermalmodels}}

\end{figure}

  The ratios of particle abundances (which are dominated by low $p_T$ particles) for central Au+Au collisions at RHIC, even for strange and multi-strange particles, 
are well described (Fig.~\ref{fig:thermalmodels}a) by fits to a thermal distribution, 
          \begin{equation}
{{d^2\sigma} \over {dp_L p_T dp_T}}\sim e^{-(E-\mu)/T} \rightarrow {\bar{p} \over p}=\frac{e^{-(E+\mu_B)/T}}{e^{-(E-\mu_B)/T}}=e^{-(2\mu_B)/T} \qquad ,
\label{eq:boltz2}
\end{equation}
 with similar expressions for strange particles, where $\mu_B$ and $\mu_S$ are chemical potentials associated with each conserved quantity: baryon number ($\mu_B$) and strangeness ($\mu_S$). 
 This should not be very surprising as the particle abundances in A+A collisions at SPS and AGS energies~\cite{PBMStachelNPA606} and in p-p~\cite{BecattiniHeinz} and $e^+ e^-$ collisions~\cite{BecattiniZPC76} are also well described by the same thermal model   ``(albeit, only by including a strangeness undersaturation factor, $\gamma_s < 1$) in p-p, $e^+ e^-$ and p+A collisions, where thermal and chemical equilibrium are thought not to  be achieved~\cite{STWP})", to such an extent that the chemical freezeout temperature $T_{ch}$ as a function of $\mu_B$ (from Eq.~\ref{eq:boltz2}) could be derived, which looks suspiciously like a phase diagram (Fig.~\ref{fig:thermalmodels}b). Of course, as the thermal equilibrium properties of the QGP are the subject of interest, it is important to understand how or if the thermal properties of the observed hadrons relate to the thermal properties of quarks and gluons in the QGP before getting too excited. Also, while these thermal models appear to be simple they have many technical details which are beyond the scope of this review~\cite{egseethermal}.
     
     This brings us to the perplexing question that if all particle ratios are explained by a thermal model then how could CERN claim from the strange particle abundances that ``all attempts to explain them using established particle interactions have failed"? This requires a somewhat more detailed examination of strange particle production. 
\begin{figure}[!thb]
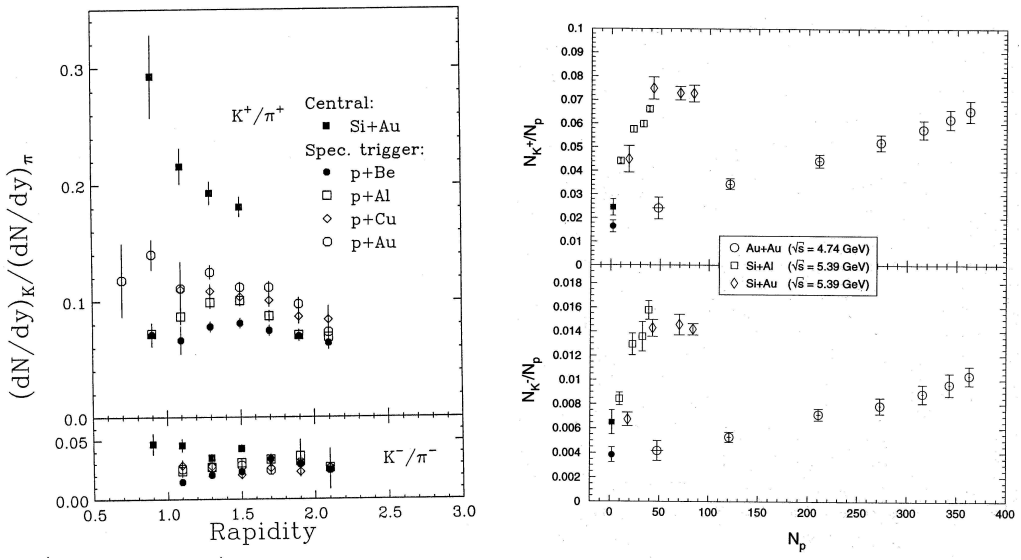

\begin{center}
\begin{tabular}{cc}
\includegraphics[scale=0.40,angle=0]{figs/SiApAscanr.epsf}&
\includegraphics[scale=0.40,angle=0]{figs/E802AAKt.epsf}
\end{tabular}
\end{center}\vspace*{-0.25in}
\caption[]{a)(left) E802 $K/\pi$ vs $y$ in p+A and Si+Au (central) collisions at $\sqrt{s_{NN}}=5.4$ GeV~\cite{AbbottPRD45}; b) (right) Integrated yield over all rapidity of $K^{\pm}$ per participant in Si+Au ($\sqrt{s_{NN}}=5.4$ GeV) and Au+Au collisons ($\sqrt{s_{NN}}=4.7$ GeV) at AGS~\cite{AhlePRC60} \label{fig:E802K}}
\end{figure}
    The strange particle production at the AGS in p+A and Si+A collisions at $14.6A$ GeV/c ($\sqrt{s_{NN}}=5.4$ GeV)~\cite{AbbottPRD45} and Au+Au collisions at $11.1A$ GeV/c ($\sqrt{s_{NN}}=4.7$ GeV)~\cite{AhlePRC60} is shown in Fig.~\ref{fig:E802K}a. In contrast to the pion yield which is roughly constant from p+Be to p+Au (not shown), the $K^+$ yield in p+A collisions increases substantially with increasing target mass $A$ and and increases in moving from the projectile to target region. Another striking feature of the data is that the $\pi^{+}/p$ and $K^+/p$ ratios both exhibit a strong rapidity dependence, but the $K^{+}/p$ ratio exhibits a remarkable target mass independence while the $\pi^+/p$ ratios do not.  As the authors noted, `` The result that the $K^+/\pi^+$ ratio increases with the target mass is somewhat surprising. At AGS energies and below, the $K^+$ yield in p+p interactions decreases faster with decreasing energy than the $\pi^+$ yield. In a naive picture of successive collisions of the projectile proton with target nucleons in heavy targets, a decreasing ratio is expected." In other words, strange particles are enhanced in p+A collisions and hadronic models can't explain this. In Si+Au collisions, the enhancement of $K^+/\pi^+$ follows the same trend with rapidity but is further enhanced by a factor of $\sim 1.5$. 
    
    The dependence on centrality and species in Si+Au and Au+Au collisions at the AGS is also very striking (Fig.~\ref{fig:E802K}b). The enhancement in the number of $K^+$ per participant in Si+A collisions at $\sqrt{s_{NN}}=5.4$ GeV turns on quickly and for Si+Au saturates at a moderate centrality. In Au+Au at $\sqrt{s_{NN}}=4.7$ GeV the number of $K^+$/participant increases slowly and smoothly as a function of centrality reaching an enhancement (compared to N-N collisions at the same $\sqrt{s_{NN}}$) the same or larger than in central Si+Au, a factor of 3-4 enhancement per participant for $K^+$ compared to N-N collisions~\cite{AhlePRC60}. The $K^+$ and $K^-$ yields track each other with centrality; and although the overall number of $K^+$ compared to $K^-$ increases from Si+Al to Si+Au to Au+Au, the $K^{+}/K^{-}$ ratio stays constant as a function of centrality for each system~\cite{Odyniec98}. The other striking feature of this measurement is that no simple description of the $K^{\pm}$ yields is given in terms of the number of participants. This tends to go against the 
simplest 
thermodynamical description, in which the volume (proportional to the number of participants) is the only extensive quantity in the problem---this is the thermodynamical (rather than quantum mechanical) explanation of why the wounded nucleon model works. 
        
\begin{figure}[!thb]
\begin{center}
\begin{tabular}{cc}

\includegraphics[scale=0.55,angle=0]{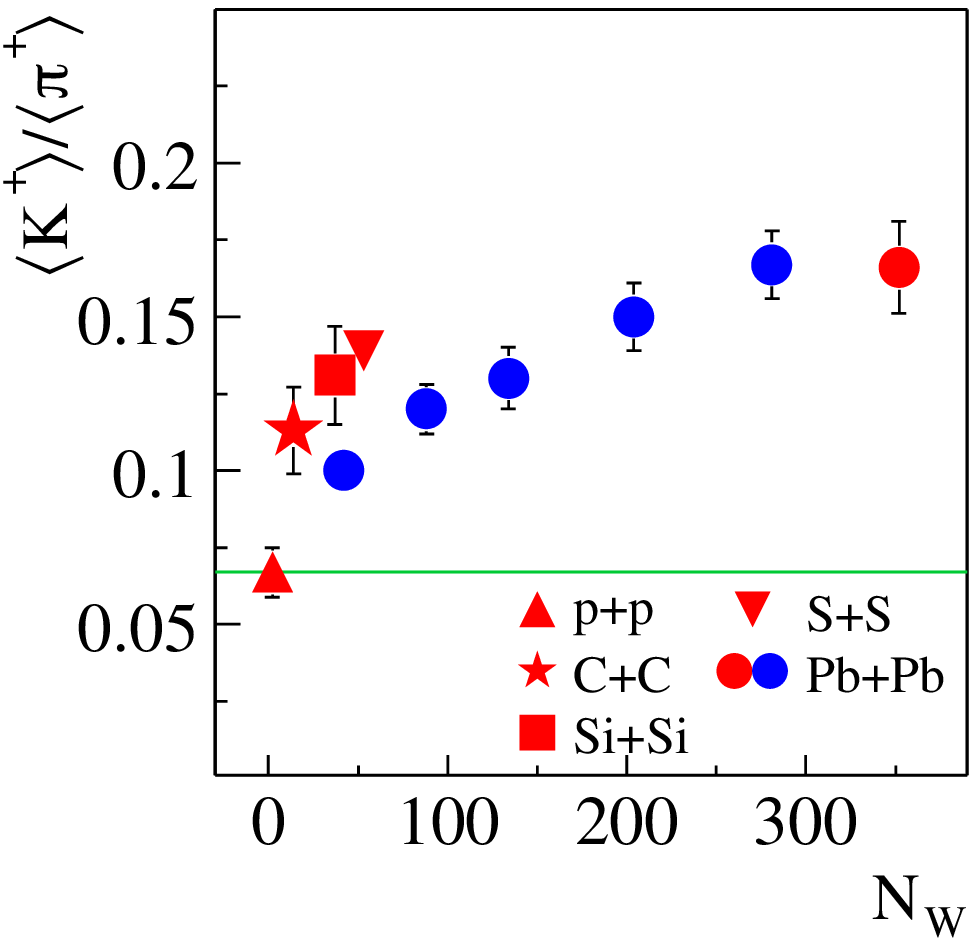}&
\includegraphics[scale=0.33,angle=0]{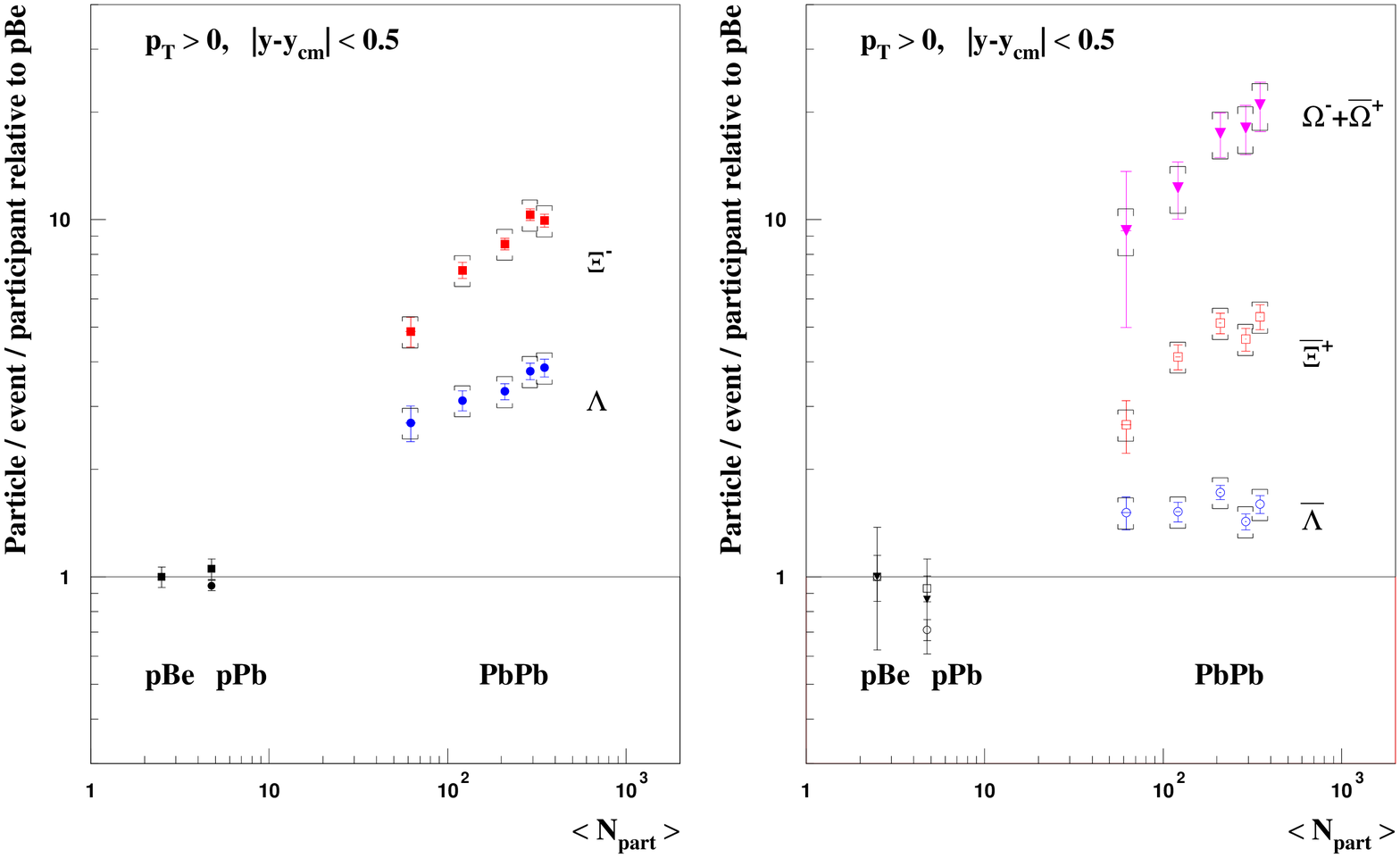}

\end{tabular}
\end{center}\vspace*{-0.25in}
\caption[]{a) (left) Total number of $K^{+}$ per $\pi^+$ (equivalent to $N_{\rm part}$) in S+S and Au+Au collisons at SPS~\cite{NA49QM99,GazdzickiQM04}; b)(right) $dN/dy$ at midrapidity as a function of $N_{part}$ for strange and multi-strange baryons~\cite{WA97QM99,NA57QM05} \label{fig:SPSstrange1}}

\end{figure}

     Interestingly, the systematics of kaon production at SPS energies show exactly the same effect (Fig.~\ref{fig:SPSstrange1}a). 
     The total yield of $K^+$ divided by the total yield of $\pi^+$ which is equivalent to $K^+$/participant, since the pions follow the wounded nucleon model, shows the same behavior at SPS energies as at the AGS, which was noted by the authors when this data was first presented at the Quark Matter conference in 1999~\cite{NA49QM99}. However, some influential theorists~\cite{HeinzQM99} at this meeting did not seem to be aware of the AGS data and this situation persisted in the ``New State of Matter" announcement~\cite{HeinzJacob}. The measurements of strange and multi-strange baryons at mid-rapidity were unique at the SPS in 1999 since no such measurements had been done at the AGS. They showed a progressive increase in enhancement by roughly a factor of 3 per unit of strangeness, so that the $\Omega$'s are enhanced by a factor of roughly 20. Due to the lack of measurements in peripheral collisions, the approximate flatness of the enhancement for $N_{\rm part} > 100$~\cite{WA97QM99} attracted some attention as evidence of a pre-hadronic state~\cite{HeinzQM99} (although the more recent data shown in  Fig.~\ref{fig:SPSstrange1}b~\cite{NA57QM05} don't look as flat) and the same effect is seen with $K^{+}$ in Pb+Pb at CERN and Si+Au at the AGS. One pre-hadronic state that was neglected was the quantum-mechanical excited nucleon responsible for the Wounded Nucleon Model for pions. Multiple excitation of the incident nucleons by successive collisions has little effect on the production of pions, but has a huge effect on the production of $K^{+}$ if it raises the excitation above the $K^+ \Lambda$ threshold. Also, this happens in the initial collisions not in the hadron gas which takes place much later in the evolution of the system, so may not be well accounted for in `hadronic' models. To quote Ref.~\cite{HeinzQM99,goodQM99quote} ``The observed strangeness enhancement pattern thus cannot be generated by hadronic final state interactions, but must be put in at the beginning of the hadronic stage."

    So, does the fact that ``models based on hadronic interaction mechanisms have consistently failed to simultaneously explain the wealth of accumulated data"~\cite{HeinzJacob}---notably the strangeness enhancement both at AGS and SPS---imply that the QGP exists at the SPS (and the AGS)? Obviously not! The fact that certain `hadronic models' do or do not explain the data is a statement about the validity of these models (which already fail for strangeness in p+A production at the AGS~\cite{AbbottPRD45}) and is certainly no justification for drawing any conclusion about the Quark Gluon Plamsa. The 
SPS 
strangeness data, either at that time or at present~\cite{JanJean04}, never supported a QGP conclusion even though strangeness enhancement was one of the original proposed signatures for the QGP (Sec.~\ref{sec:strange}). 
Strangeness is enhanced in nuclear collisions at all c.m. energies~\cite{Senger04,Oeschler04}, even in p+A collisions. Hence strangeness enhancement is an ordinary feature of nuclear collisions, not a unique feature of the QGP.   
    
\begin{figure}[!thb]
\begin{center}
\begin{tabular}{cc}
\includegraphics[scale=0.55,angle=0]{figs/BrahmsHornl.epsf}&
\includegraphics[scale=0.4,angle=0]{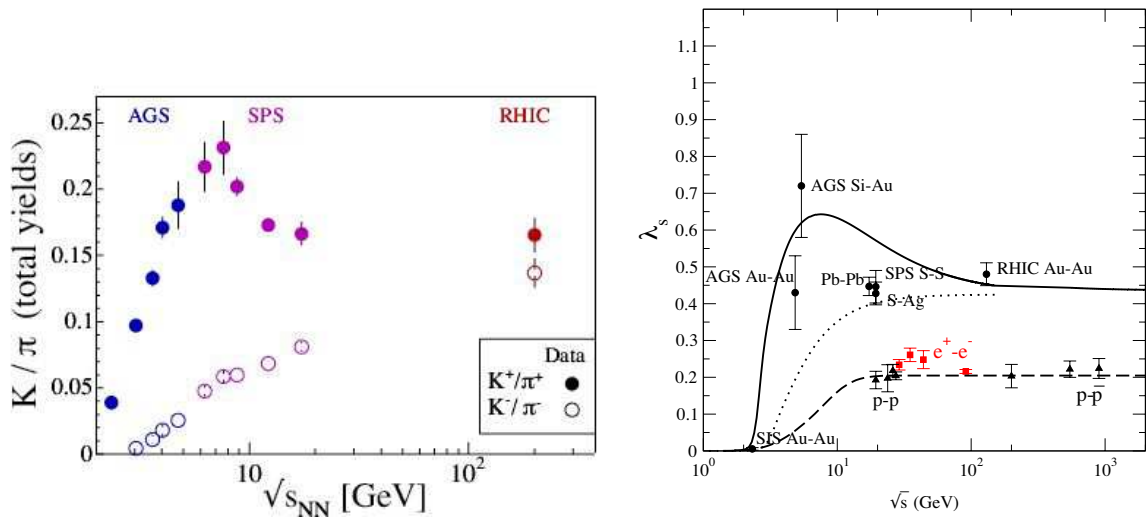}

\end{tabular}
\end{center}\vspace*{-0.25in}
\caption[]{a) (left) Ratio of total yields of $K^+/\pi^+$ and $K^-/\pi^-$ as a function of $\sqrt{s_{NN}}$~\cite{BRWP}; b) (right) Ratio of $s\bar{s}$ to $u\bar{u}$ and $d\bar{d}$ quarks, $\lambda$~\cite{Anisovich73}, as a function of $\sqrt{s_{NN}}$ from Ref.~\cite{BecattiniPRC64} with predictions from Ref.~\cite{PBMCleymans02}. The dotted line is the effect of the $T$ variation, with $\mu_B=0$, and the solid line includes the $\mu_B$ variation.
\label{fig:thehorn} }
\end{figure}

    Interestingly, strangeness production and the region between the SPS and AGS energies has drawn renewed interest lately~\cite{GazdzickiQM04,Maiani05}. The integrated yield (over all phase space, rapidity, azimuth, $p_T$) is shown in Fig.~\ref{fig:thehorn}a~\cite{BRWP}. 
As noted above, there is an enrichment of $K^+$ in regions of large baryon density due to associated production via e.g $pp\rightarrow p K^+ \Lambda$. At the AGS, where the net proton density ($\mu_B$) is high at midrapidity (recall Fig.~\ref{fig:PHOBOSQM05}) the number of $K^+$ strongly exceeds the number of $K^-$, which have no baryon-associated production channel and are mostly produced in a $K^- K^+$ pair. The pair production channel increases smoothly with $\sqrt{s_{NN}}$  while the $K^{+}$ production exhibits a maximum at $\sqrt{s_{NN}}\sim 7$ GeV. This maximum is explained~\cite{BecattiniPRC64,PBMCleymans02} (see Fig.~\ref{fig:thehorn}b) by the interplay between the natural rise in strangeness production (e.g. $K^-$, Fig.~\ref{fig:thehorn}a) with increasing $\sqrt{s_{NN}}$ (or $T$) combined with the reduction of associated production due to the reduction of $\mu_B$ with increasing $T$ (Fig.~\ref{fig:thermalmodels})  as the nuclear transparency reduces with increasing $\sqrt{s_{NN}}$ and the valence protons move away from midrapidity  (Fig.~\ref{fig:PHOBOSQM05}). Of course, the burning issue at this time is whether or not the maximum  (colloquially the `horn'~\cite{GazdzickiQM04}) at $\sqrt{s_{NN}}\sim 7$ GeV is a discontinuity (a real signature of a phase transition) or just an example of the very interesting but not QGP physics of A+A collisions. Clearly, more precise measurements in this region are required. 

\subsection{Flow}
   A distinguishing feature of A+A collisions compared to either p-p or p+A collisions is the collective flow observed. This effect is seen over the full range of energies studied in heavy ion collisions, from incident kinetic energy of $100A$ MeV to c.m. energy of $\sqrt{s_{NN}}=200$ GeV~\cite{LaceyQM05}. Collective flow, or simply flow, is a collective effect which can not be obtained from a superposition of independent N-N collisions.  It comes in three varieties: directed flow, radial flow and elliptical flow.

   \begin{figure}[!thb]
\begin{center}

\includegraphics[scale=0.66,angle=0]{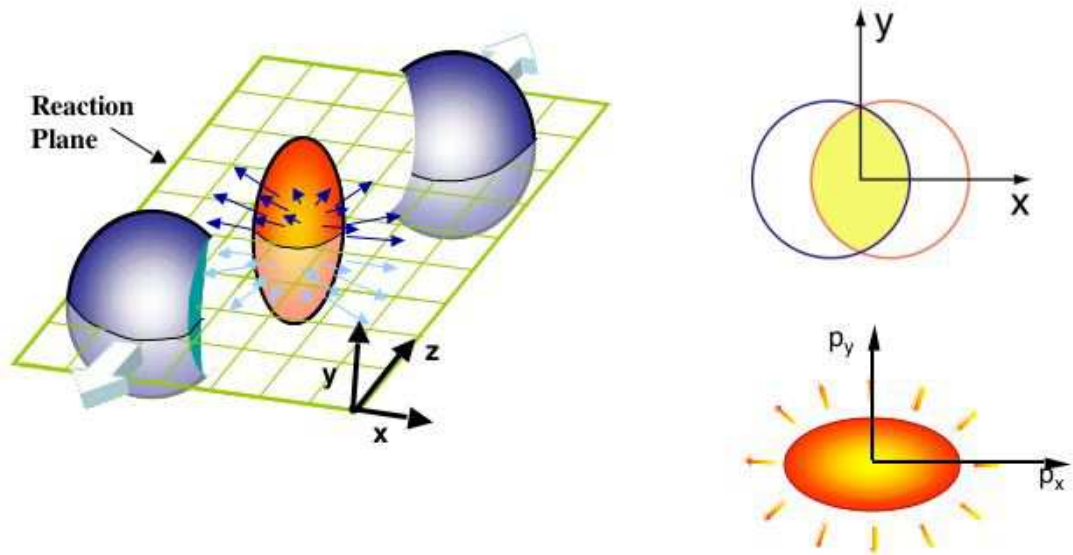}
\end{center}\vspace*{-0.25in}
\caption[]{a) (left) Almond shaped overlap zone generated just after an A+A collision where the incident nuclei are moving along the $\pm z$ axis, and the reaction plane, which by definition contains the impact parameter vector (along the $x$ axis) Thanks to Masashi Kaneta for the figure~\cite{KanetaQM04}. b) (right) View of the collision down the $z$ axis: (top) spatial distribution (bottom) momentum distribution after elliptic flow ($v_2$) develops 
\label{fig:MasashiFlow}}

\end{figure}
   Immediately after an A+A collision, the overlap region defined by the nuclear geometry is almond shaped (see Fig~\ref{fig:MasashiFlow}) with the shortest axis along the impact parameter vector. Due to the reaction plane breaking the $\phi$ symmetry of the problem, the semi-inclusive single particle spectrum is modified from Eq.~\ref{eq:siginv} by an expansion in harmonics~\cite{Ollitrault} of the azimuthal angle of the particle with respect to the reaction plane, $\phi-\Phi_R$~\cite{HeiselbergLevy}, where the angle of the reaction plane $\Phi_R$ is defined to be along the impact parameter vector, the $x$ axis in Fig.~\ref{fig:MasashiFlow}: 
  \begin{eqnarray}
{Ed^3 N \over dp^3}&=&{d^3 N\over p_T dp_T dy d\phi}\\[0.2cm]
&=&{d^3 N\over 2\pi\, p_T dp_T dy} [1+2 v_1 \cos(\phi-\Phi_R)+ 2 v_2\cos2(\phi-\Phi_R)+ \cdots ] .\nonumber
\label{eq:siginv2} 
\end{eqnarray} 
The expansion parameter $v_1$ is called the directed flow and $v_2$ the elliptical flow. If no collective behavior takes place, i.e. the interaction is merely a superposition of independent nucleon-nucleon collisions, then the outgoing momentum distribution of the particles would be isotropic in azimuth. However, since the leading participating nucleons in the forward region $+z$ (Fig.~\ref{fig:MasashiFlow}a) will interact with many other nucleons in the ``almond", they will be pushed away from the rest of the participants, into the $+x$ direction, while the $-z$ going participants are pushed towards $-x$. This is what causes the directed flow, $v_1$, which was discovered at the Bevalac~\cite{PlasticBall} and is clearly sensitive to the Equation of State. For instance if one imagines the almond to be composed of billiard balls requiring lots of pressure for a small deformation (hard EOS) a larger $v_1$ would result than if the almond suddenly melts, perhaps turning into a `perfect fluid', with a much softer EOS~\cite{egseeDLLsci}. 

    The same principles apply to $v_2$, the parameter of $\cos 2(\phi-\Phi_R)$, which (unlike $v_1$) doesn't change sign with rapidity, and hence is non-zero at midrapidity. If thermal equilibrium is reached, then the pressure gradient is directed mainly along the direction of the impact parameter ($x$ axis in Fig.~\ref{fig:MasashiFlow}b) and collective flow develops along this direction. If all the particles are approximately at rest in the fluid and thus move with the fluid velocity, the transverse momentum distribution will reflect the fluid profile. Hence the anisotropic spatial distribution is carried over to an anisotropic momentum distribution through the pressure gradient~\cite{Ollitrault}.   
   
\begin{figure}[!thb]
\begin{center}
\begin{tabular}{cc}

\includegraphics[scale=0.4,angle=0]{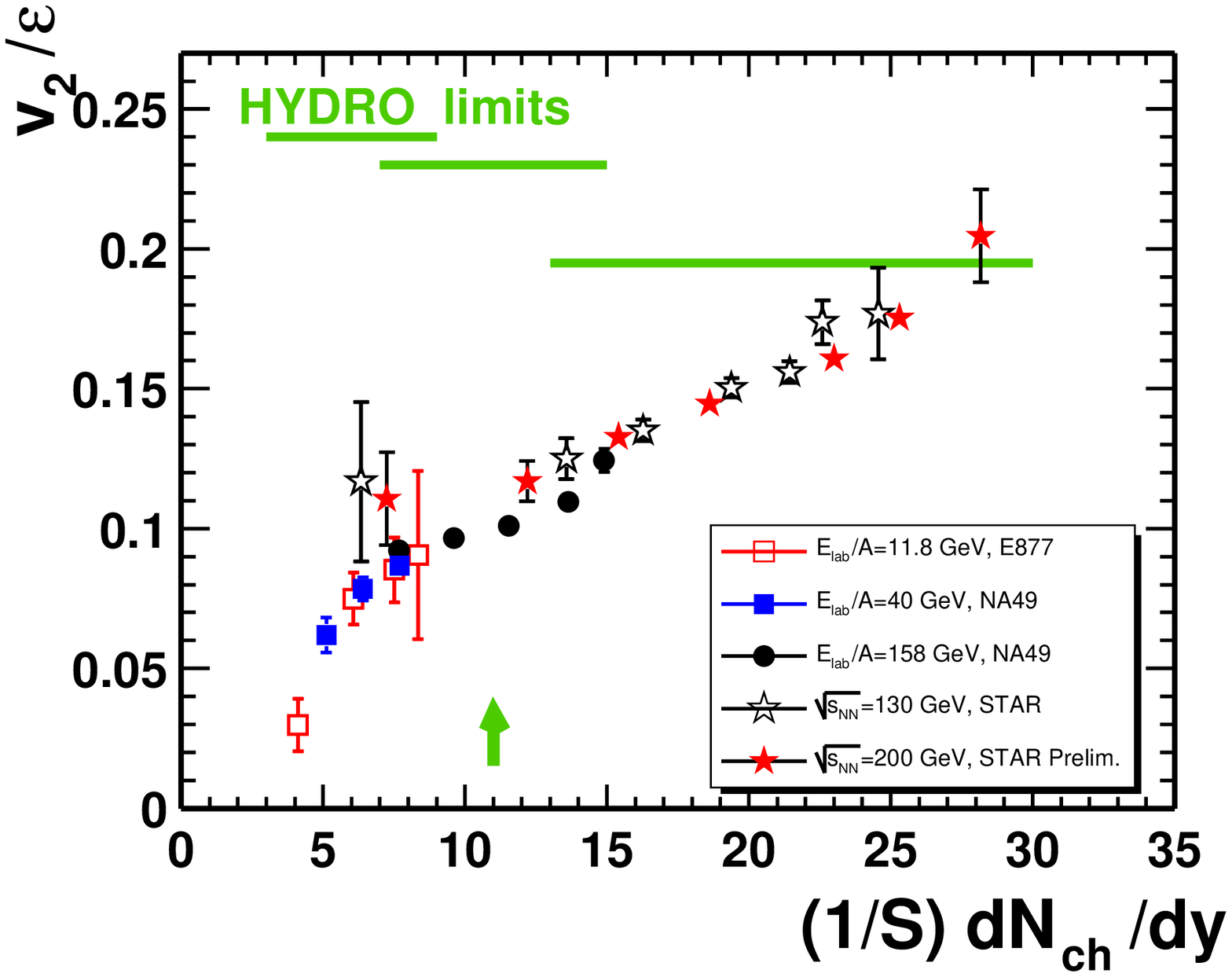}\hspace*{-0.5cm}&
\includegraphics[scale=0.4,angle=0]{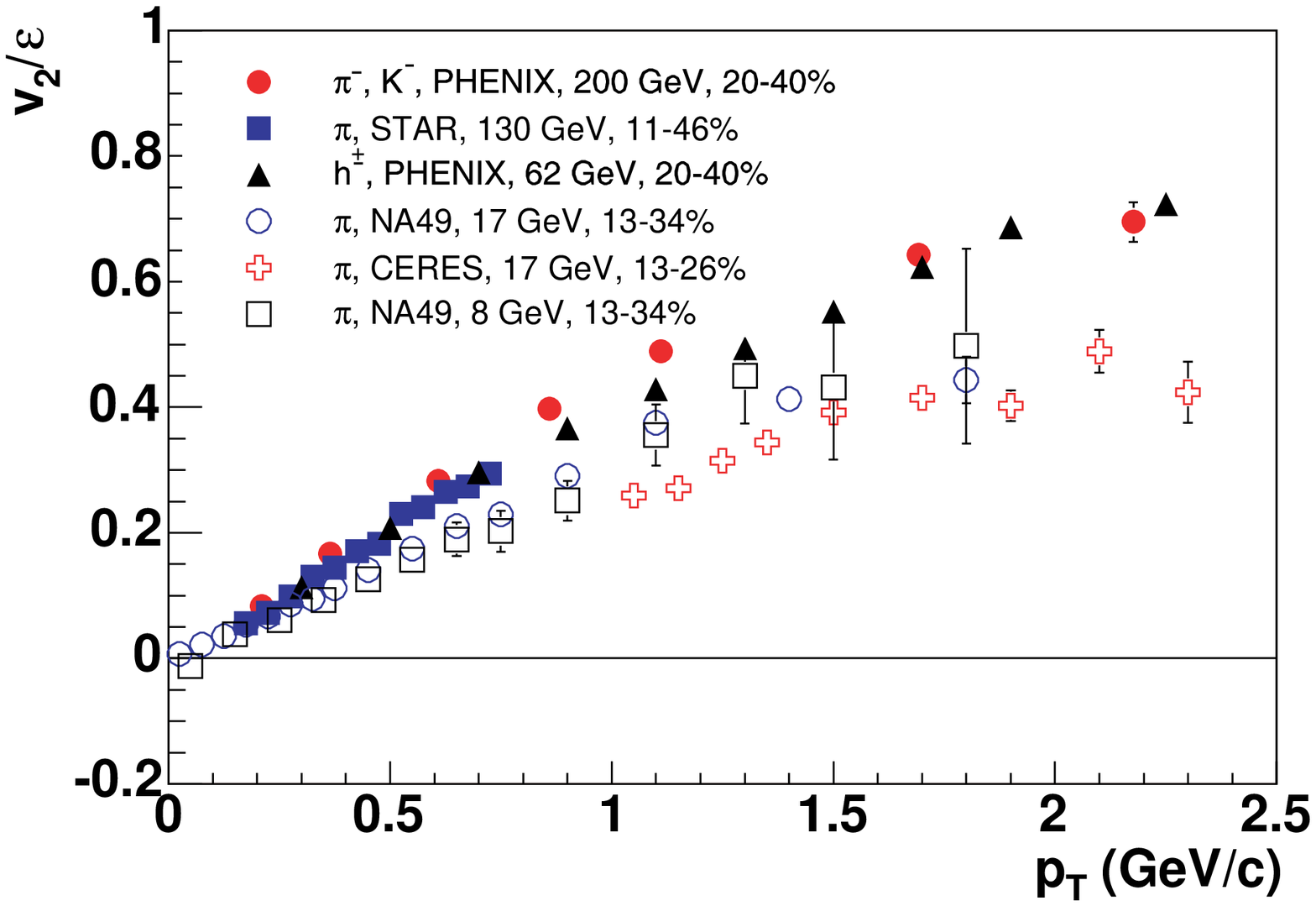}

\end{tabular}
\end{center}\vspace*{-0.25in}
\caption[]{a)(left) $v_2/\varepsilon$ as a function of particle density $\rho=(1/\pi R_x R_y)\, dN_{\rm ch}/dy$ as compiled by Ref.~\cite{NA49PRC68} for midcentral collisions ($\sim 13-34$\% upper percentile). b) (right) $v_2/\varepsilon$ as a function of $p_T$ for midcentral collisions at RHIC as compiled by Ref.~\cite{PXWP}    \label{fig:flow1}}

\end{figure}

    It is important to emphasize that the spatial anisotropy turns into an momentum anisotropy only if the outgoing particles or partons interact with each other~\cite{HeiselbergLevy}. Thus the momentum anisotropy is proportional to the spatial anisotropy of the almond, represented by the eccentricity, $\varepsilon=(R^2_y-R^2_x)/(R^2_y +R^2_x)\simeq (R_y-R_x)/(R_y +R_x)$, at the time ($t_0$) of thermalization. This is due to the fact that the mean number of scatterings in the transverse plane is different along the $x$ and $y$ axes~\cite{VoloshinPoskanzer,SorgePRL82,HeiselbergLevy}.  The mean number of scatterings  is proportional to the particle density, $\rho=(1/\pi R_x R_y)\, dn/dy$ (similar to Eq.~\ref{eq:eBj}) times the interaction cross section ($\sigma$) times the distance traversed:
    \begin{equation}
    v_2\propto R_y\,\sigma {1 \over {\pi R_x R_y}}\, {dn \over dy} - 
     R_x\,\sigma {1\over {\pi R_x R_y}}\, {dn \over dy} \propto 
    \varepsilon\,\sigma {1 \over {\pi R_x R_y}}\, {dn \over dy} \qquad ,
    \label{eq:VP}
    \end{equation}
    where $R_x=\sqrt{\mean{x^2}}$, $R_y=\sqrt{\mean{y^2}}$. Hence one test for hydrodynamic evolution~\cite{VoloshinPoskanzer} is to plot $v_2/\varepsilon$ as a function of $\rho=(1/\pi R_x R_y)\, dn/dy$ (see Fig.\ref{fig:flow1}).
     The data follow this simple scaling law very nicely. The hydro-limits indicated are for full thermalization of the system at the value of $\varepsilon$ given by the initial nuclear geometry of the almond at the time of overlap. If the system doesn't thermalize rapidly, the flow tends to vanish because the eccentricity reduces as the system expands~\cite{PXWP,KolbPLB459}. The value of the hydro limit is also determined by the  EOS through the speed of sound and whether the state is hadronic, partonic or undergoes a phase transition~\cite{RuuskanenZPC38,HydroReview}. The speed of sound, $c_s^2=\partial P/\partial \epsilon$ (e.g. a simple EOS being $\epsilon=P/c_s^2$), determines how long it takes information about the initial spatial anisotropy to propagate to the whole system, $\sim R_y/c_s$~\cite{Bhalerao}. The hydro-limits shown on Fig.~\ref{fig:flow1} are for a particular choice of an EOS with no phase transition~\cite{NA49PRC68,STWP}. However, recent, more thorough calculations indicate that the hydro-limit is reached at RHIC but that $v_2$ at the SPS is considerably below it~\cite{KolbPLB500,HydroReview}. Because perfect fluid hydrodynamics depends on so few parameters, there are many simple scaling tests~\cite{Bhalerao} (analogous to Eq.\ref{eq:VP} and Fig.~\ref{fig:flow1}) that can be performed to test its validity~\cite{LaceyQM05}.
    
    It is also possible that the increase of $v_2/\varepsilon$ with increasing $\sqrt{s_{NN}}$ (averaged over $p_T$) is due to the harder $p_T$ spectra at larger $\sqrt{s_{NN}}$. However the $p_T$ dependence of $v_2/\varepsilon$ (Fig.~\ref{fig:flow1}b) clearly increases more rapidly with increasing $\sqrt{s_{NN}}$. 
    
\begin{figure}[!htb]
\begin{center}
\begin{tabular}{cc}
\includegraphics[scale=0.84,angle=0]{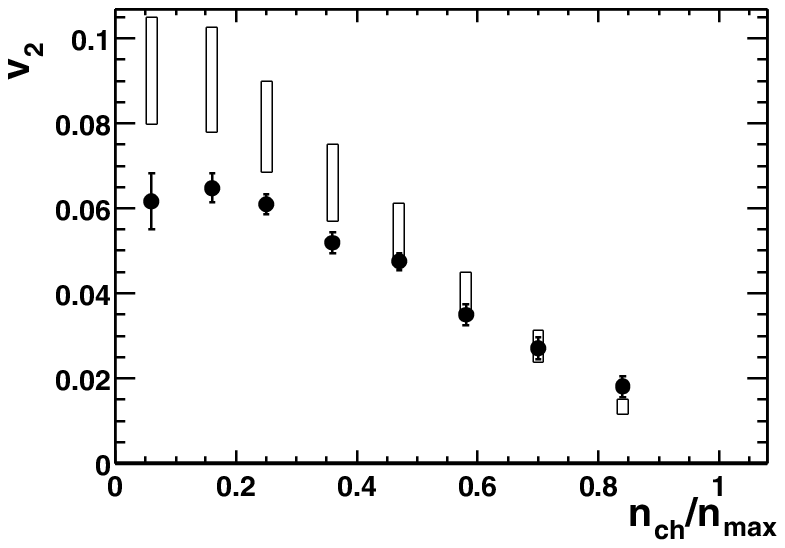}&\hspace*{-0.1cm}
\includegraphics[scale=0.43,angle=0]{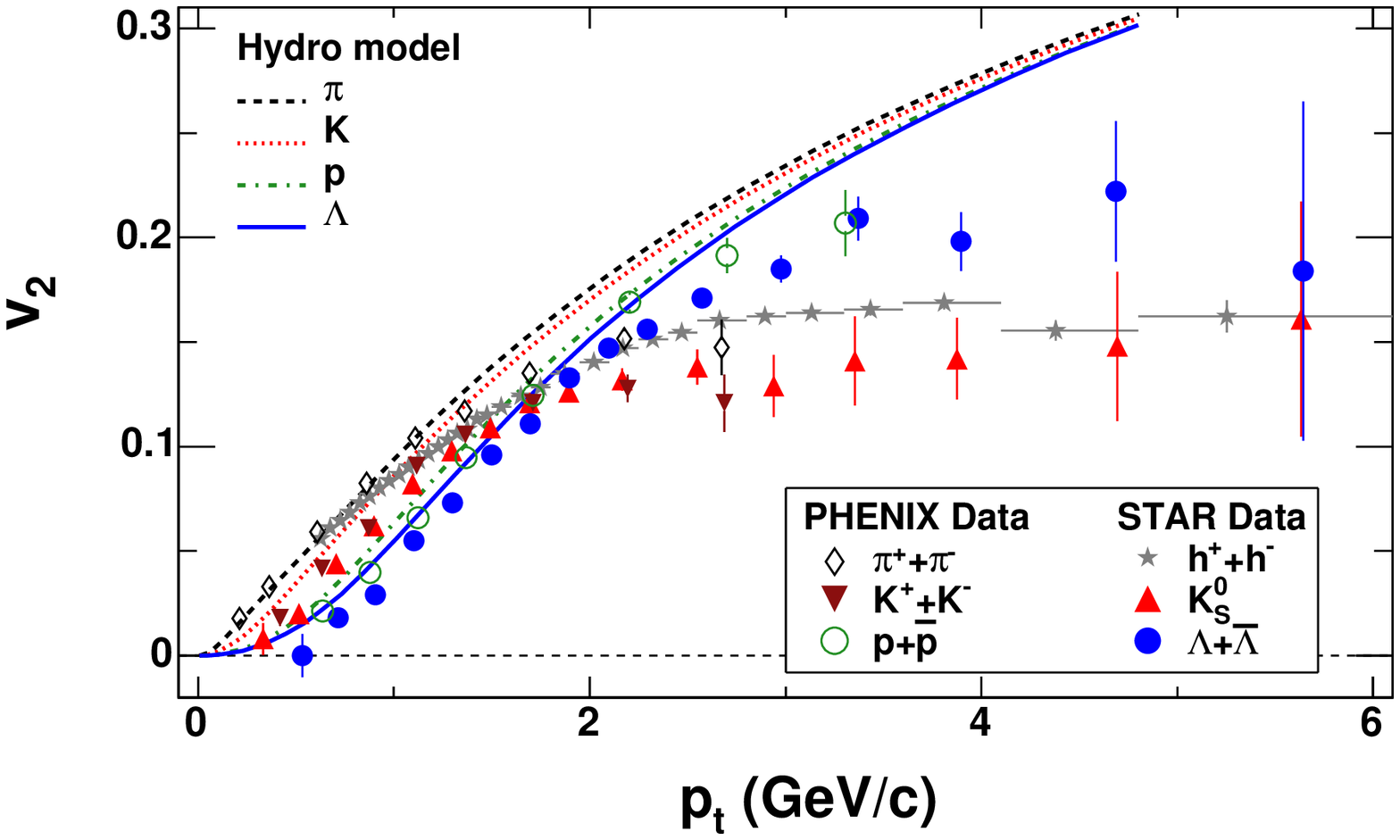}
\end{tabular}
\end{center}\vspace*{-0.25in}
\caption[]{a) (left) $v_2$ as function of centrality measured as a fraction of the total number of charged particles~\cite{STAR130}. The boxes represent the expected hydro-limit with $v_2/\varepsilon=0.19$ (lower edge) and 0.25 (upper edge). b) (right) $v_2$ as a function of $p_T$ for identified particles in minimum bias Au+Au collisions at $\sqrt{s_{NN}}=200$ GeV together with a hydro calculation~\cite{STARPRC72}. \label{fig:flow2}}
\end{figure}

\begin{figure}[!htb!htb]
\begin{center}
\begin{tabular}{cc}
\includegraphics[scale=0.38,angle=0]{figs/TeaneyPRC68.epsf}&
\includegraphics[scale=0.36,angle=0]{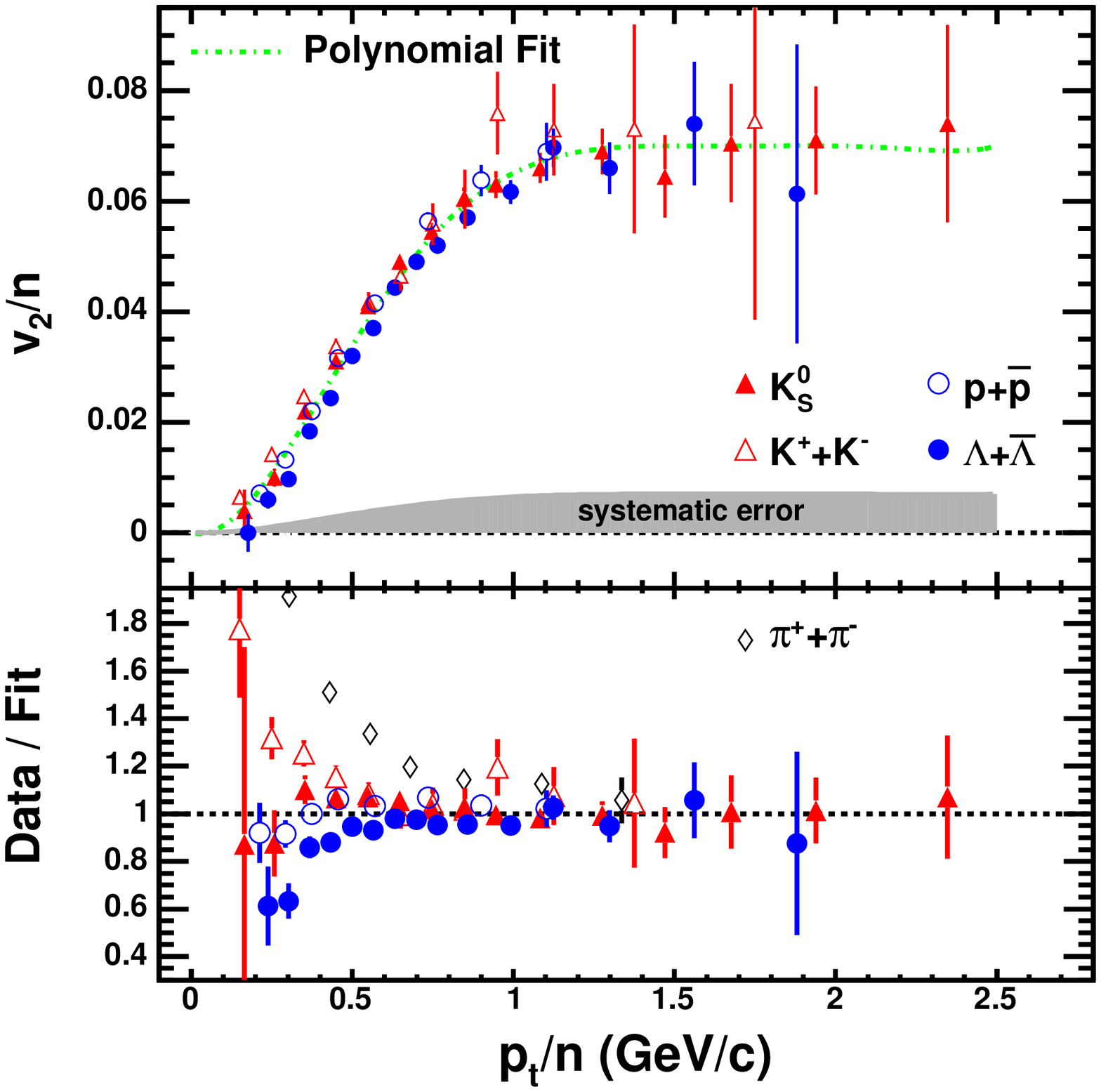}
\end{tabular}
\end{center}\vspace*{-0.25in}
\caption[]{a) (left)$v_2$ as a function of $p_T$ in a hydro calculation~\cite{TeaneyPRC68} for mid-central collisions for different values of $\Gamma_s/\tau_0$, the ``sound attenuation length" which is zero for a ``perfect fluid" and increases linearly with the viscosity. b) (left) $v_2/n$ vs $p_T/n$ for identified particles, where $n$ is the number of constituent quarks~\cite{STARPRC72}   \label{fig:flow3}}
\end{figure}

    Since the eccentricity $\varepsilon$ is much larger for peripheral than for central collisions, the dependence of $v_2$ on centrality has a characteristic shape (Fig.~\ref{fig:flow2}a)~\cite{STAR130}. This was one of the first publications from RHIC and showed that $v_2$ was surprisingly large and near the hydro-limits. Another surprise~\cite{STARPRL90} (Fig.~\ref{fig:flow2}b)~\cite{STARPRC72} was that the $v_2$ followed the hydro prediction out to $p_T\sim 2$ GeV/c and then plateaued at a constant value to much higher $p_T$. This was one of the principal arguments for the ``perfect fluid" because any modest value of viscosity~\cite{TeaneyPRC68} would cause the $v_2$ to decrease towards zero near $p_T\sim 1.7$ GeV/c (Fig.~\ref{fig:flow3}a). 

	As hydrodynamics appears to work in both p-p and A+A collisions, and collective flow is observed in A+A collision over the full range of energies studied, a key question is what is flowing at RHIC and is it qualitatively different from the flow observed at lower $\sqrt{s_{NN}}$ ? One interesting proposal in this regard is that the constituent quarks flow~\cite{VoloshinQM02}, so that the flow should be proportional to the number of constituent quarks $n_q$, in which case $v_2/n_q$ as a function of $p_T/n_q$ would represent the constituent quark flow as a function of constituent quark $p_T$ and would be universal.  
Interestingly, the RHIC data (Fig.~\ref{fig:flow3}b) seem to support this picture, although the fact that the $\pi^+ +\pi^-$ deviate most from the universal curve should raise some suspicions as the pion is the only particle whose mass is much less than that of its constituent quarks. 
\begin{figure}[!htb]
\begin{center}
\begin{tabular}{cc}
\includegraphics[scale=0.4,angle=0]{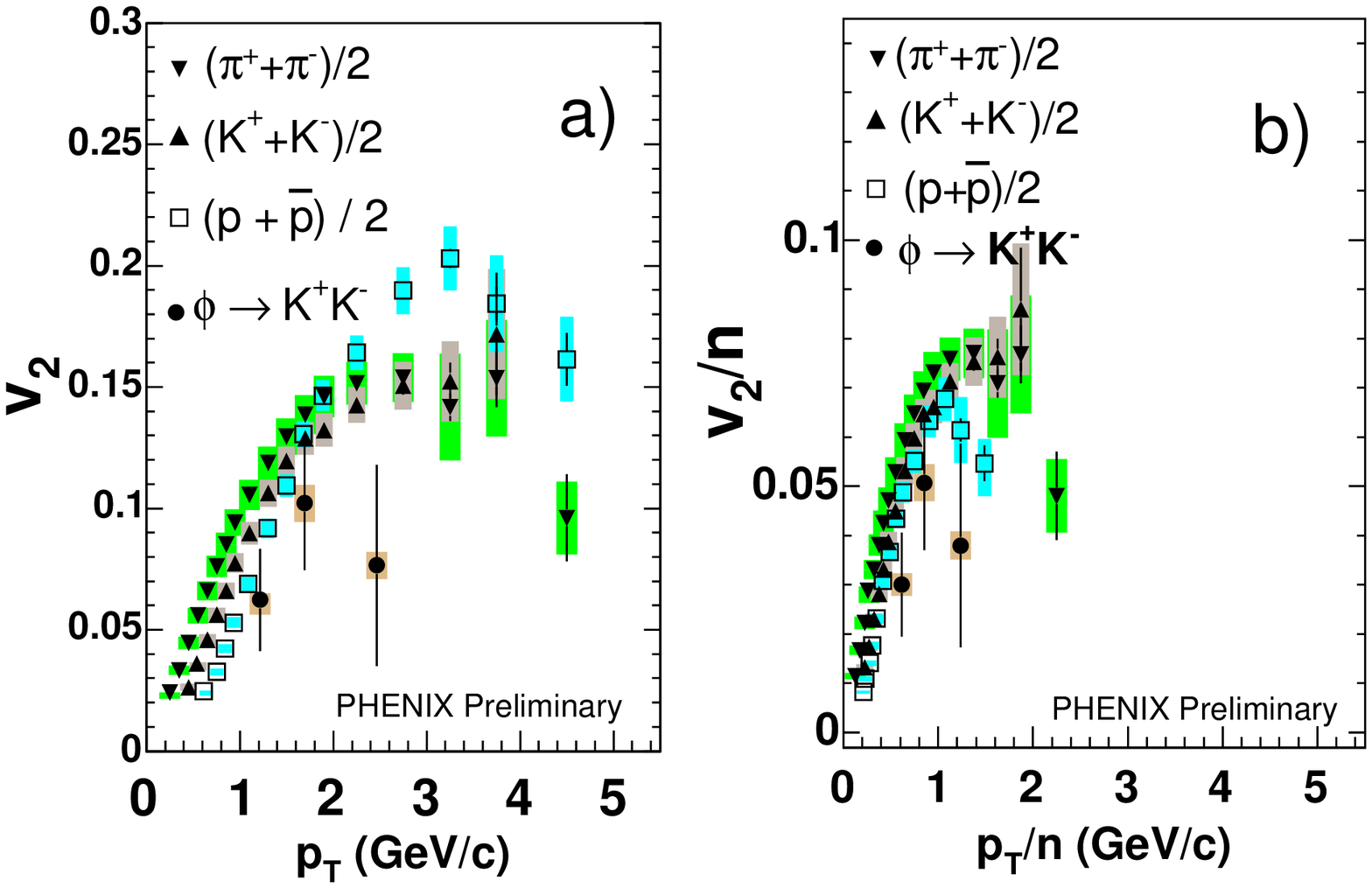}&\hspace*{-0.4cm}
\includegraphics[scale=0.4,angle=0]{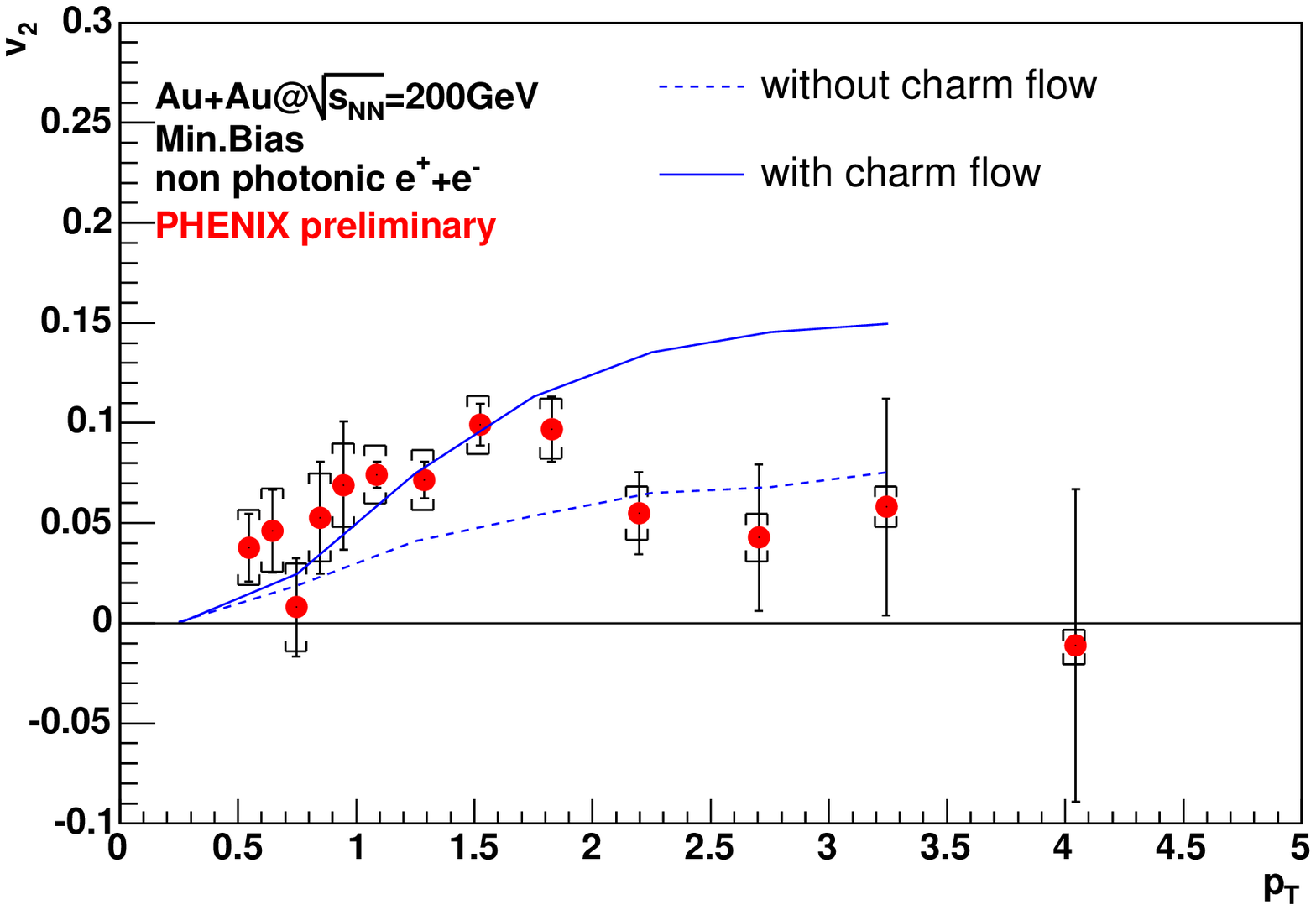}
\end{tabular}
\end{center}\vspace*{-0.25in}
\caption[]{a) $v_2$ vs $p_T$ for $\phi\rightarrow K^+ K^-$ and other particles as indicated in $\sqrt{s_{NN}}=200$ GeV minimum bias Au+Au collisions~\cite{PXphiQM05}; b) same data plotted as $v_2/n$ vs $p_T/n$ where $n$ is the number of constituent quarks. c) (right) $v_2$ of non-photonic $e^+ + e^-$ from semi-leptonic heavy flavor decay~\cite{PXcharmQM05}. The solid curve is if both charm and light quarks flow, while the dashed curve is if only the light quark flows~\cite{GrecoKoRapp}. \label{fig:whatflows}}
\end{figure}
Another striking hint as to what is flowing at RHIC is given in Fig.~\ref{fig:whatflows} where the $\phi$ meson (Fig.~\ref{fig:whatflows}a)~\cite{PXphiQM05} and charm particles (Fig.~\ref{fig:whatflows}c) detected by their large semi-leptonic decay~\cite{CCRS,HinchLL,BG,MJTS96,PXPRL88} exhibit the same $v_2$ as other particles~\cite{PXPRC72charm,PXcharmQM05} indicating for the $\phi$ that the flow is partonic because the hadronic interaction cross section of the $\phi$ meson is much smaller than for the other hadrons both in the constituent quark model and as measured in photoproduction~\cite{Lipkin,Joos,KT67}. For the charm particles, the $v_2$ of the decay electrons follows the $v_2$ of the $D (c\bar{d},c\bar{u})$ mesons~\cite{Batsouli,GrecoKoRapp}, but due to their different masses the $c$ and $\bar{u}, \bar{d}$ quarks have different momenta for the same velocity required for formation by coalescence, so the $v_2$ of the $D$ mesons is reduced to that of the light quark at lower $p_T$ if the $c$ quark itself does not flow. The data (Fig.~\ref{fig:whatflows}b) favor the flow of the $c$ quark, but clearly lots more work remains to be done to improve both the measurement and the theory. As no data for either $\phi$ or charm flow exist at the SPS it is difficult to know whether the observation of what must be partonic flow for these particles at RHIC is qualitatively different from flow at the SPS. 

    The last variety of flow is radial flow, which was illustrated in Fig.~\ref{fig:radialflow}. Particles or partons which travel with the flow velocity, $\vec{u}$, acquire increasingly larger kinetic energies with increasing mass so that the $\mean{p_T}$ increases with increasing mass and centrality which is manifested by the flattening of the $p_T$ distribution at low $p_T$~\cite{HeiselbergLevy,BorghiniPreprint}, an apparent increase in the inverse-slope:
    \begin{equation}
    {\mean{p_T}\over 2}\approx T_{\rm apparent}\approx T+{1\over 2} m \mean{u_{\perp}^2}
    \qquad . \label{eq:Tapp} \end{equation}
     While the elliptic flow is only sensitive to the geometry of the flow profile, the radial flow is sensitive to the velocity and hence puts a requirement on hydro models to explain the $p_T$ spectra for identified particles as well as the observed $v_2$. While some hydro models rise to the challenge, there is no hydro model which also explains the size of the interaction volume measured by Bose-Einstein correlations (HBT)~\cite{egPXWPHBT}, though all agree that the flow velocity $\mean{u_{\perp}}\sim 0.5$ (in units of the speed of light).  Also there is no definitive explanation of the decrease of $v_2$ away from mid-rapidity~\cite{PHWP,STWP} which was not discussed here.  

\subsubsection{Hydrodynamics and Bose-Einstein correlations}
When two identical bosons (usually $\pi^+ \pi^+$ or $\pi^- \pi^-$) occupy nearly the same coordinates in phase space, constructive interference occurs due to symmetry of the wave-function imposed by Bose-Einstein statistics, a quantum-mechanical effect. This is measured with relatively soft particles ($p_T \leq 1$ GeV/c) using 
a two-particle correlation function, ${\cal C}_2^{\rm BE}$~\cite{forexperts}, which represents the probability of detecting two identical pions with momenta $\vec{p}_1$, $\vec{p}_2$, $\vec{q}=\vec{p}_1 -\vec{p}_2$, divided by the product of the probabilities for detecting the individual pions (which is the probability of detecting two such pions if they are uncorrelated). For a classical chaotic source, ${\cal C}_2^{\rm BE}=1$, while for a quantum source of identical bosons, ${\cal C}_2^{\rm BE}=2$~\cite{Pratt}. For a gaussian source,  ${\cal C}_2^{\rm BE}$ takes the form~\cite{Pratt,Bertsch}:
\begin{equation}
{\cal C}_2^{\rm BE}(q)= { {{\cal P}(\vec{p}_1, \vec{p}_2)} \over { {\cal P}(\vec{p}_1)\,{\cal P}(\vec{p}_2)} } =1+\lambda \exp(-R^2_{\rm side} q^2_{\rm side} -R^2_{\rm out} q^2_{\rm out} -R^2_{\rm long} q^2_{\rm long} ) 
\end{equation} 
where the coordinate system is chosen so that $q_{\rm long}$ is along the beam direction, $q_{\rm out}$ is parallel to the average transverse momentum of the pair, $\vec{k}_T=(\vec{p}_{T1} +\vec{p}_{T2})/2$, and $q_{\rm side}$ is the other transverse component, perpendicular to $\vec{k}_T$. The parameter $\lambda$ measures the strength of the correlation and would have the value $\lambda= 1$  for a fully chaotic quantum source. The extracted source dimensions are commonly called HBT radii, after a similar technique developed by radio astronomers to measure the angular size of radio sources~\cite{HBT}. 

	For a time-dependent source, the energy difference, $q_0=E_1 - E_2$, of the two pions must be taken into account. The radii then become functions of $\vec{k}_T$; and $R_{\rm out}$ acquires an additional dependence on the time duration of particle emission, $\tau$, such that for a cylindrical source, 
$R^2_{\rm out}=R^2_{\rm side} + \beta_T^2 \tau^2$, where $\beta_T$ is the average velocity of the pair, corresponding to $\vec{k}_T$. It is important to note that dynamical effects due to final state interactions can be large, which  makes the interpretation of these  measurements a very specialized subject~\cite{specialized}. One such example is the case of a source which expands due to collective flow.  The HBT radii do not increase as a function of velocity but instead decrease, since they correspond to ``lengths of homogeniety'', or regions of the source which emit particles of similar momentum~\cite{Sinyukov}. 

\begin{figure}[!htb]
\begin{center}
\includegraphics[scale=0.70,angle=0]{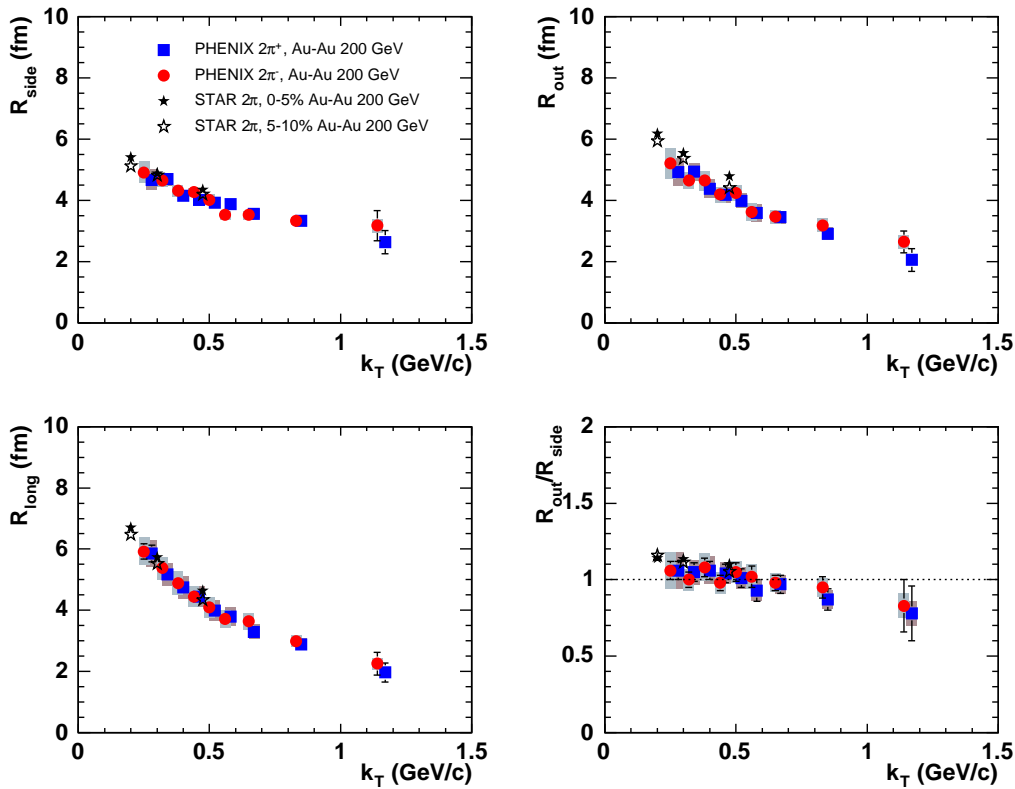}
\end{center}\vspace*{-0.25in}
\caption[]{The $k_T$ dependence of HBT radii~\cite{PXWP} for PHENIX~\cite{Adler:2004rq} 0--30\% most central, and STAR~\cite{STARPRC71}, as labeled, in Au+Au collisions at $\sqrt{s_{NN}}=200$ GeV  \label{fig:HBTdata}}
\end{figure}
    HBT measurements in search of the QGP were primarily motivated by theoretical predictions of a large source size and/or a long duration of  particle emission which would result from the presence of a long-lived mixed phase during a first-order phase transition from a QGP to a gas of hadrons~\cite{PXWP,Bertsch,PrattQGP,DRMG96}. This would be indicated by large values of $R_{\rm out}/R_{\rm side} \geq 1.5-3$.  However, beautiful measurements of Bose-Einstein correlations with pions at RHIC~\cite{Adler:2004rq,STARPRC71} show no such effect (see Fig.~\ref{fig:HBTdata})~\cite{PXWP}. Although a clear decrease of all the HBT radii with increasing $k_T$ is seen, indicating an expanding source, $R_{\rm out}/R_{\rm side}$ hovers close to or below 1, in disagreement with the hydro models~\cite{egPXWPHBT,HydroReview}. This is the ``RHIC HBT puzzle''~\cite{MGrev01}. 
    
    Does the HBT puzzle indicate a problem with hydrodynamics calculations, or the absence of a first-order phase transition, or both? With regard to the former,  hydro-inspired parameterized fits seem to be able to explain the data~\cite{Buda-Lund}. However, the exact solution of 3+1 dimensional relativistic hydrodynamics, using the correct initial conditions, EOS, and realistic hadronization is a non-trivial, non-linear mathematical problem which has not yet been solved~\cite{Weiner05}. Clearly, the RHIC HBT puzzle is a message to the theoretical community that the time has come to make the effort. Regarding the absence of a first-order phase transition, this seems to be consistent with the latest thinking~\cite{Krishna99} (see Fig.~\ref{fig:phaselat}a) as well as with measurements of event-by-event fluctuations. 
    
\begin{figure}[!htb]
\begin{center}
\begin{tabular}{cc}
\includegraphics[scale=0.9,angle=0]{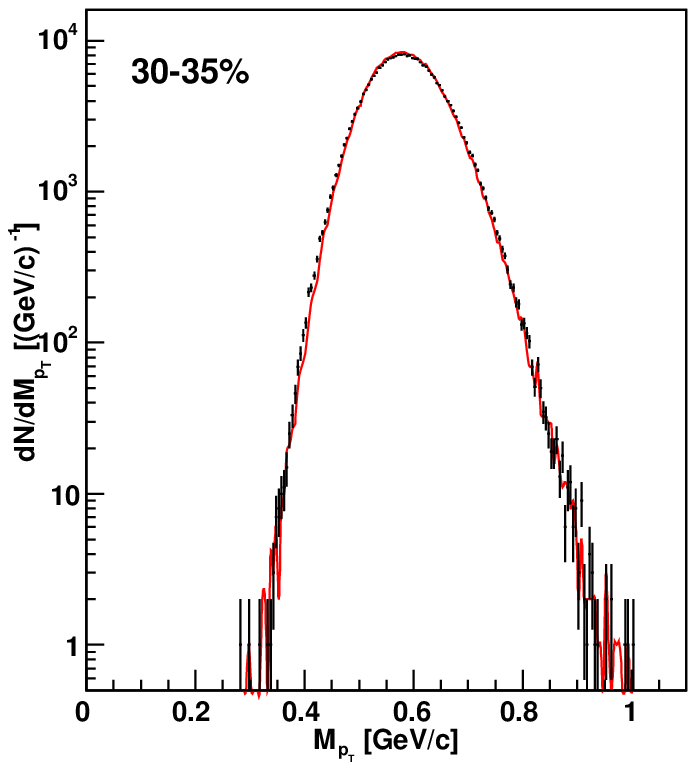}&\hspace*{0.25in}
\includegraphics[scale=0.4,angle=0]{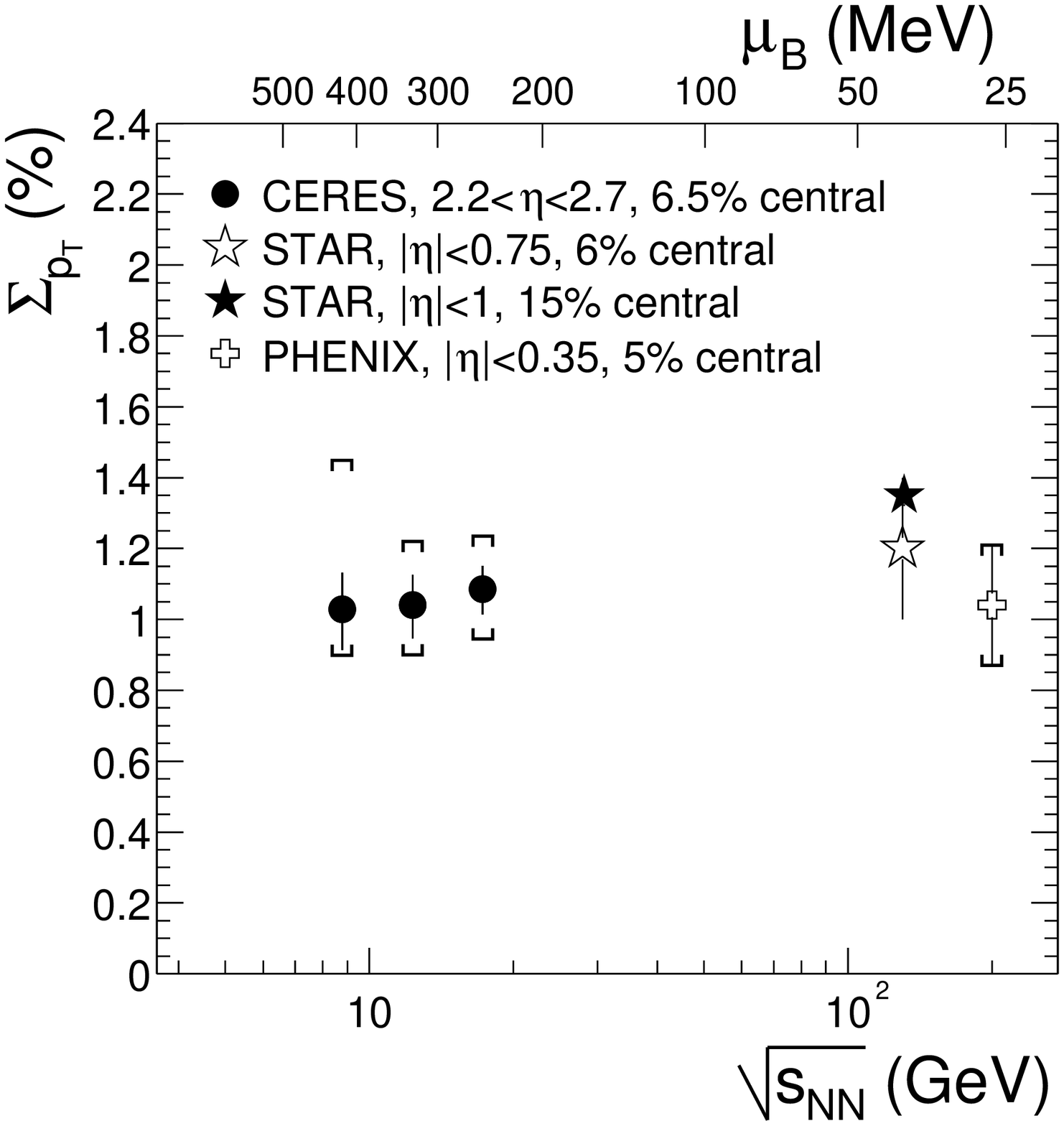}
\end{tabular}
\end{center}\vspace*{-0.25in}
\caption[]{ a) (left) $M_{p_T}$ for Au+Au collisions at $\sqrt{s_{NN}}=200$ GeV with 30--35\% centrality~\cite{Adler:2003xq}: measured events (\fullsquare\ ), mixed events (histogram). b) (right) Compilation~\cite{CERESQM04} of $\Sigma_{p_T}$ (in per-cent)  for SPS (Pb+Au) and RHIC (Au+Au) experiments, as indicated, as a function of $\sqrt{s_{NN}}$ and $\mu_B$ \label{fig:MpTfluct} }
\end{figure}

    Measurements of the distribution of the event-by-event average transverse momentum of charged particles provide severely small limits on possible non-random fluctuations due to a phase transition. For events with $n$ detected charged particles with magnitudes of
transverse momenta, $p_{T_i}$, the event-by-event average $p_T$, denoted $M_{p_T}$~\cite{NA49PLB459}, is defined as:   
\begin{equation}
M_{p_T}=\overline{p_T}={1\over n} \sum_{i=1}^n p_{T_i}
\qquad .\label{eq:defMpT}
\end{equation}
A typical $M_{p_T}$ distribution for a measurement in one 
centrality class of $\sqrt{s_{NN}}=200$ GeV Au+Au
collisions in PHENIX~\cite{Adler:2003xq} is shown in Fig.~\ref{fig:MpTfluct}a (data points) compared to a mixed-event distribution (histogram) which defines a baseline for random fluctuations. 
The difference between the data and the
mixed-event random baseline distribution is barely visible to the
naked eye. The non-random fluctuation is quantified by the difference between the measured and random values of the variance divided by the squared mean:  
\begin{equation}
\Sigma^2_{p_T}={ \sigma^2_{M_{p_T}} \over {\mu^2} } - \left ( { \sigma^2_{M_{p_T}} \over {\mu^2} }\right )_{\rm random} \qquad ,
\label{eq:sigsq}
\end{equation}
where $\mu=\mean{M_{p_T}}$  and $\sigma^2_{M_{p_T}}=\mean{M_{p_T}^2}-\mean{M_{p_T}}^2$. 
If the the entire non-random fluctuation 
were due to fluctuations of the temperature ($T$) of the initial state~\cite{Korus:2001au}, with
r.m.s. variation relative to the mean, $\sigma_{T}/\langle
T\rangle$, then 
$\Sigma_{p_T}=\sigma_{T}/\mean{T}$. Fig~\ref{fig:MpTfluct}b~\cite{CERESQM04}  shows measurements of $\Sigma_{p_T}$ for central collisions at SPS (Pb+Au) and RHIC (Au+Au) energies, with similar small values ($\sim 1$\%) observed in all cases. These results put stringent limits on the critical fluctuations that were
expected for a sharp phase transition, both at SPS
energies and at RHIC, 
but are consistent with the expectation from lattice QCD that the transition is a smooth crossover~\cite{Krishna99}.
\subsubsection{Summary and conclusions on flow}
          
     In summary, collective flow is observed in A+A collisions at all c.m. energies; but, uniquely at RHIC, hydrodynamics with full thermalization appears to describe the $v_2$ and $p_T$ spectra, with constituent quarks as the flowing objects. Since hydrodynamics also provides a decent description of p-p data~\cite{Weiner05}, much experimental and theoretical work  remains to be done before the smoking gun of the QGP can be claimed from the success of hydrodynamics in Au+Au collisions at RHIC. Suggestions for further  study include measurements of $v_2$ with U+U collisions, where the highly deformed Uranium nuclei should provide a strong $v_2$ signal even for central collisions~\cite{HydroReview} and the search for collective effects in p-p collisions, where due to the difficulty of defining the reaction plane, the search for radial flow in central (high multiplicity) p-p collisions might be the most promising. Also, there is still no unified hydrodynamic description at RHIC of $v_2$, $p_T$ spectra and spatial sizes determined by HBT. Even if it should turn out that constituent quarks flow at RHIC but not at the SPS, one must heed the warning of Ref.~\cite{STWP}, ``it must be kept in mind that constituent quarks are not partons: they are effective
degrees of freedom normally associated with chiral symmetry breaking and confinement,
rather than with the deconfinement of a QGP.'' Nevertheless, the description of the medium produced at RHIC as a ``perfect fluid"~\cite{THWPs,Thoma05} seems appropriate. 
\subsection{Jet Quenching}

\begin{figure}[!htb]
\begin{center}
\begin{tabular}{cc}
\includegraphics[scale=0.36,angle=0]{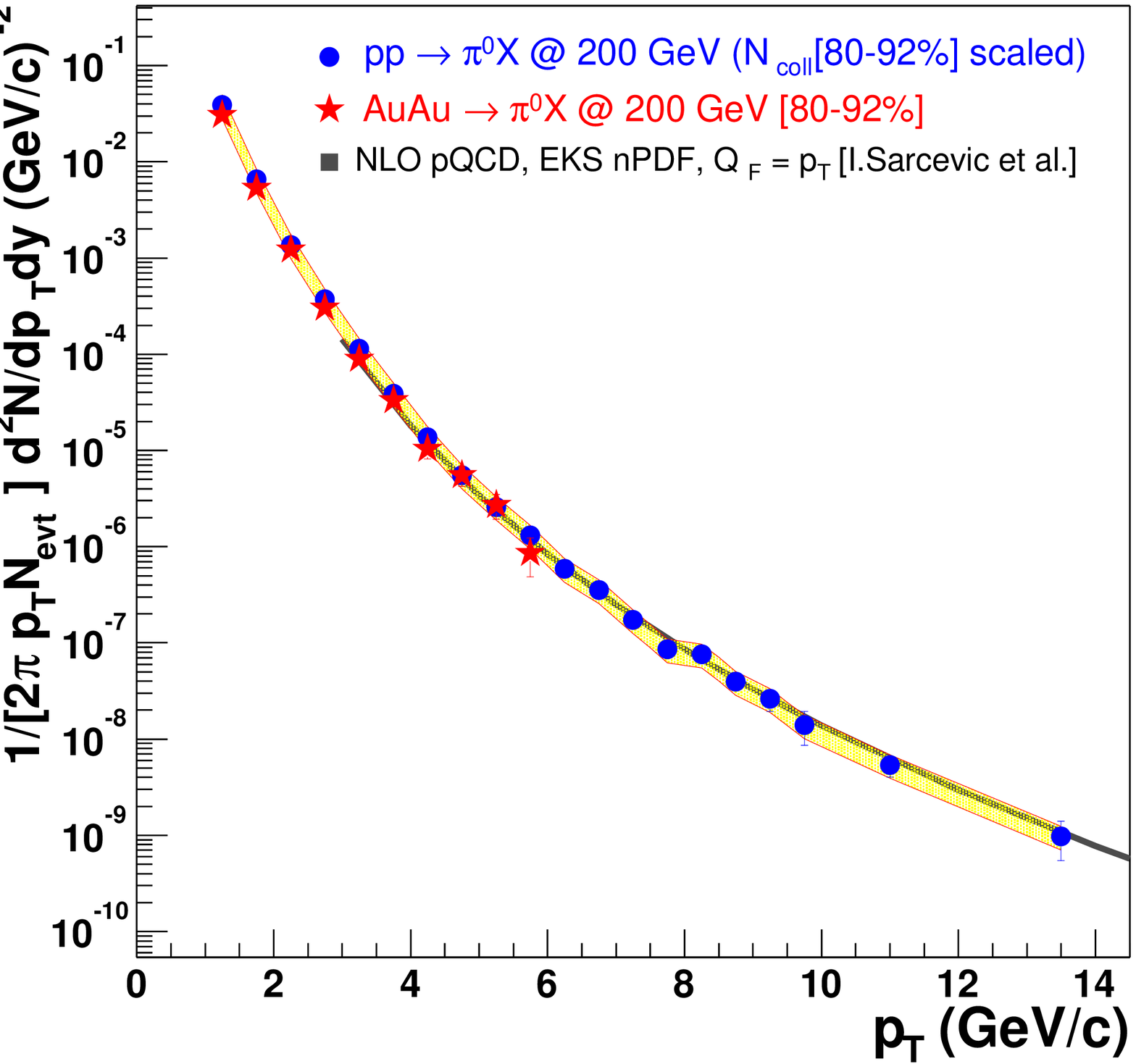}&
\includegraphics[scale=0.36,angle=0]{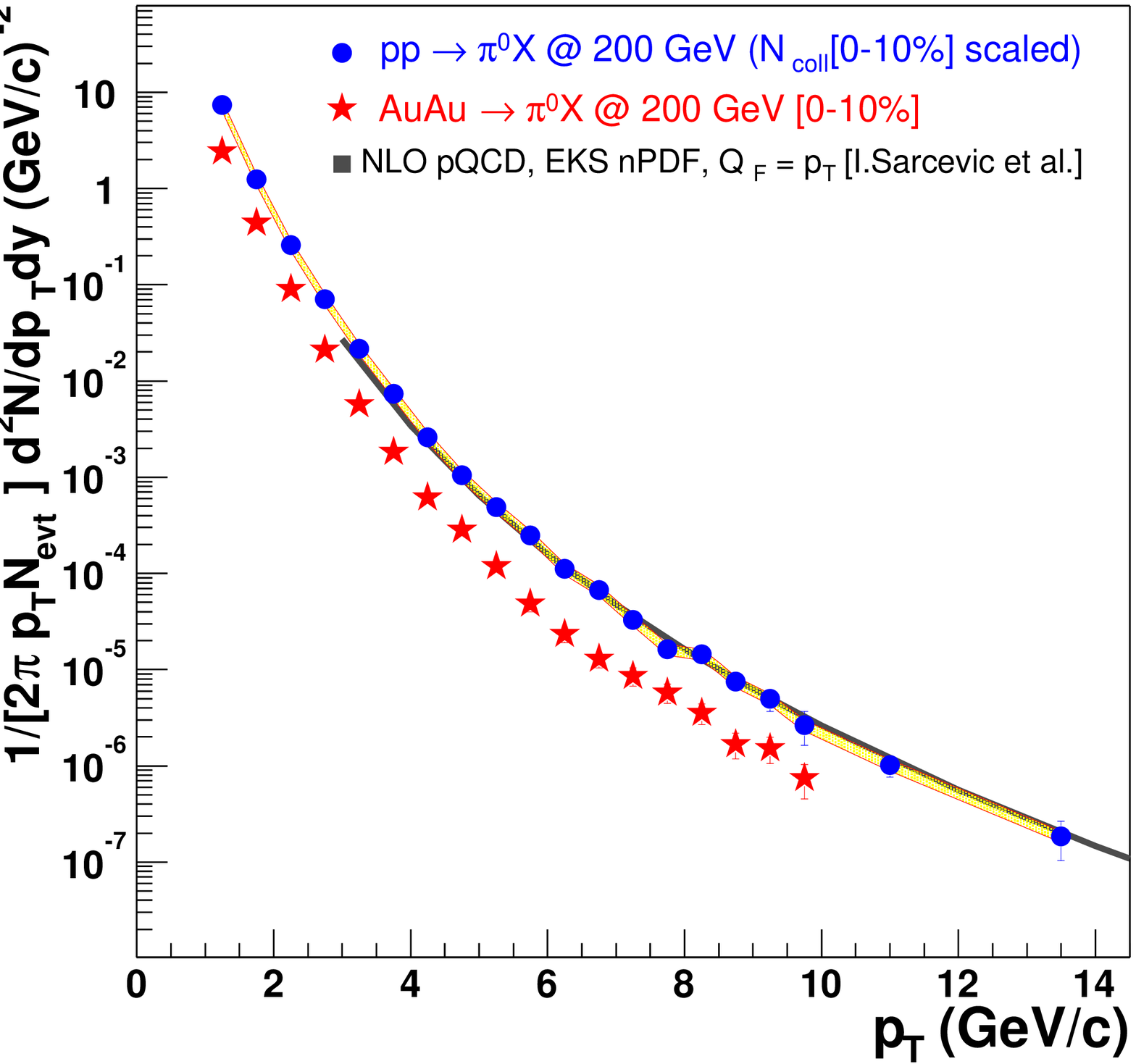}
\end{tabular}
\end{center}\vspace*{-0.25in}
\caption[]{a)(left) Invariant yield of $\pi^0$ in Au+Au peripheral collisions at $\sqrt{s_{NN}}=200$~GeV (stars) and compared to p-p collisions, scaled by  $N_{coll}$ (circles)~\cite{PXpi0AuAu200,PXpppi0}. The black line is a pQCD calculation.~\cite{Ina03} The yellow band around the scaled p-p points represents the overall normalization uncertainties. b) (right) The same for central collisions.    \label{fig:PXpi0spectra}}
\end{figure}

    One of the major, arguably 
{\it the} major discovery at RHIC, was the observation of jet quenching~\cite{PXppg003,seealsoQM01} (Sec.~\ref{sec:quenching}) by the suppression of $\pi^0$ and non-identified charged hadrons in Au+Au collisions at mid-rapidity for large  transverse momenta, $p_T > 2$ GeV/c. In p-p collisions, particles with $p_T\geq 2$ GeV/c at mid-rapidity (perpendicular to the collision axis) are produced from states with two roughly back-to-back jets which are the result of hard-scattering of the constituents of the nucleon (current-quarks and gluons) as described by QCD (Sec.~\ref{sec:hardscattering}). The suppression of high $p_T$ particles  was observed by all four RHIC  experiments~\cite{PXWP,BRWP,PHWP,STWP} and is well calibrated by using measurements in p-p, d+Au and Au+Au collisions at $\sqrt{s_{NN}}=200$ GeV.   
    The $p_T$ spectra of $\pi^0$ from p-p collisions at $\sqrt{s}=200$ GeV~\cite{PXpppi0} and Au+Au collisions at $\sqrt{s_{NN}}=200$ GeV~\cite{PXpi0AuAu200} are shown in Fig.~\ref{fig:PXpi0spectra}.  
Note that the p-p measurements for $p_T \geq 3$ GeV/c are in excellent agreement with a pQCD calculation~\cite{Ina03}.

    Since hard scattering is point-like, with distance scale $1/p_T< 0.1$ fm, the cross section in p+A (B+A) collisions, compared to p-p, should be simply proportional to the relative number of possible point-like encounters~\cite{MMay}, a factor of $A$ ($BA$) for p+A (B+A) minimum bias collisions. For semi-inclusive reactions in centrality class $f$ at impact parameter $b$, the scaling is proportional to $T_{AB}(b)$, the overlap integral of the nuclear thickness functions~\cite{Vogt99}, where $\mean{T_{AB}}_{f}$ averaged over the centrality class is:
    \begin{equation}
\langle T_{AB}\rangle_{f}=\frac{\displaystyle\int_{f} T_{AB}(b)\,d^2b}{\displaystyle\int_{f} (1- e^ {-\sigma_{NN}\,T_{AB}(b)})\, d^2 b}=\frac{\langle N_{coll}\rangle_f}{\sigma_{NN}} \quad, 
\label{eq:TABf}
\end{equation}
and where $\langle N_{coll}\rangle_f$ is the average number of binary nucleon-nucleon
inelastic collisions, with cross section $\sigma_{NN}$, in the centrality class $f$.
This leads to the description of the scaling for point-like processes as binary-collision
(or $N_{coll}$)  scaling. This description is convenient, but confusing, because the scaling has nothing to do with the inelastic hadronic collision probability, it is proportional only to the geometrical factor $\mean{T_{AB}}_{f}$ (Eq.~\ref{eq:TABf}). 

    Effects of the nuclear medium, either in the initial or final state, may modify the point-like scaling. This is shown rather dramatically in Fig.~\ref{fig:PXpi0spectra}, where for peripheral collisions, Fig.~\ref{fig:PXpi0spectra}a, the  AuAu data are in excellent agreement with the point-like scaled p-p data while for central collisions, Fig.~\ref{fig:PXpi0spectra}b, the Au+Au data are suppressed relative to the scaled p-p data by a factor of $\sim 4-5$ for $p_T\geq 3$ GeV/c. A quantitative evaluation of the  suppression is made using the ``nuclear modification factor'', $R_{AB}$, the ratio of the measured semi-inclusive yield to the point-like scaled p-p cross section: 
\begin{equation}
R_{AB} = \frac{dN_{AB}^P}{\langle T_{AB} \rangle_{f} \times d\sigma_{NN}^P}
       = \frac{dN_{AB}^P}{\langle N_{coll} \rangle_{f} \times dN_{NN}^P} \label{eq:RAB}
\end{equation}
where $dN_{AB}^P$ is the differential yield of a point-like process $P$
in an $A+B$ collision and $d\sigma_{NN}^P$ is the cross section of $P$ in an $NN$ (usually p-p)  collision. For point-like scaling, $R_{AB}=1$.

\begin{figure}[!thb]
\begin{center}
\begin{tabular}{cc}
\includegraphics[scale=0.36,angle=0]{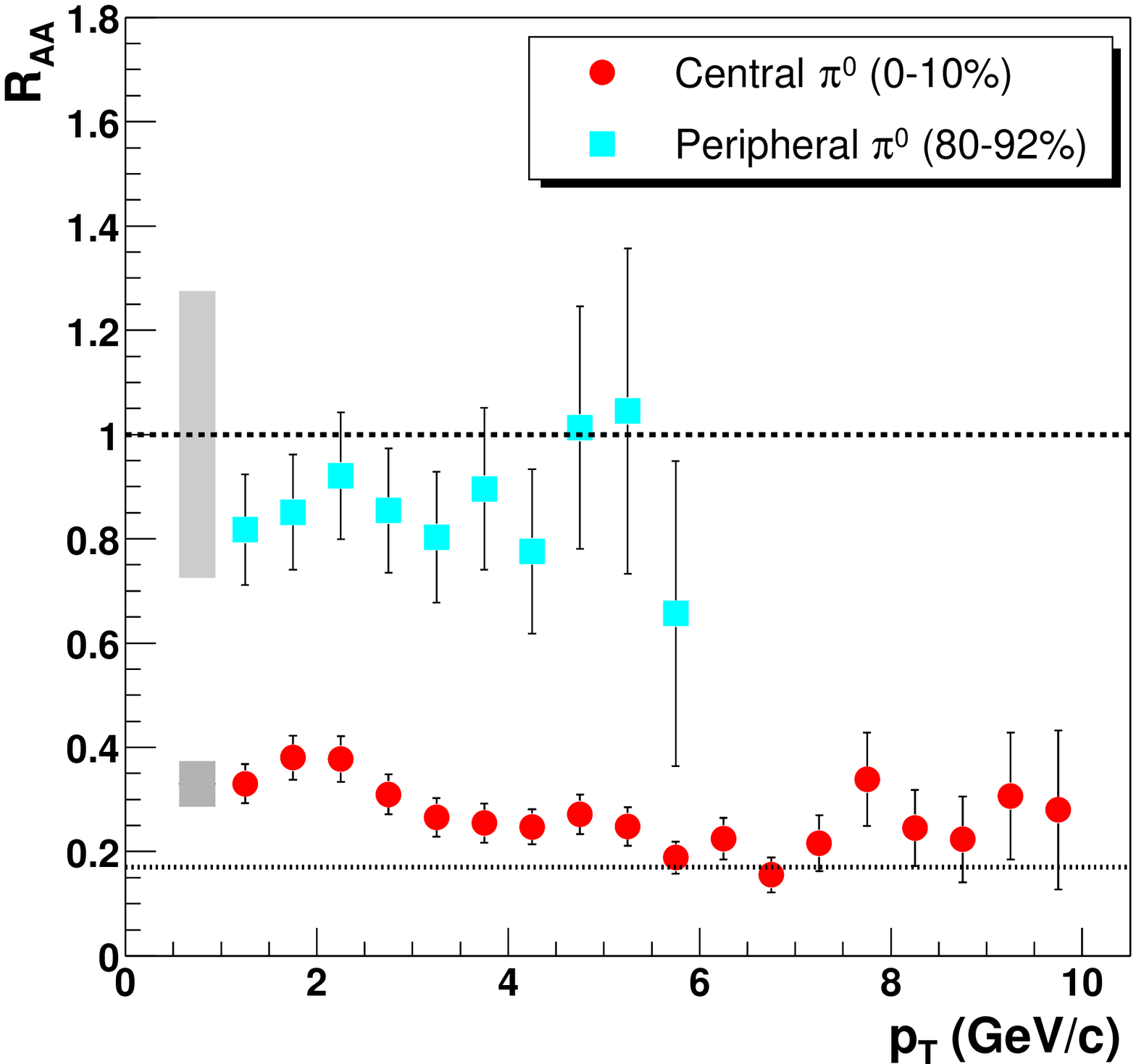}&
\includegraphics[scale=0.40,angle=0]{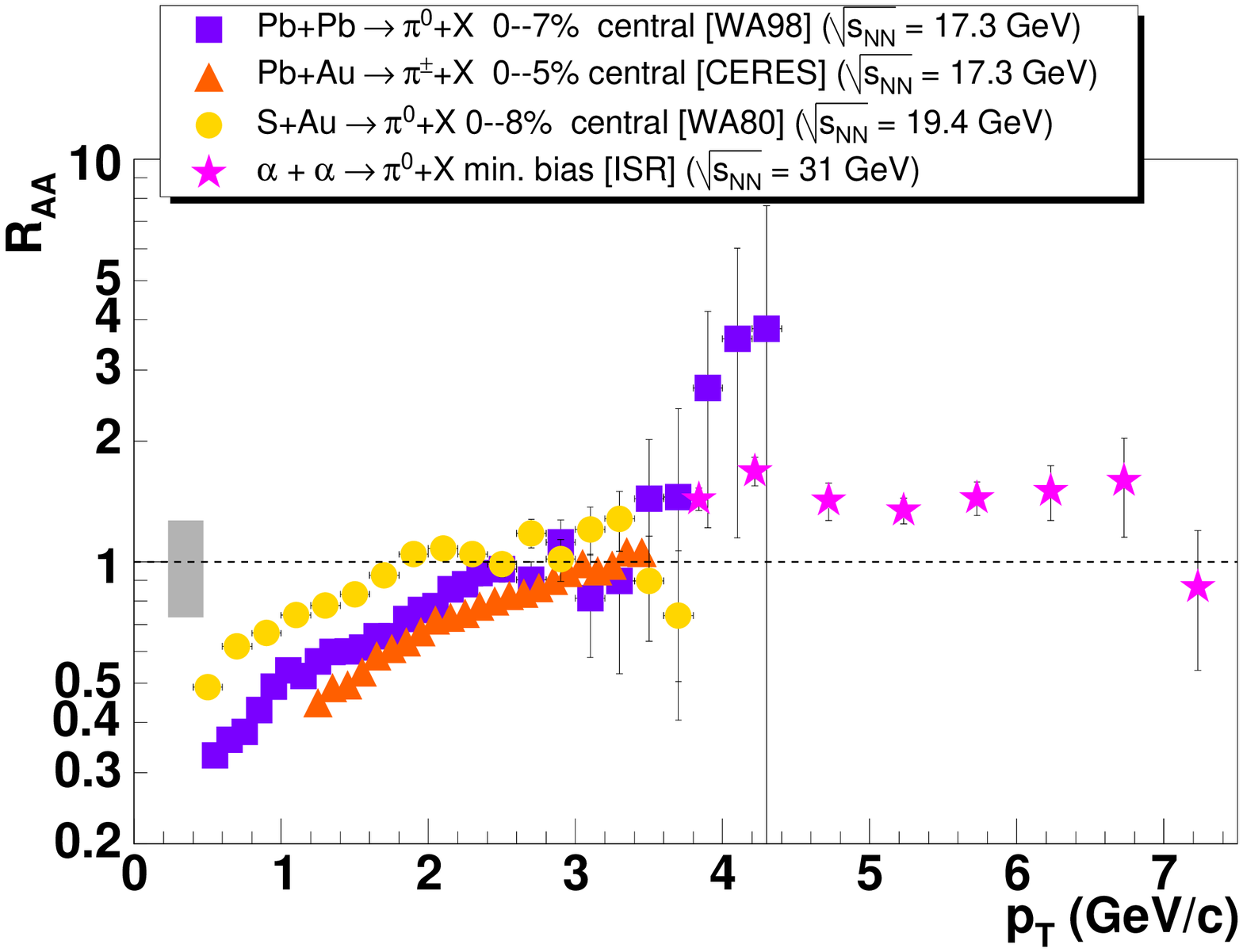}
\end{tabular}
\end{center}\vspace*{-0.25in}
\caption[]{a) (left) Nuclear modification factor $R_{AA}(p_T)$ for $\pi^0$ in central \fullcircle and peripheral \fullsquare~Au+Au at $\sqrt{s_{NN}}=200$ GeV.~\cite{PXpi0AuAu200}. The error bars include all point-to-point experimental (p-p, Au+Au) errors. The shaded bands represent the fractional systematic uncertainties for each centrality which can move all the points at that centrality up and down together. b) (right) Compilation of $R_{AA} (p_T)$ for $\pi^0$ production in A+A collisions from Refs.~\cite{PXWP,DdEPLB04}. \label{fig:RAA1}}
\end{figure}

	The suppression, $R_{AA}$, of $\pi^0$ in central Au+Au collisions at RHIC (see Fig.~\ref{fig:RAA1}a), although quite dramatic in its own right, is even more dramatic when compared to previous data. All previous measurements of nuclear effects at high $p_T\geq 2$ GeV/c in p+A and A+A collisions at lower $\sqrt{s_{NN}}$ have given results which are larger than point-like scaling (Fig.~\ref{fig:RAA1}b), a situation called the `Cronin Effect'~\cite{Cronin} and thought to be due to the multiple scattering of the incident partons in the nuclear matter before the hard-collision~\cite{Krzy,Lev}. The suppression observed at RHIC is a totally new effect.

\begin{figure}[htb]
\begin{center}
\begin{tabular}{cc}
\includegraphics[scale=0.34,angle=0]{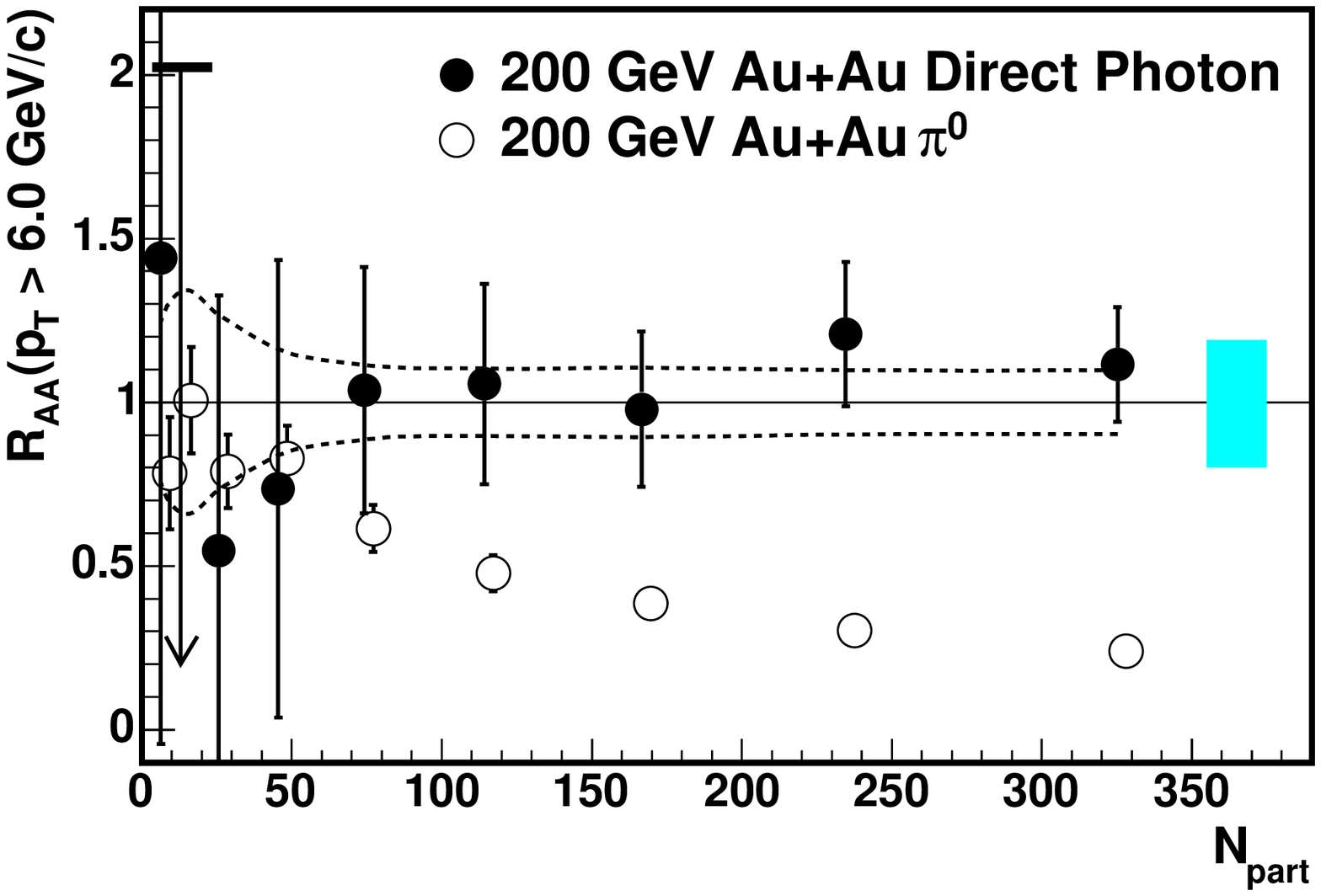}&
\includegraphics[scale=0.375,angle=0]{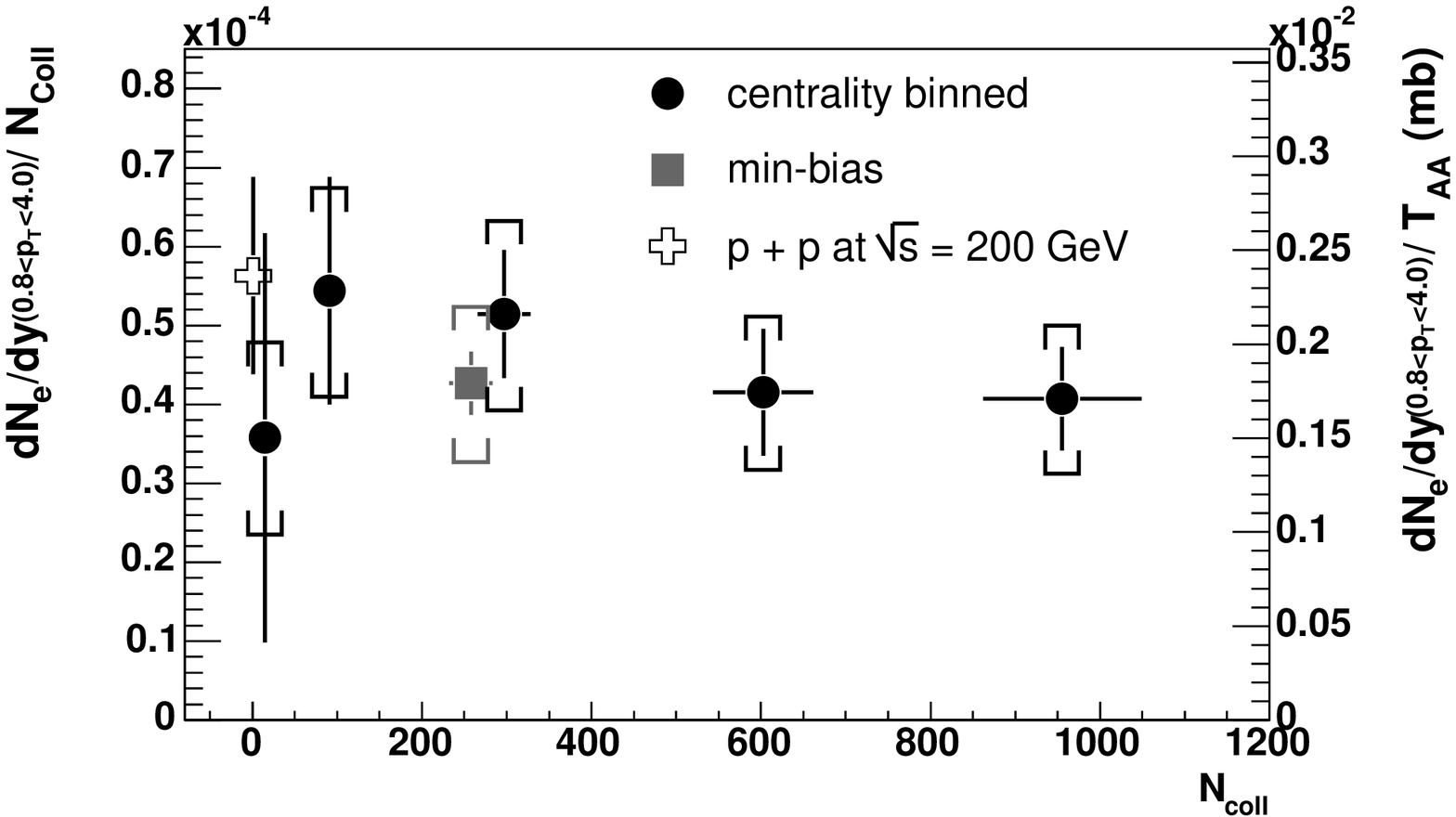}
\end{tabular}
\end{center}\vspace*{-0.25in}
\caption[]{a) (left) $R_{AA}$ for $p_T > 6$ GeV/c for prompt photon  \fullcircle and $\pi^0$ \opencircle production as a function of centrality ($N_{\rm part}$) in Au+Au collisions at $\sqrt{s_{NN}}=200$ GeV~\cite{PXdirg}. Experimental systematic errors for prompt photon are shown by dashes while the shaded region on the right indicates the theoretical scale error from the NLO calculation of direct photons in p-p~\cite{PXdirg}. b) Non-photonic electron yield ($0.8 < p_T < 4.0$~GeV/c), dominated by semi-leptonic charm decays, measured at $\sqrt{s_{NN}}=200$ GeV for Au+Au (scaled by $N_{\rm coll}$) and p-p collisions as a function of centrality ($N_{\rm coll}$). The right-hand scale shows the corresponding electron cross section per $N-N$ collision in the above $p_T$ range~\cite{PXcharmPRL94}. \label{fig:RAA2}}
\end{figure}

     Naturally, the first question asked about the RHIC suppression was whether it is an initial state effect, produced, for instance,  by `shadowing' of the structure functions in nuclei, or a final state effect produced by the medium. Although originally answered by all 4 RHIC experiments by the observation of no suppression in d+Au collisions~\cite{BRWP,PHWP,STWP,PXWP}, a clearer answer came from later measurements of QCD hard-photon production ($gq\rightarrow \gamma q$)~\cite{PXdirg}, Fig.~\ref{fig:RAA2}a, and the total yield of charm particles ($gg\rightarrow c\bar{c}$)  deduced from measurements of non-photonic $e^{\pm}$~\cite{PXcharmPRL94}  (Fig.~\ref{fig:RAA2}b) in Au+Au collisions.  
     Both these reactions are sensitive to the same initial state partons as $\pi^0$ production, but the photons, which are themselves elementary constituents  which participate in and emerge directly from the hard-scattering, do not interact with the final state medium; and the number of $c \bar{c}$  pairs produced does not depend on whether or not the $c$ or $\bar{c}$ later interact with the medium. Hence the fact that $R_{AA}=1$ as a function of centrality for these two reactions, which is dramatically different from the suppression of $\pi^0$, indicates that the $\pi^0$ suppression is produced by the interaction of the outgoing hard-scattered parton with the medium, losing energy before it fragments into a $\pi^0$. The energy loss is indicated by the shift to lower $p_T$ of the spectrum (Fig.~\ref{fig:PXpi0spectra}b) for a given yield. Another possiblity is that the energy loss is hadronic rather than partonic (to be discussed below). 
\subsubsection{More Surprises}          
\begin{figure}[!htb]
\begin{center}
\includegraphics[scale=0.75,angle=0]{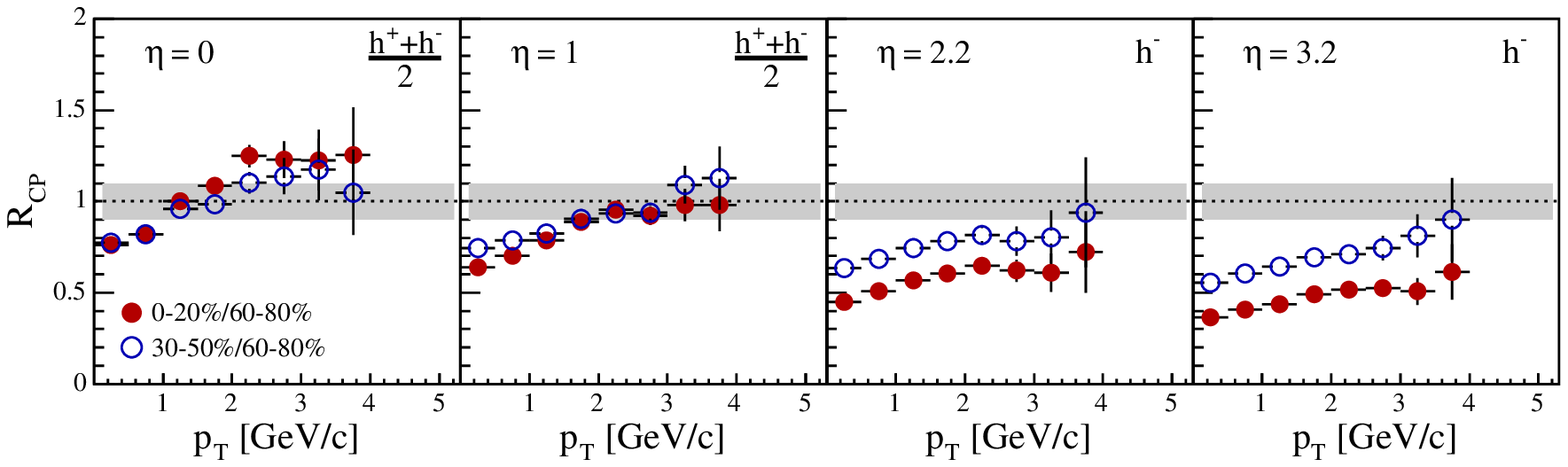}\\
\includegraphics[scale=0.75,angle=0]{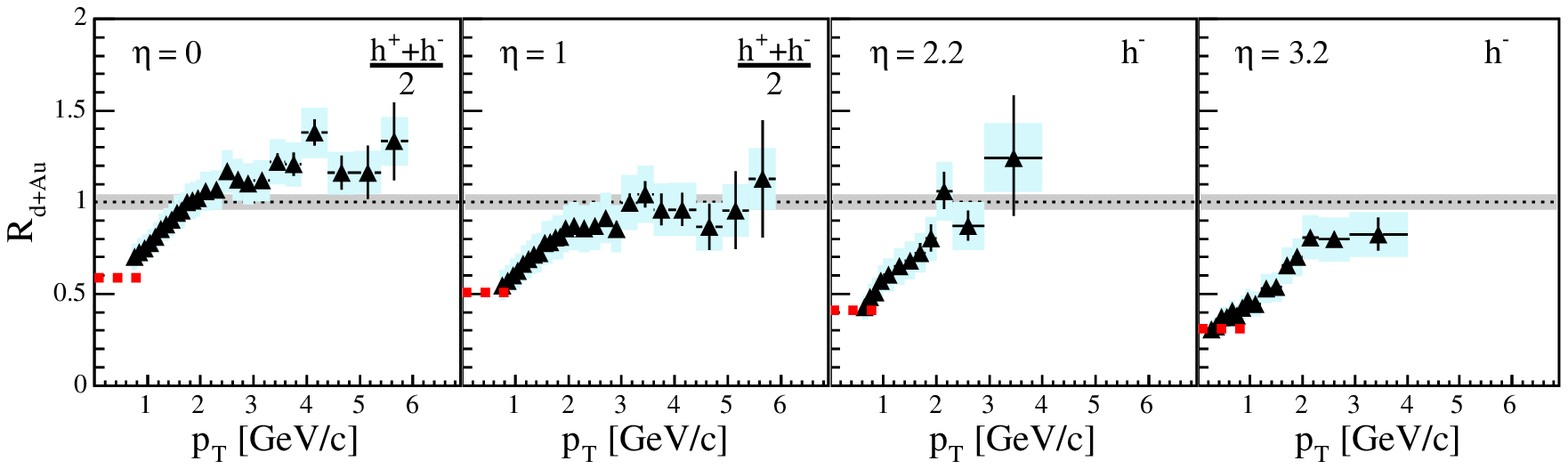}
\end{center}\vspace*{-0.25in}
\caption[]{a) (top) Central \fullcircle and semicentral \opencircle $R_{CP}(p_T)$ ratios for charged hadrons at pseudorapidities $\eta=0,1.0.2.2,3.2$ in d+Au collisions at $\sqrt{s_{NN}}=200$ GeV~\cite{BrahmsdAuPRL93}. b) (bottom)$R_{\rm dAu}(p_T)$ for same data. Statistical errors are shown with error bars. Systematic errors are shown with shaded boxes around the data points. The shaded band around unity  is the systematic error on $N_{\rm coll}$. \label{fig:BrahmsdAu}}
\end{figure}

     It turned out that the d+Au measurement was less clear-cut than sketched out in Fig.~\ref{fig:dAuAuAu}. For particles detected at 90$^\circ$ from the collision axis (at mid-rapidity, $\eta=0$), the d+Au results showed no suppression, possibly a Cronin enhancement~\cite{BrahmsdAuPRL93} (see Fig.~\ref{fig:BrahmsdAu}a). 
However for more forward emission, larger $\eta$, corresponding to lower values of $x$ (Eq.~\ref{eq:QCDabscat}), $R_{CP}$, the ratio of binary-scaled semi-inclusive yields of central compared to peripheral collisions, showed a huge suppression. This was exactly what was predicted if the initial state of the Au nuclei at RHIC were a color glass condensate (CGC), in which the gluon structure function at  small $x$ is not an incoherent superposition of nucleon structure functions but is limited with increasing $A$ by non linear gluon-gluon fusion ($gg\rightarrow g$) resulting from the overlap of gluons from several nucleons~\cite{GMc}. A CGC can produce either a Cronin-like enhancement or a suppression depending on the initial conditions and the kinematic range covered~\cite{KKTPRD68}. However, the suppression, if any, at large $\eta$ is much less pronounced when a measured p-p reference is used and $R_{\rm d+Au}$ is plotted (Fig.~\ref{fig:BrahmsdAu}b). 

	There are two lessons from this example. (i) It is important to obtain a p-p reference spectrum, since $R_{CP}$ just shows the relative change of A+A collisions from more peripheral to more central which may be confused by a Cronin effect for peripheral collisions. (ii) There may be yet another ``new state of nuclear matter'', the CGC, to deal with. This may be particularly important at the LHC where all the physics moves to much lower values of $x$.

\begin{figure}[!htb]
\begin{center}
\begin{tabular}{cc}
\includegraphics[scale=0.34,angle=0]{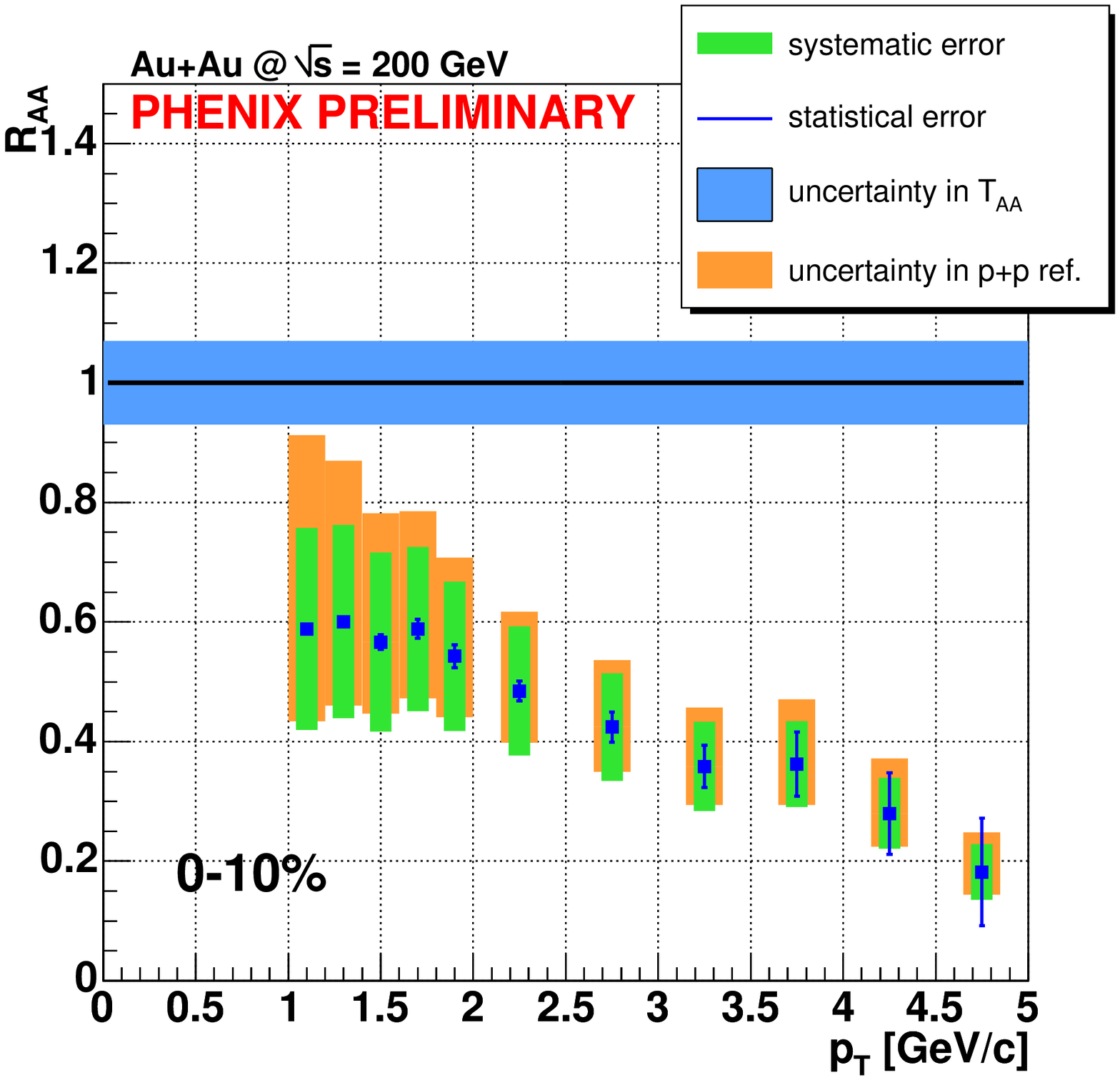}&
\includegraphics[scale=0.44,angle=0]{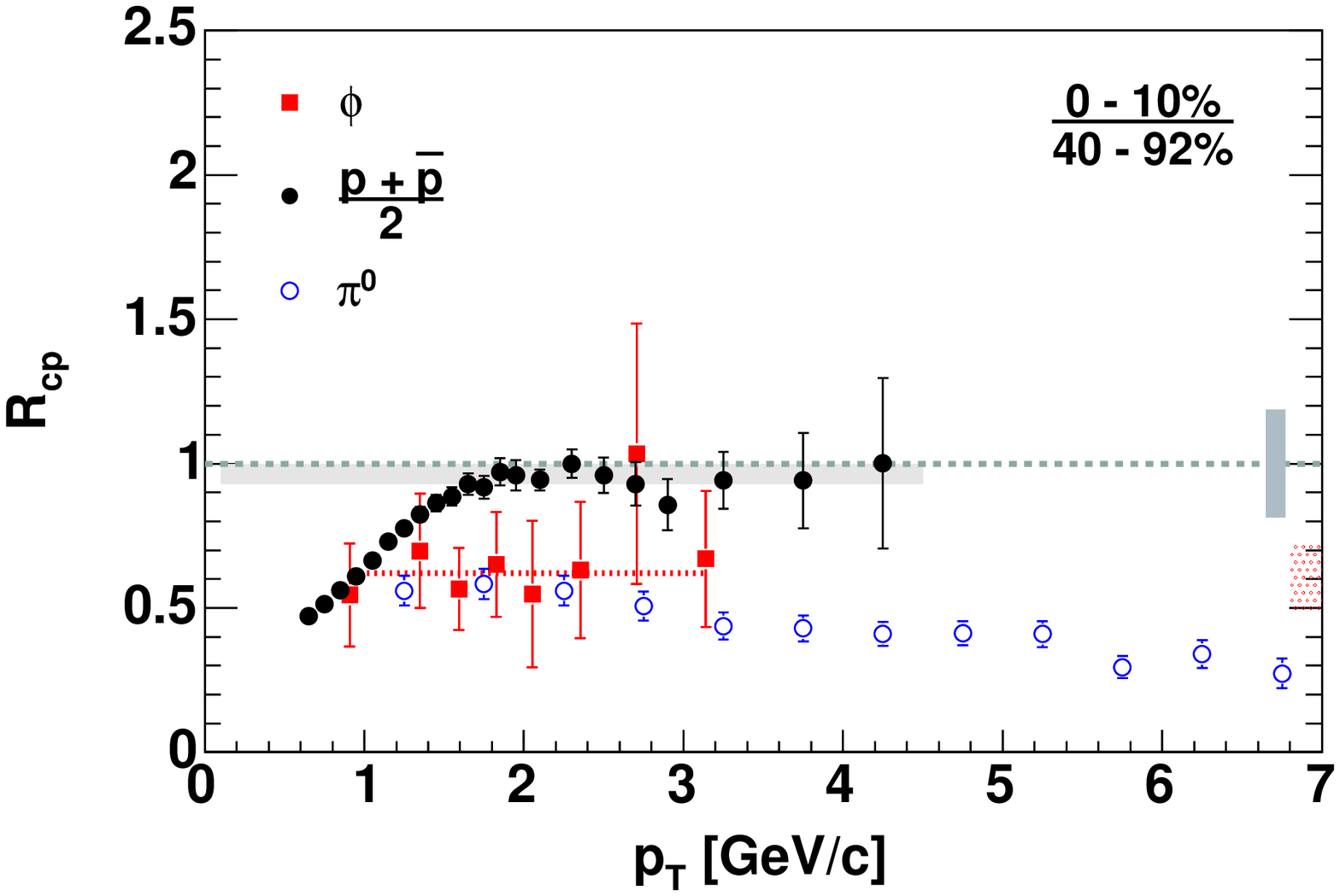}
\end{tabular}
\end{center}\vspace*{-0.25in}
\caption[]{a) (left) $R_{AA}(p_T)$ for non-photonic electrons in central collisions (0--10\% upper percentile) of Au+Au at $\sqrt{s_{NN}}=200$ GeV~\cite{PXcharmQM05}. b) (right) $R_{CP}$ for $\pi^0$ and $\phi$ mesons and for $p+\bar{p}$ in Au+Au at $\sqrt{s_{NN}}=200$ GeV~\cite{PXphiPRC72}. The vertical dotted bar on the right represents the systematic error in $N_{\rm coll}$. The shaded solid bar around $R_{CP}=1$ represents the systematic error by which the $\phi$ can move with respect to the other data. \label{fig:RAAphicharm}}
\end{figure}

     The d+Au data was not the only surprise at RHIC. An even bigger, and totally unpredicted, surprise was revealed with the measurement of $R_{AA}$ for identified particles other than the $\pi^0$ in the range $2\leq p_T\leq 5$ GeV/c   (Fig.~\ref{fig:RAAphicharm}). 
Charm particles, as measured by non-photonic $e^+ +e^-$, exhibit a suppression comparable to $\pi^0$ (Fig.~\ref{fig:RAAphicharm}a), indicating a surprisingly strong interaction with the medium~\cite{PXcharmQM05}. This is is a very recent result and has caused quite a commotion since the heavy $c$ quark should lose much less energy to the medium than light quarks and gluons~\cite{DjorPRL94}.  Another totally surprising result is that the protons and anti-protons (Fig.~\ref{fig:RAAphicharm}b) are not suppressed at all, but follow point-like scaling. (The $N_{coll}$ scaling of the $p_T$ spectra can be seen in Fig.~\ref{fig:radialflow}b). 

\begin{figure}[!htb]
\begin{center}
\begin{tabular}{cc}
\includegraphics[scale=0.35,angle=0]{figs/pizeroesrule.epsf}&
\includegraphics[scale=0.4,angle=0]{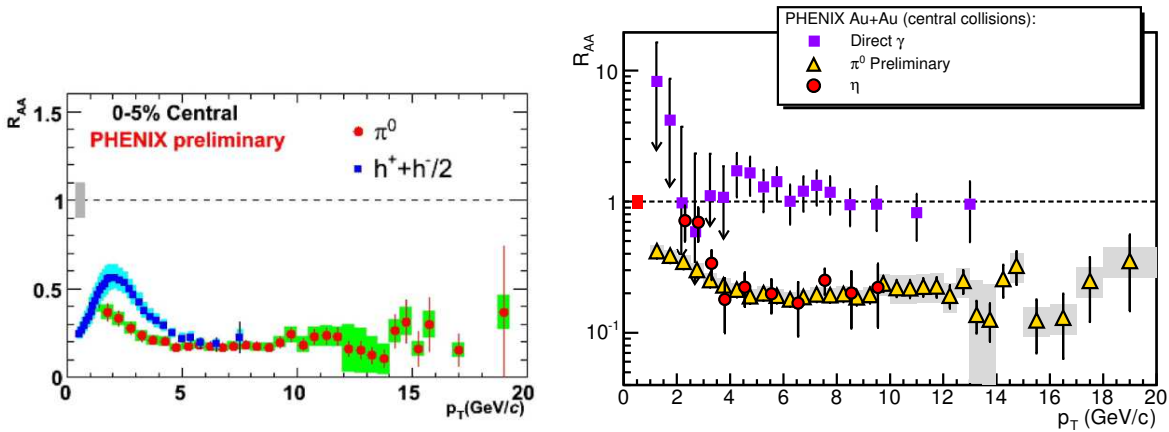}
\end{tabular}
\end{center}\vspace*{-0.25in}
\caption[]{a) (left) $R_{AA}(p_T)$ for $\pi^0$ and non-identified charged hadrons in Au+Au central collisions at $\sqrt{s_{NN}}=200$~GeV~\cite{PXQM05Maya}. b) (right) $R_{AA}(p_T)$ for $\pi^0$, prompt $\gamma$ and $\eta$ in the same reaction.~\cite{PXQM05Akiba} \label{fig:pizeroesrule}}
\end{figure}

	These two results show that particle identification is crucial to unraveling all aspects of the physics of heavy ion collisions, even for hard-scattering. In particular, the early measurements~\cite{PXppg003} showed a difference of suppression between $\pi^0$ and non-identified charged particles for $1.5\leq p_T\leq 4$ GeV/c (see Fig.~\ref{fig:pizeroesrule}a) which led to confusion.  However, that effect is now understood to be due to the proton `anomaly', which appears to vanish for $p_T\geq 6$ GeV/c where the $R_{AA}$ from non-identified charged particles comes into agreement with the value of $R_{AA}$ for $\pi^0$, which, notably, remains essentially flat at a value of $R_{AA}\sim 0.2$ from $3\leq p_T\leq 20$ GeV/c. Clearly, something is going on in the region $2\leq p_T\leq 6$ GeV/c (which has come to be called ``intermediate $p_T$") which needs to be understood. 

		One interesting proposal to explain the proton anomaly concerns coalescence of constituent quarks~\cite{Greco,Fries,Hwa}. If the light quarks have a thermal distribution, $\exp(-bp_T)$, then production of protons via coalesence in the intermediate $p_T$ region is favored over fragmentation from a power-law spectrum. Also, the natural $p/\pi$ ratio is close to 1 because a proton of a given $p_T$ is produced by the coalescence of 3 quarks of $p_T/3$, so there is no penalty compared to forming a pion from two quarks of $p_T/2$ because $[\exp(-bp_T/3)]^3=[\exp(-bp_T/2)]^2$. This model nicely predicted that fragmentation would become dominant above $\sim 5$ GeV/c so that the $p/\pi$ ratio would return to its normal fragmentation value (Fig.~\ref{fig:ratios}b) and that the suppression would become the same for $\pi^0$ and $p$, as nicely shown by the data (Fig.~\ref{fig:pizeroesrule}a). 
				The model also predicts, as observed~\cite{PXphiPRC72} (Fig.~\ref{fig:RAAphicharm}b), that the $\phi$ meson $(s\bar{s})$ would act like a $\pi^0$ meson, which has a similar constituent quark configuration, rather than like a proton which has a similar mass but a different quark configuration~\cite{PXphiPRC72}. Similarly, $R_{AA}$ for the $\eta$ meson~\cite{PXQM05Maya}, nicely tracks the $\pi^0$ (Fig.~\ref{fig:pizeroesrule}b). The model has an added attraction which is worth a direct quote~\cite{Fries}, ``Finally, our scenario requires the assumption of a thermalized partonic phase characterized by an exponential momentum spectrum. Such a phase may be appropriately called a quark-gluon plasma." Unfortunately,  other facets of the data do not fit this model, as the intermediate $p_T$ protons and mesons both seem to come from jet-like configurations rather then from soft processes (see later discussion).  
\subsubsection{Jet properties from two-particle correlations}

    The study of jet properties via two-particle correlations, pioneered at the CERN-ISR~\cite{egseeMJTHP04}, is used at RHIC rather than reconstructing jets using all particles within a cone of size $\Delta r=\sqrt{(\Delta \eta)^2 + (\Delta \phi)^2)}$.  This is because the large multiplicity in A+A collisions (recall Fig.~\ref{fig:collstar}b,c) results in a huge energy $\pi
(\Delta r)^2 \times {1 \over {2\pi}} {{dE_T} \over {d\eta}} \sim 375$ GeV in the standard cone, $\Delta r=1$, for Au+Au central collisions. Elliptical flow further complicates the jet measurement because the large non-jet background is modulated by $\cos2\phi$, which is comparable in width to the jets studied so far. 
\begin{figure}[!thb]
\begin{center}
\begin{tabular}{cc}
\includegraphics[scale=0.4,angle=0]{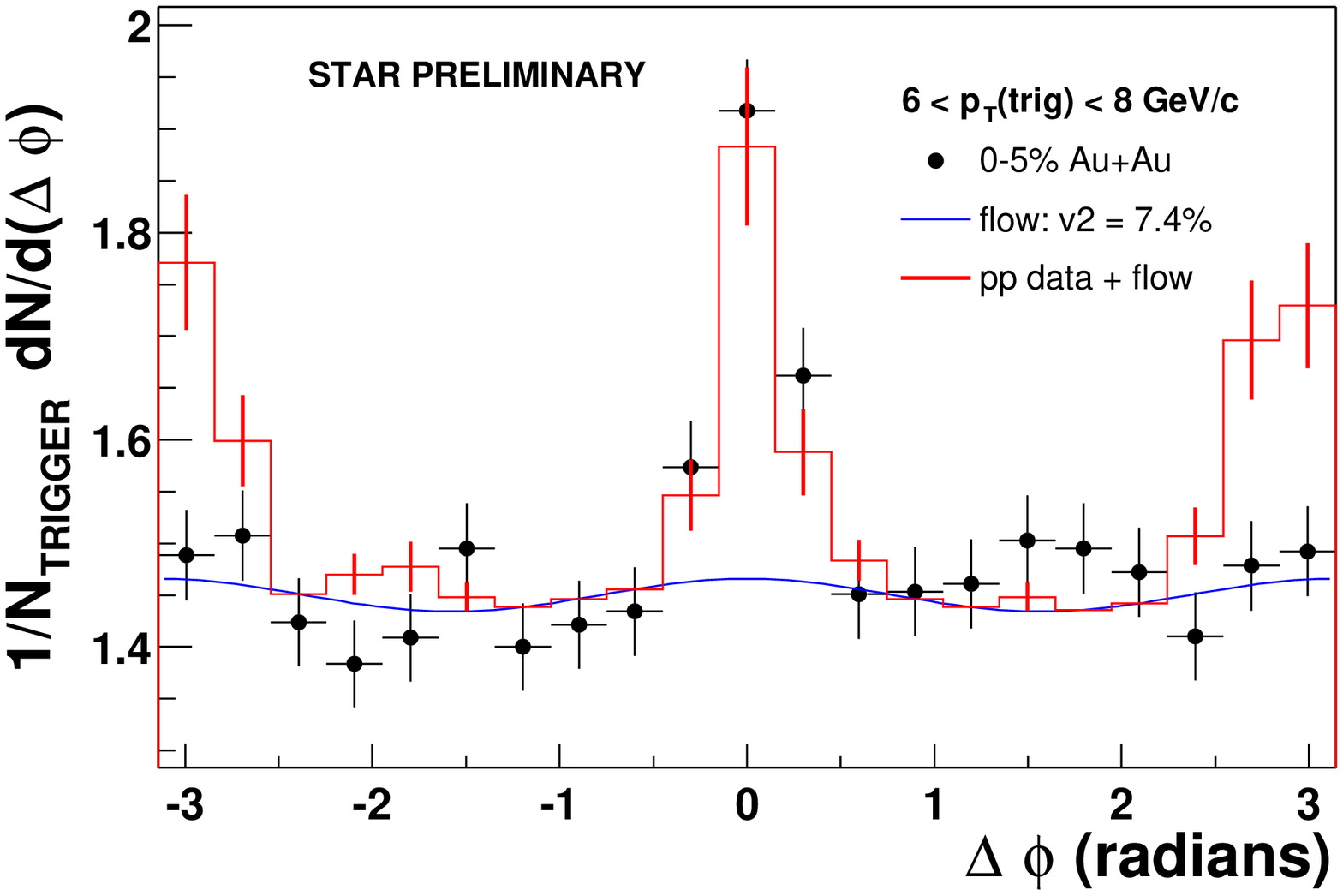}&
\includegraphics[scale=0.36,angle=0]{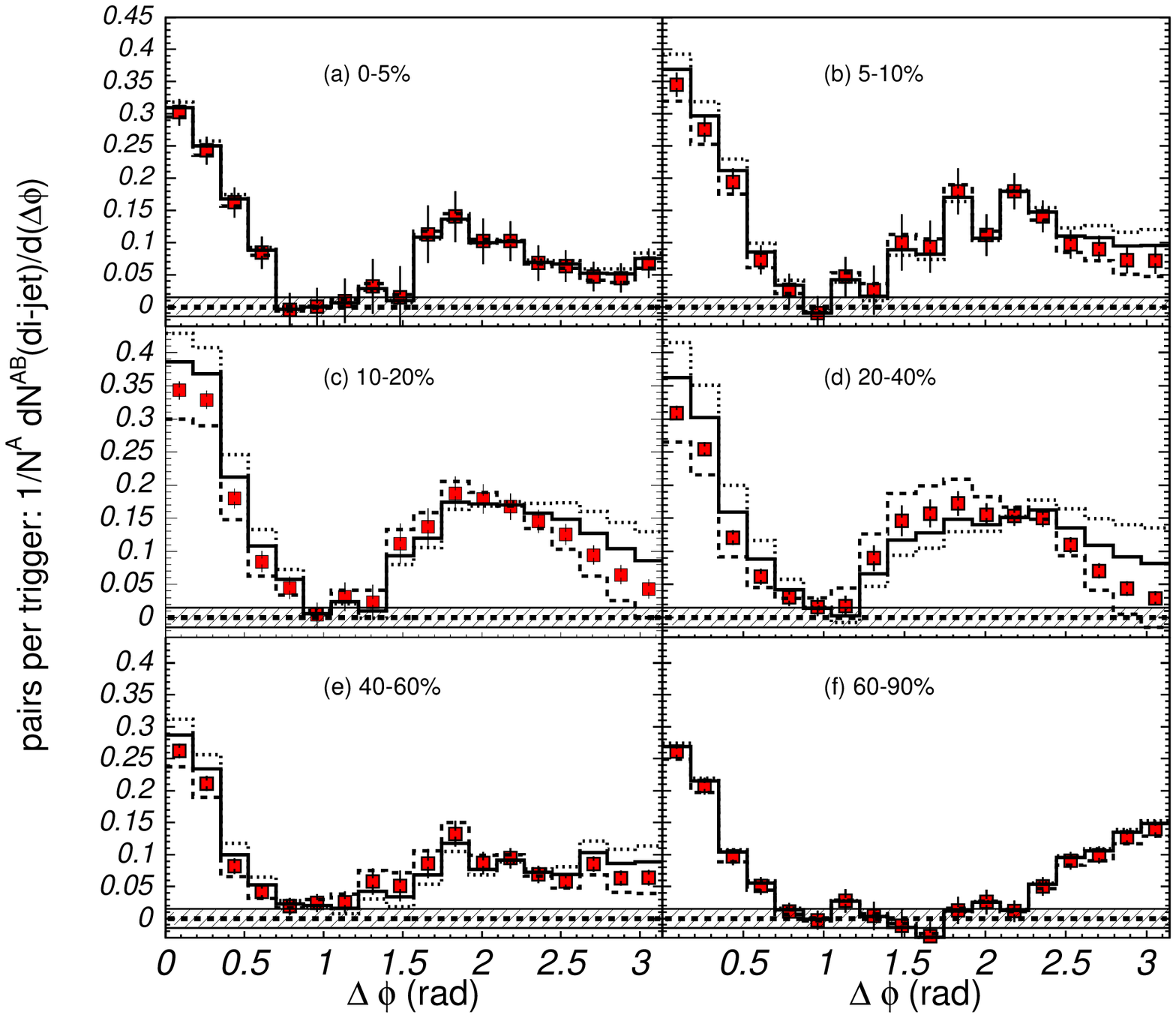}
\end{tabular}
\end{center}\vspace*{-0.25in}
\caption[]{a)(left) Associated charged tracks with $2\leq p_T\leq p_{T_t}$ GeV/c per trigger charged particle with $6\leq p_{T_t}\leq 8$ GeV/c for central Au+Au collisions at $\sqrt{s_{NN}}=200$ GeV as a function of the angle $\Delta \phi$ between the tracks in the range $|\eta|<1.4$ \fullcircle compared to the data in p-p collisions added to the flow modulated Au+Au background~\cite{STARHardkeQM02}; b) (right) Associated charged tracks with $1\leq p_T\leq 2.5$ GeV/c per trigger charged particle with $2.5\leq p_{T_t}\leq 4.0$ GeV/c after subtraction of flow-modulated background. The dashed (solid) curves are the distributions that would result from increasing (decreasing) the flow modulation by one unit of the systematic error; the dotted curve would result from decreasing by two units~\cite{PXppg032}. Note that only the jet correlation, after background subtraction, is shown. \label{fig:jets}}
\end{figure}
In Fig.~\ref{fig:jets}a, the conditional probability of finding an associated charged particle with $2\leq p_T\leq p_{T_t}$ per trigger charged particle with $6\leq p_{T_t}\leq 8$ GeV/c is shown for p-p and Au+Au central collisions, where the p-p data have been added to the large flow-modulated Au+Au background. (Note the offset zero.) For the p-p data there two peaks, a same-side peak at $\Delta\phi=0$ where associated particles from the jet cluster around the trigger particle and a peak at $\Delta\phi=\pi$ radians, from the away jet. For Au+Au central collisions, the same side peak is virtually identical to that in p-p collisions, while the away peak, if any, is masked by the $v_2$ modulation, and, in any case, is much smaller than observed in p-p collisions~\cite{STARHardkeQM02}. The  `vanishing' of the away jet is consistent with jet quenching in the medium due to energy loss---the away parton loses energy, and perhaps stops, so that there are fewer fragments in a given $p_T$ range.  The fact that the number of associated particles in a cone around the trigger particle is the same in Au+Au as in p-p collisions is a strong argument against hadronic absorption as the cause of jet quenching~\cite{Cassing04}. Since all hadrons would be absorbed roughly equally, the associated peak would be suppressed as much as the inclusive spectrum in a hadronic scenario, which is clearly not seen. The only escape from this conclusion is if the partons or hadrons were so strongly absorbed in the medium that only jets emitted from the surface were seen. Of course, since pions at mid-rapidity with $p_T > 1.4$ GeV/c, $\gamma > 10$, are not formed before 14fm, due to the quantum-mechanical formation time, they fragment outside the medium,  even taking account of the flow velocity. Thus hadronic absorption in the medium is not possible for pions.  

    The away jet reappears if the $p_T$ of the associated charged particles is lowered to the range $1\leq p_T\leq 2.5$ GeV/c for trigger charged particles  with $p_{T_t}$ in the range $2.5\leq p_{T_t}\leq 4.0$ GeV/c (Fig.\ref{fig:jets}b)\cite{PXppg032,seealsoFQW}. In the most peripheral collisions,  the shape of the trigger and away jets looks the same as in p-p collisions (Fig.~\ref{fig:jets}a). However with increasing centrality, the away jet becomes much wider (Fig.~\ref{fig:jets}b) and possibly develops a dip at $\Delta \phi=\pi$. Since the outgoing partons travel much faster than $c_s$ in the medium, a sonic-boom or mach-cone might develop, as suggested by the dip~\cite{ShuryakMIT}. There are many other ideas to explain the apparent dip, not the least of which is to get a better understanding of how exactly to extract the flow effect. 
    
   Study of jet correlations in A+A collisions is much more complicated than the same subject in p-p collisions and one can expect a long learning curve. However even at this early stage, there is one definitive result from jet correlations\cite{PXPRC71}, in the sense that it is adequate to cast serious doubt on the validity of coalescence models to explain the anomalous $p/\pi$ ratio at intermediate $p_T$.  \begin{figure}[!htb]
\begin{center}
\begin{tabular}{cc}
\includegraphics[scale=0.39,angle=0]{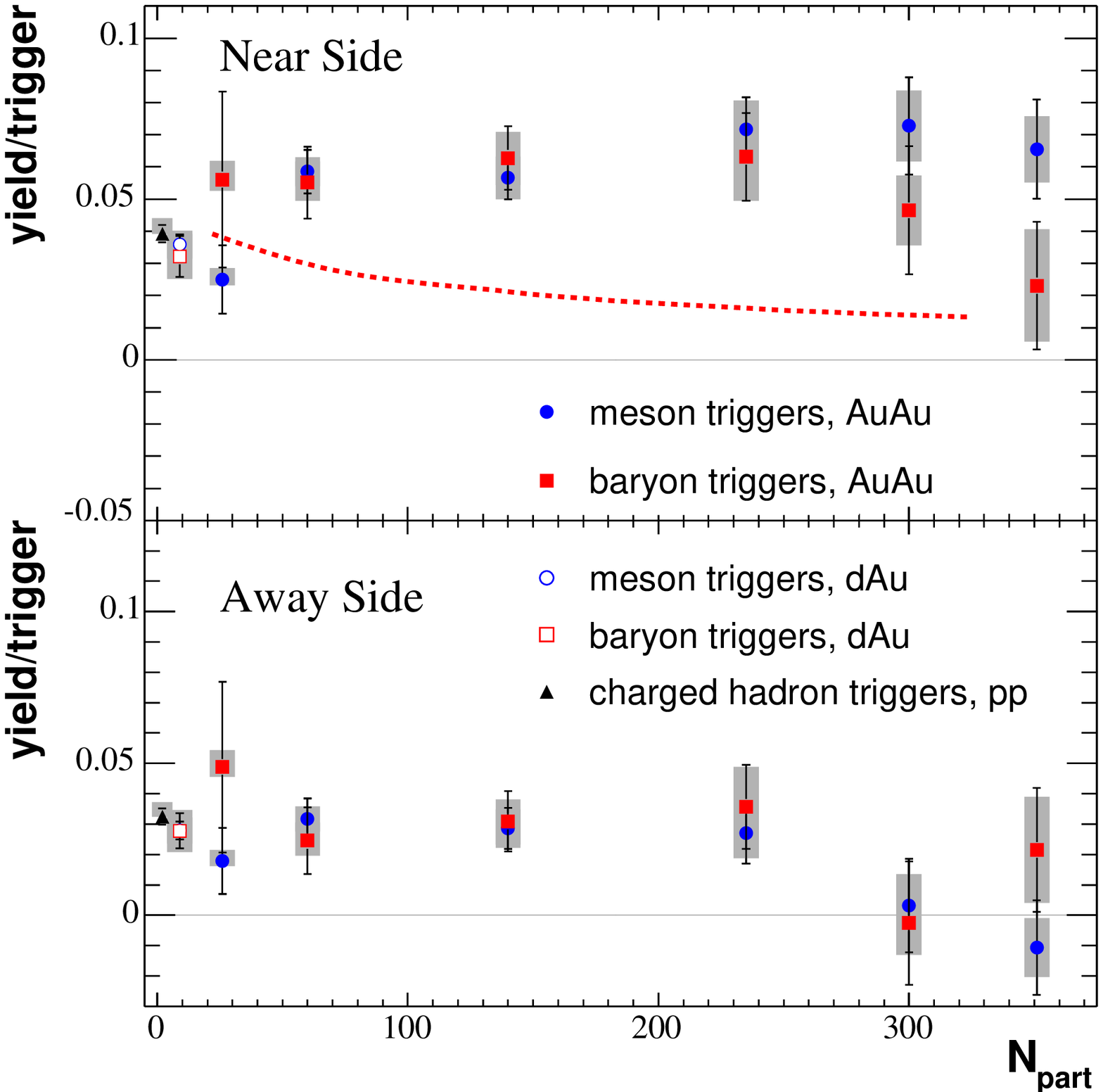}&
\includegraphics[scale=0.36,angle=0]{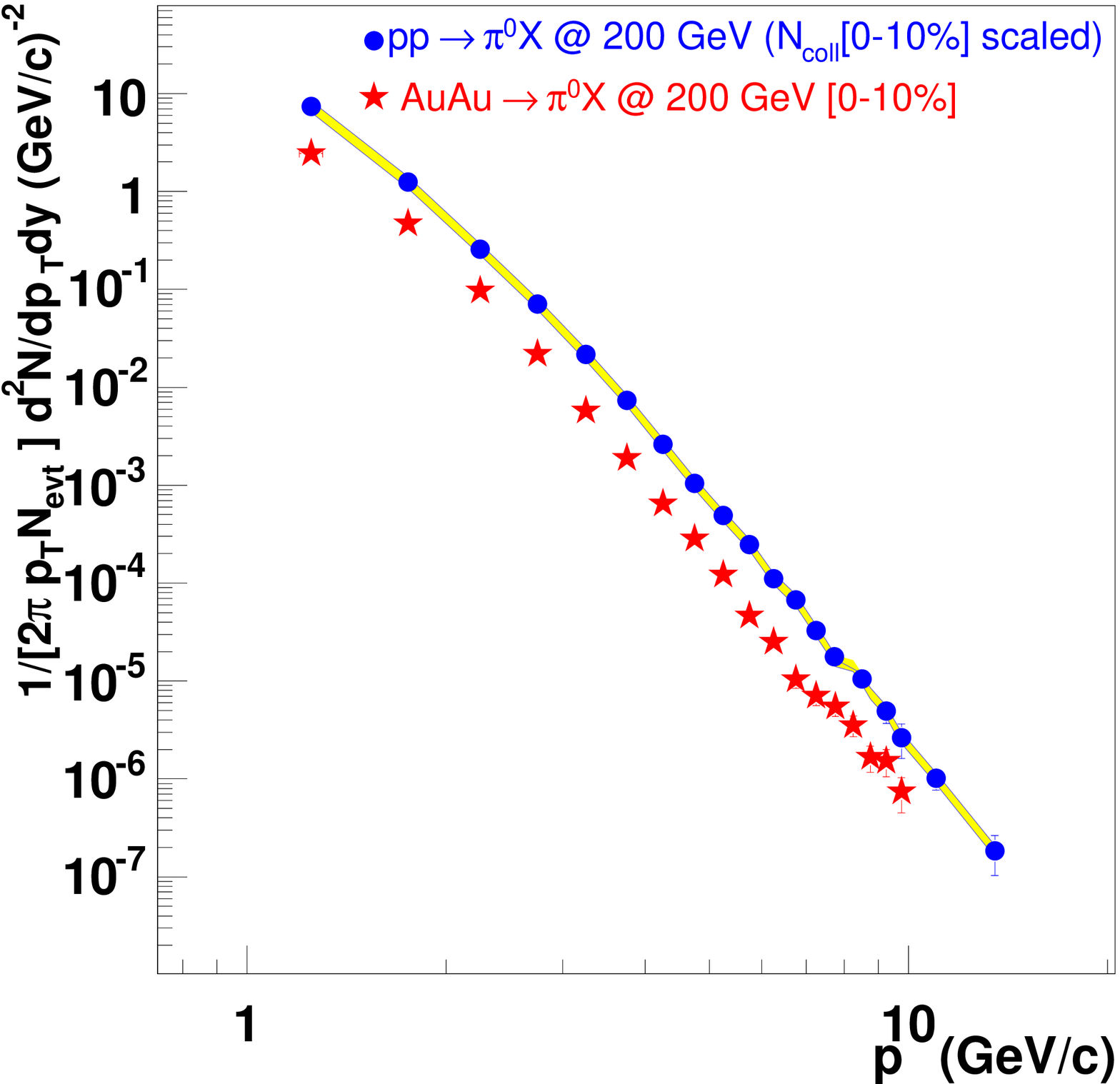}
\end{tabular}
\end{center}\vspace*{-0.25in}
\caption[]{a) (left) Conditional yield per trigger meson (circles), baryon (squares) with $2.5< p_T < 4$ GeV/c, for associated charged hadrons with $1.7 < p_T < 2.5$ GeV/c integrated within $\Delta \phi=\pm 0.94$ radian of the trigger (Near Side) or opposite azimuthal angle,    for Au+Au (full), d+Au (open) collisions at $\sqrt{s_{NN}}=200)$ GeV~\cite{PXPRC71}. Shaded boxes indicate centrality dependent systematic errors. An overall systematic error which moves all the points by 12\% is not shown. p-p data are shown for non-identified charged hadron triggers. b) (right) log-log plot of central $\pi^0$ data from Fig.~\ref{fig:PXpi0spectra}b \label{fig:AnnePower}}
\end{figure}
Fig.~\ref{fig:AnnePower}a shows the integrated associated particle yields/trigger for p-p, d+Au and AuAu collisions from correlations, as in Fig.~\ref{fig:jets}b, where the trigger is identified as either a meson or a baryon in the range $2.5\leq p_{T_t}\leq 4.0$ GeV/c and the associated particles, in the range $1.7\leq p_T\leq 2.5$ GeV/c, are not identified. The yield of associated particles/per trigger on the near side, from the same jet as the trigger hadron, is the same for meson and baryon triggers as a function of centrality, except perhaps in the most central bin; and the same effect is seen for the away-side yields. The red-dashed curve indicates the expected trigger-side conditional yield if all the anomalous protons in Au+Au collisions were produced by coalescence. This shows that meson and baryons at intermediate $p_T$ are produced by hard-processes with the same di-jet structure, and not by soft coalescence. 
A recent paper~\cite{FriesPRL94} claimed to demonstrate that two-parton correlations of order of 10\% would be sufficient to move the red-dashed curve of Fig.~\ref{fig:AnnePower}a into agreement with the data for the near side correlations. However the away side correlations, which are the same for meson and baryon triggers and show a clear hard-scattering di-jet structure, were not addressed.    
Clearly, the baryon anomaly remains a puzzle. 
\subsubsection{Detailed tests of the energy loss formalism}
   At face value, suppression of high $p_T$ particles at RHIC appears to support the jet quenching description. Unfortunately, as noted in Sec.~\ref{sec:quenching}, the properties of the medium can not be derived straightforwardly from the data but must be extracted using models~\cite{GLV,Wang,Wiedemann}. All the models~\cite{GMc,VitevQM04,WangPRC70,WangNPA750,DaineseEPJC38,ArmestoPRD71,DjorPRL94} agree that the medium in Au+Au central collisions exhibits a color charge density, expressed as the gluon density in rapidity, of $dN^g/dy\approx 1100$, corresponding to an energy density $\epsilon\approx 30-10$ GeV/fm$^3$ at an initial thermalization time $t_0\approx 0.2-0.5$ fm, with a time averaged $\mean{\hat{q}}\approx 4-14$ GeV$^2$/fm, all roughly 100 times that of cold nuclear matter, or roughly 10 times that of a nucleon, probed in a cylinder around the outgoing parton of radius $r_D=1/\mu=0.4$~fm, where $\mu=0.5$~GeV is the `screening scale'~\cite{GLV}, which corresponds to $\sqrt{t_{min}}$ for the gluon bremsstrahlung.   
    
   Experimentalists cannot simply accept these numbers pronounced by theorists drawing curves through various data points, but must have a way to verify the validity of the theory directly from the measurements. Since the theory involves the complicated interplay of 3 nuclear effects (Cronin+shadowing+quenching)~\cite{VGPRL89} the best we can do at present is to try to test formulas provided by the theorists to approximate their results. For instance, although the total energy loss $\Delta E$ in a static  medium is supposed to be independent of the energy of the parton, $E$, and proportional to $L^2$\cite{BDMPS2}, once expansion is added, ``the radiative spectrum and the mean energy loss can be related to the soft gluon rapidity density"\cite{VitevQM04}, and $\mean{\Delta E}$ becomes proportional to $L$ but remains independent of $E$ apart from logarithms\cite{VitevQM04}:
   \begin{equation} 
   \mean{\Delta E}\approx -\frac{9 C_R \pi \alpha_s^2}{4} {1\over A_{\perp}} {{dN^g}\over {dy}} L \ln{{{2E} \over {\mu^2 L}}} + \cdots \qquad ,
   \label{eq:VitevFormula}
   \end{equation} 
   where $C_R$ = 4/3 (3) is the Casimir invariant for quarks (gluons) and $A_{\perp}\approx\pi R_x R_y$ is the effective area of the collision.  
      
   The average energy loss $\mean{\Delta E}$ can be measured from the shift in the $p_T$ spectrum at constant Yield~\cite{BaierQM02} (see Fig.\ref{fig:AnnePower}b). It is apparent from this log-log plot of the binary-scaled p-p spectrum and Au+Au central collision spectrum for $\pi^0$, that the p-p spectrum is a pure power law for $p_T\geq 3$ GeV/c, $Ed^3\sigma/dp^3=Ap_T^n$,  with $n=8.1\pm0.1$~\cite{ColeHP04}, and that the Au+Au spectrum is shifted down in $p_T$ from the scaled p-p spectrum by a constant fraction, $-\Delta p_T/p_T\equiv S_{\rm loss}$, since the spectra are parallel. This explains why $R_{AA}$ is a constant (see Fig~\ref{fig:pizeroesrule}b), since   the relation between $S_{\rm loss}$ and $R_{AA}$ is particularly simple for a power law: 
   \begin{equation}
   S_{\rm loss}(p_T)=1-R_{AA}(p_T)^{1/(n-2)} \qquad .
   \label{eq:ColeTannenbaum}
   \end{equation}
   
\begin{figure}[!htb]
\begin{center}
\begin{tabular}{cc}
\includegraphics[scale=0.38,angle=0]{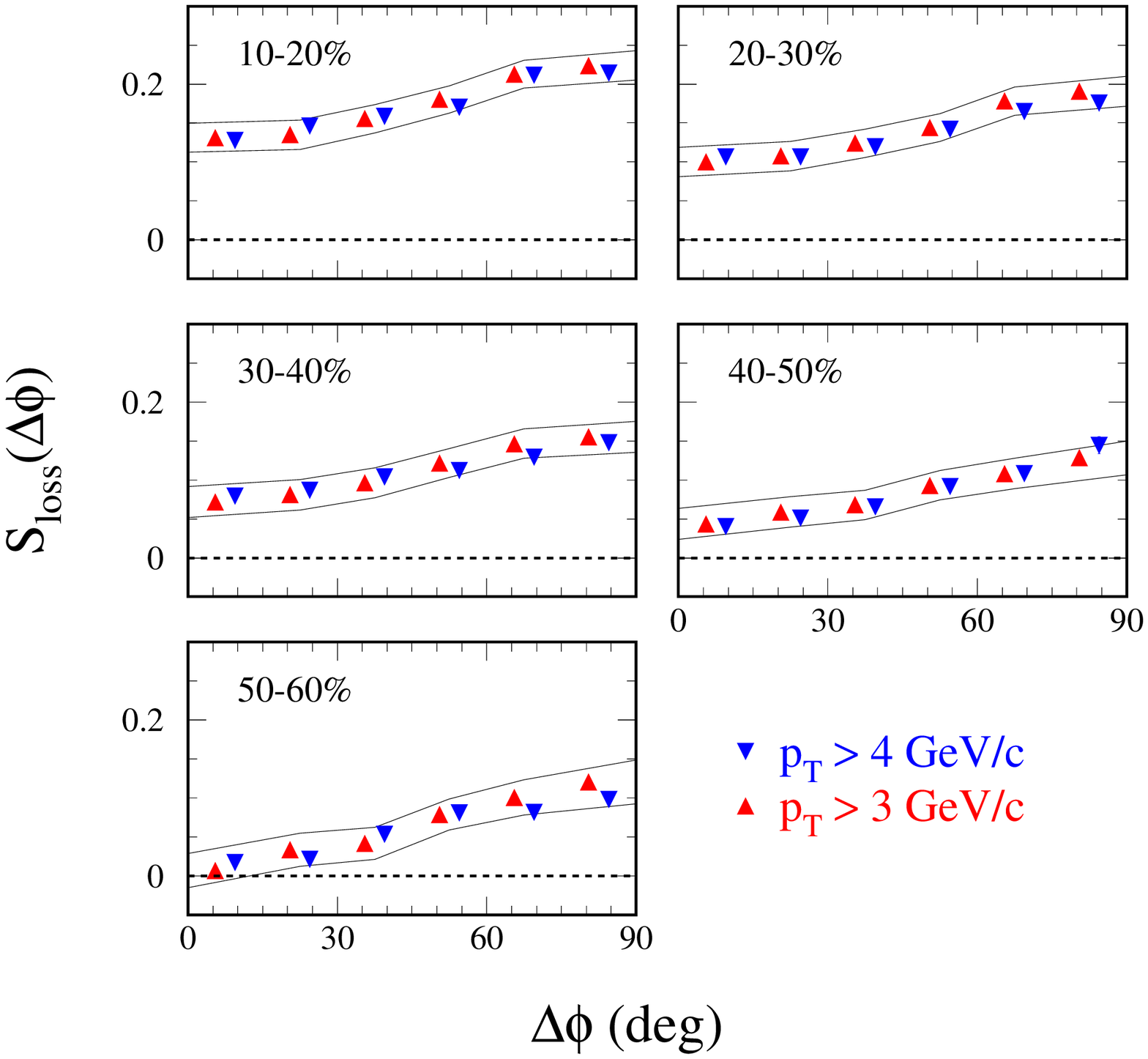}&
\includegraphics[scale=0.37,angle=0]{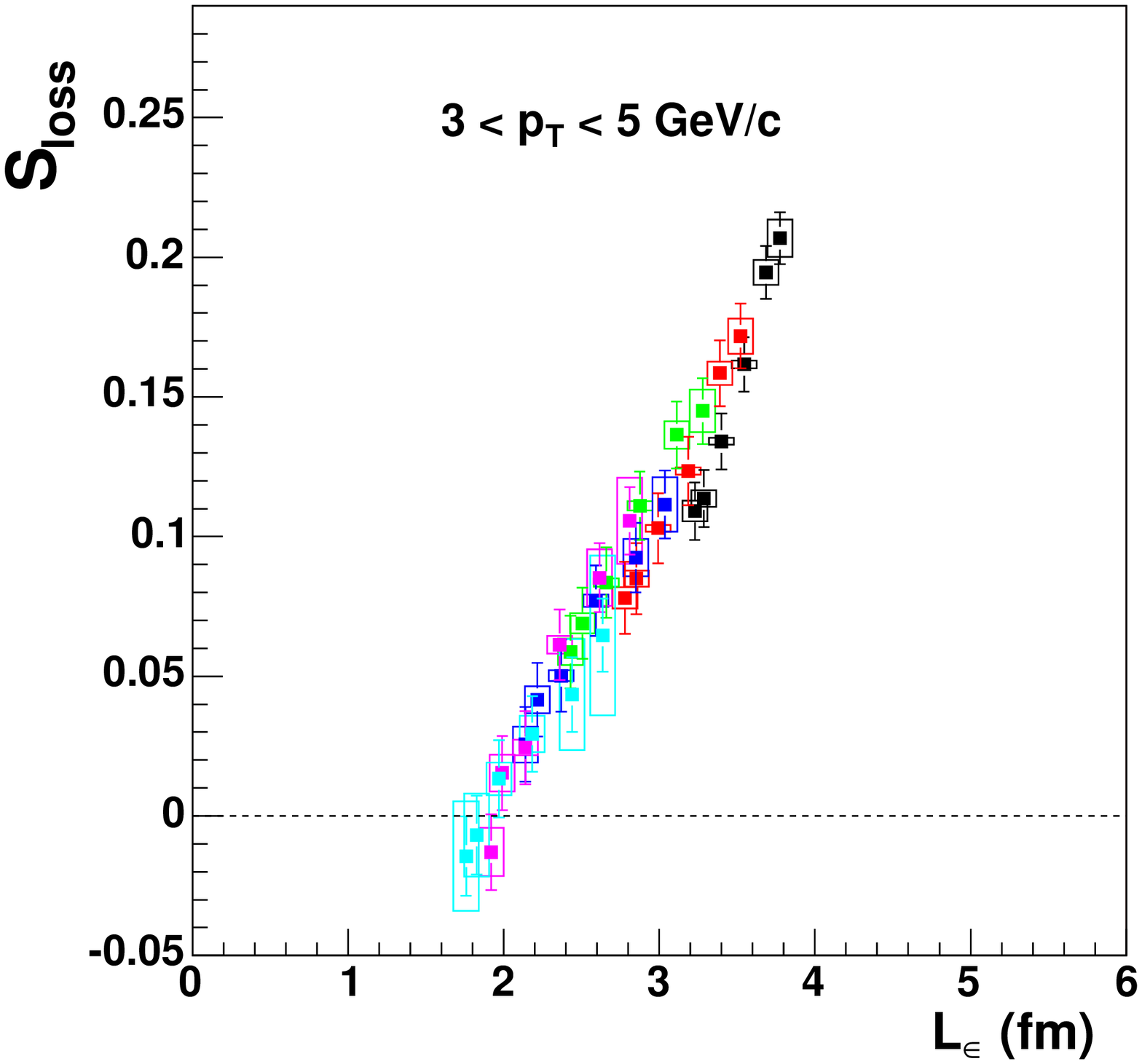}
\end{tabular}
\end{center}\vspace*{-0.25in}
\caption[]{a)(left) $S_{\rm loss}$ as a function of angle $\Delta\phi=\phi-\Phi_R$ with respect to the reaction plane in $\sqrt{s_{NN}}=200$GeV Au+Au collisions as a function of centrality~\cite{ColeHP04}. b)(right) $S_{\rm loss}$ versus $L_{\varepsilon}$, the length from the center of the almond, combining the various data points from (a), with different centrality bins represented by different symbols. \label{fig:Sloss}}
\end{figure}

   The dependence of Eq.~\ref{eq:VitevFormula} on the gluon density can be tested by  measurement of $S_{\rm loss}$ as a function of centrality (recall  Fig.~\ref{fig:RAA2})~\cite{PXWP}. The $L$ dependence of Eq.~\ref{eq:VitevFormula} for constant initial condition can be studied by measurement of $S_{\rm loss}$ as a function of angle with respect to the reaction plane, where the average $L$ of the system can be calculated~\cite{ColeHP04} (Fig.~\ref{fig:Sloss}a).
The compilation of $S_{\rm loss}$ vs $L$ by this method (Fig.~\ref{fig:Sloss}b) shows a reasonably linear behavior with $L$ with the totally striking result that $S_{\rm loss}$ appears to go to zero at finite $L\sim 2$ fm, suggestive of a formation time.  This may explain why attempts to explain the azimuthal anisotropy $v_2$ at large $p_T$ solely in terms of energy loss failed~\cite{DreesFengJia,Pantuev0506}. Again, it should be noted that this type analysis is in its infancy and a long learning curve is to be expected. 
\subsubsection{The smoking gun?}
 
   The jet suppression observed at RHIC is unique in that it has never been seen in either p+A collisions or in A+A collisions at lower $\sqrt{s_{NN}}$ and it probes the color charge density of the medium. Many questions and unsolved problems remain which are under active investigation, but this effect comes closest of all, in the author's opinion, to meeting the criteria for declaring the medium a Quark Gluon Plasma: 
   \begin{itemize}
   \item There is no such effect in p+A collisions at any $\sqrt{s_{NN}}$
\item It is not the `ordinary physics' of A+A collisons since it only occurs for $\sqrt{s_{NN}} \geq 30$ GeV
\item In all discussions of the effect, the operative `charge' is color and the operative degrees of freedom are quarks and gluons.
   \end{itemize}
   
\subsection{$J/\Psi$ Suppression}
  
\begin{figure}[!htb]
\begin{center}
\begin{tabular}{cc}
\includegraphics[scale=0.44,angle=0]{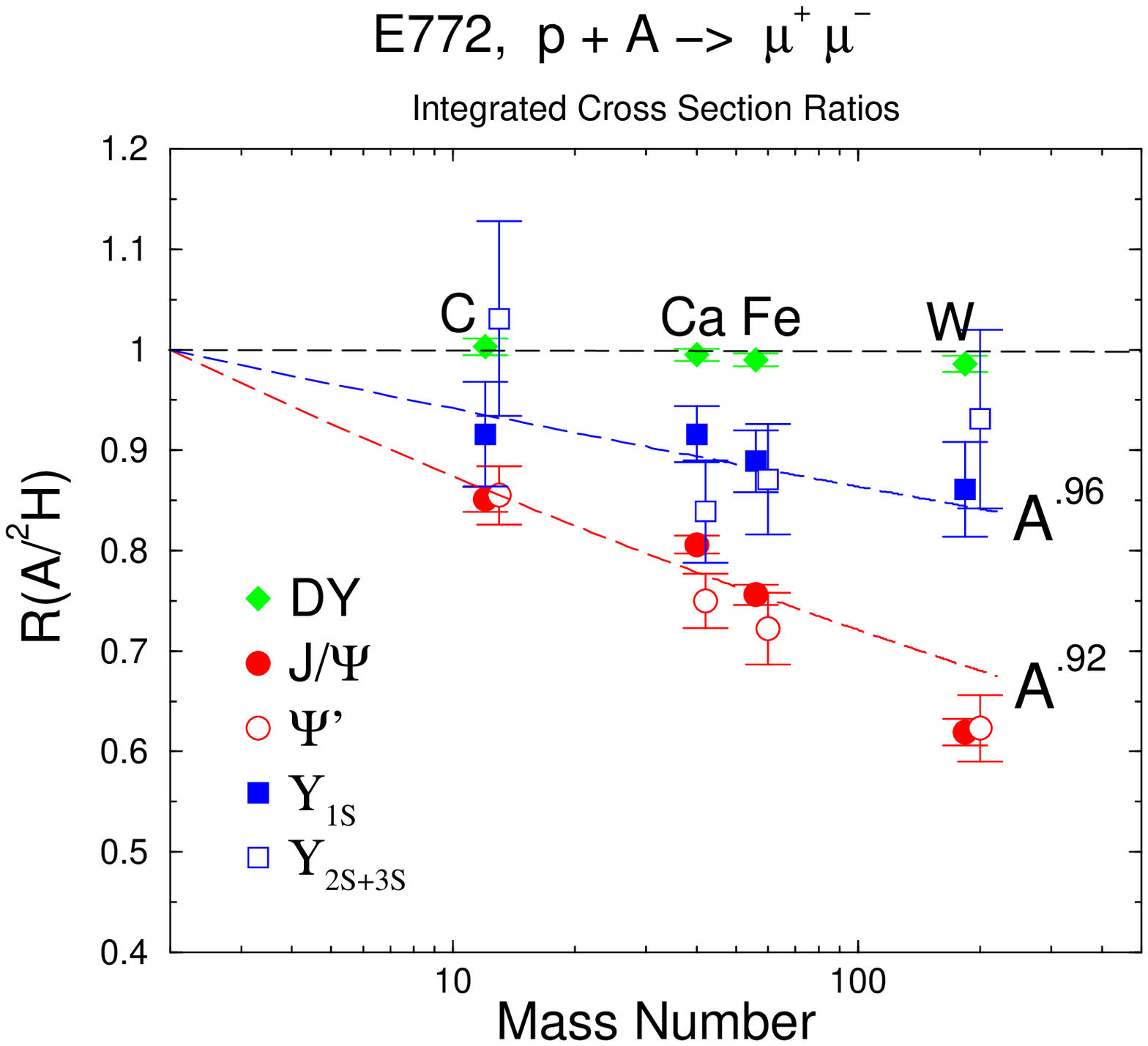}&\hspace*{0.2cm}
\includegraphics[scale=0.36,angle=0]{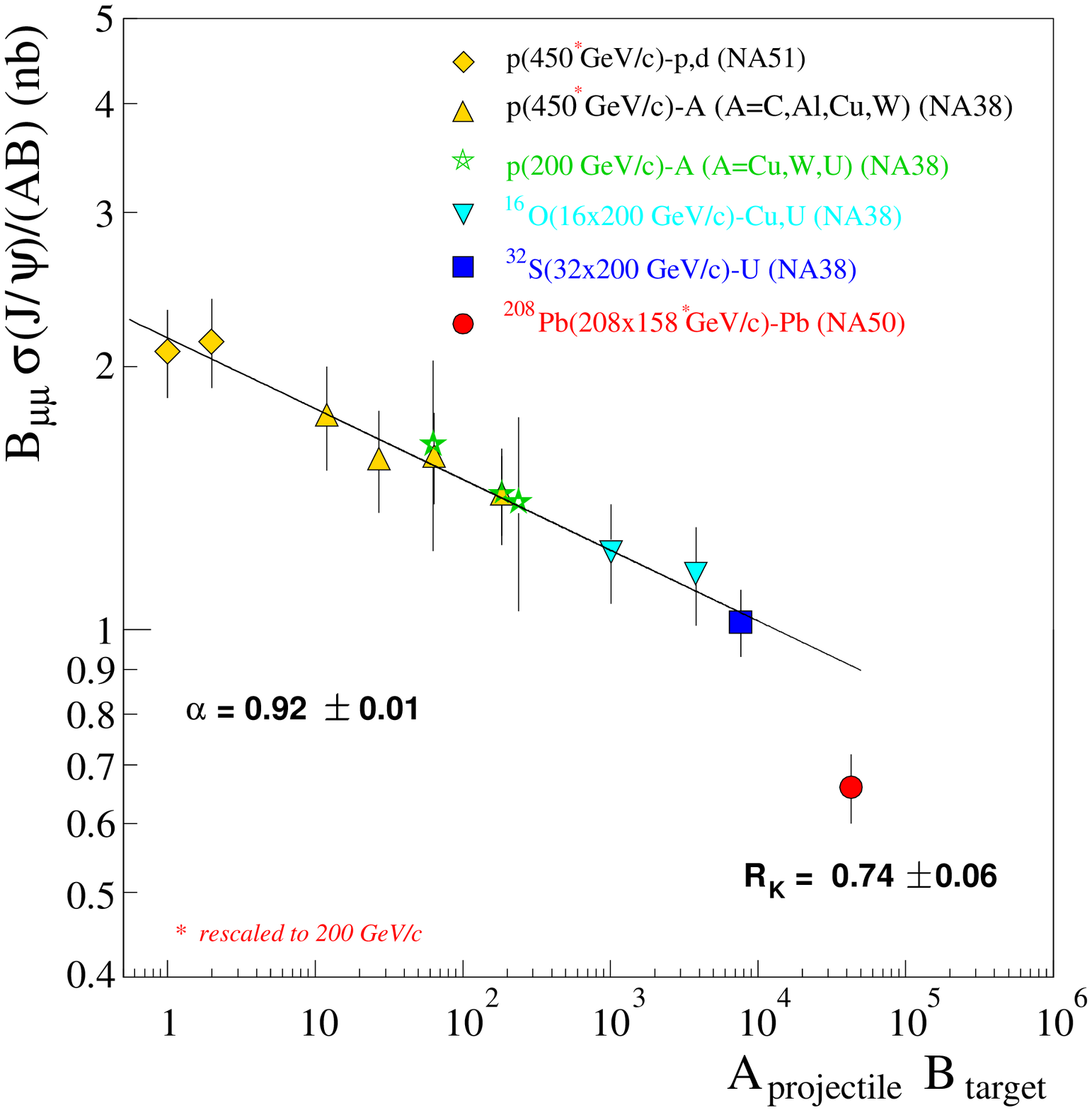}
\end{tabular}
\end{center}\vspace*{-0.25in}
\caption[]{a) (left) $A$ dependence of total cross section for $\Psi$, $\Psi^{'}$, $\Upsilon$ production in 800 GeV p+A collisions~\cite{E772}. The per nucleon ratio between heavy nuclei and deuterium is shown versus the mass number, $A$. b) (right) Total cross section for $J/\Psi$ production divided by $AB$ in A+B collisions at 158--200$A$ GeV~\cite{NA50PLB450}  \label{fig:JPsiAB}}
\end{figure}

\begin{figure}[!htb]
\begin{center}
\begin{tabular}{cc}
\includegraphics[scale=0.5,angle=0]{figs/NA50RAAJPSI.epsf}&
\includegraphics[scale=0.39,angle=0]{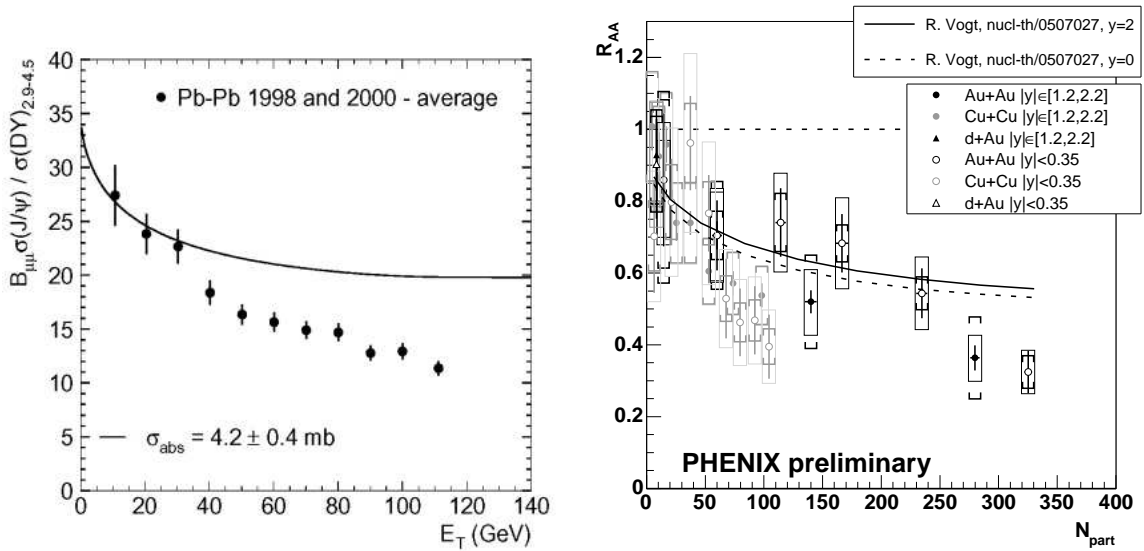}
\end{tabular}
\end{center}\vspace*{-0.25in}
\caption[]{a) (left) Cross section for $J/\Psi$ divided by Drell-Yan as a function of centrality measured by $E_T$ for $158A$ GeV Pb+Pb collisions~\cite{NA50EPJC39}. The solid line represents the expected suppression for cold nuclear matter. The plot may be converted to $R_{AA}$ by dividing by 34. b) $R_{AA}$ for $J/\Psi$ production as a function of centrality ($N_{\rm part}$) at $\sqrt{s_{NN}}=200$~GeV for the systems and rapidity ranges indicated~\cite{PXJPsiQM05}.  The solid and dashed lines represent the expected suppression in cold nuclear matter for Au+Au~\cite{Vogt0507027} \label{fig:NA50PX}}
\end{figure}

Although $J/\Psi$ suppression was thought to be the `gold-plated' signature of deconfinement since 1986~\cite{MatsuiSatz86}, there were problems from the very beginning. The principal problem is that the $J/\Psi$ is suppressed in p+A collisions (see Fig.~\ref{fig:JPsiAB}). Point-like processes such as hard-scattering scale relative to p-p collisions like $A (AB)$ for p+A (A+B) minimum bias collisions, but unlike Drell-Yan which exhibits an $A^{1.00}$ dependence, the $J/\Psi$ is suppressed by $A^{0.92}$ in p+A collisions~\cite{E772} (Fig.~\ref{fig:JPsiAB}a). The suppression continues as $(AB)^{0.92}$ for minimum bias A+B collisions~\cite{NA50PLB450} (Fig.~\ref{fig:JPsiAB}b) at the SPS until the heaviest system, Pb+Pb, where the suppression increases by $\sim 25$\%. However the suppression is much more impressive as function of centrality (Fig.~\ref{fig:NA50PX}a), and it was claimed that the expected discontinuity~\cite{NA50PLB450} due to the phase transition, and the sequential suppression of the $\chi_c$, $\Psi^{'}$ and direct $J/\Psi$ had been observed~\cite{SatzQM99}. 

	To say the least, the interpretation of the suppression in terms of a QGP was controversial, and the effect at the SPS could be explained by hadronic~\cite{CapellaFerreiro05,BKCS04} and even thermal models~\cite{GGPRL83}. The clincher would be the measurement of the $J/\Psi$ at RHIC. All the initially produced $J/\Psi$, the ones not suppressed at CERN, would be totally suppressed in the much hotter denser QGP at RHIC, which would prove that the $J/\Psi$ suppression at CERN was indeed the result of deconfinement. However, 
in the ensuing years a new   
problem developed when it was realized that if a QGP were indeed produced, the thermal $c,\bar{c}$ quarks would recombine~\cite{PBMStachelPLB490,ThewsPRC63} to form $J/\Psi$. Thus, if the $J/\Psi$ suppression were the same at RHIC as at CERN, this would imply that RHIC, not CERN, had discovered the QGP since all the initially produced $J/\Psi$, which were suppressed at CERN by whatever mechanism, would be totally suppressed at RHIC, leaving only the thermal $J/\Psi$ produced by recombination in a QGP.    The 
new problem is that nobody would believe this explanation. Incredibly, this is exactly what happened (see Fig.~\ref{fig:NA50PX}), the $J/\Psi$ suppression, expressed as $R_{AA}$ turned out to be the same at RHIC~\cite{PXJPsiQM05} as in the famous SPS measurement~\cite{NA50EPJC39}

\begin{figure}[!htb]
\begin{center}
\begin{tabular}{cc}
\includegraphics[scale=0.36,angle=0]{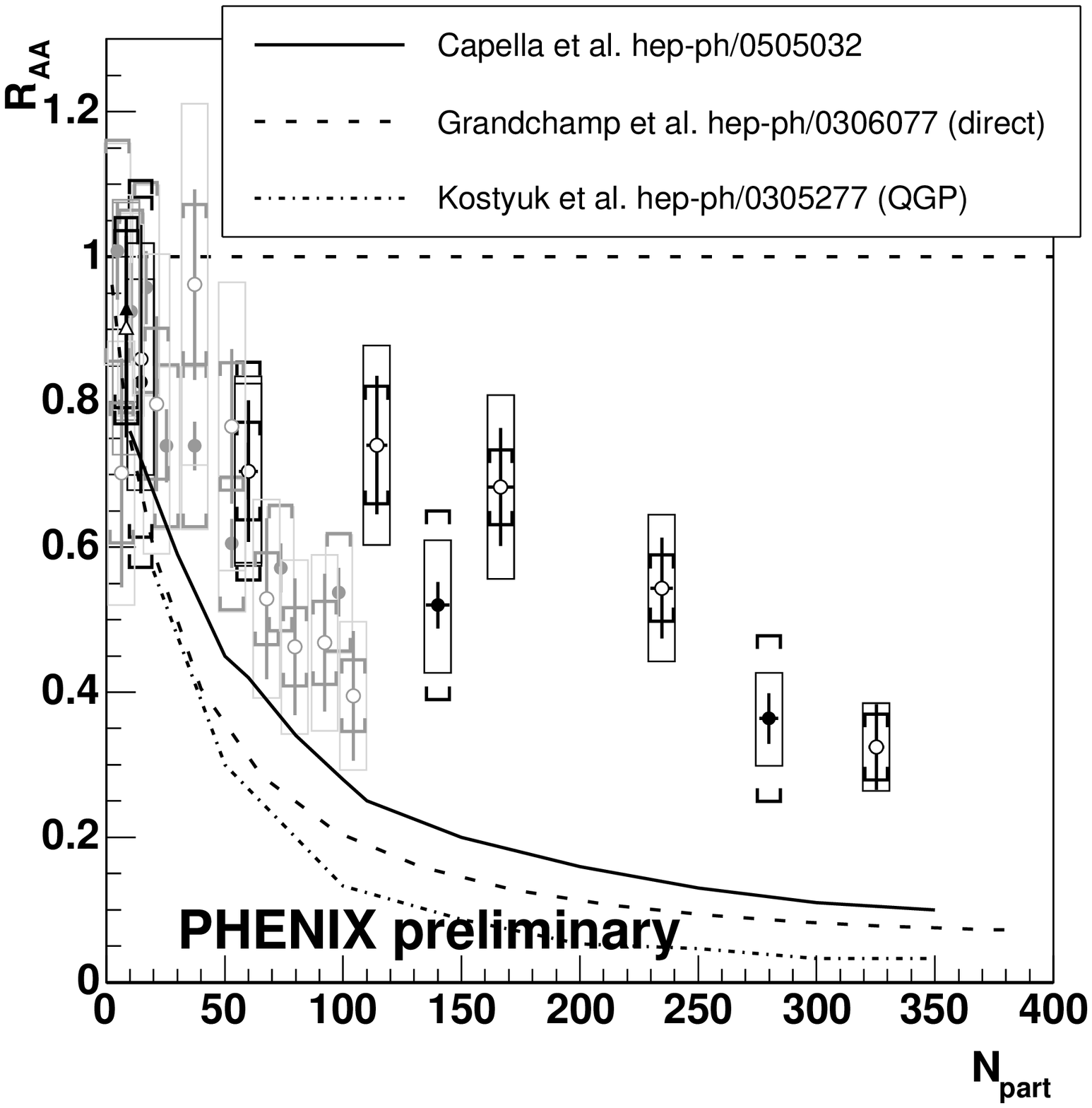}&
\includegraphics[scale=0.36,angle=0]{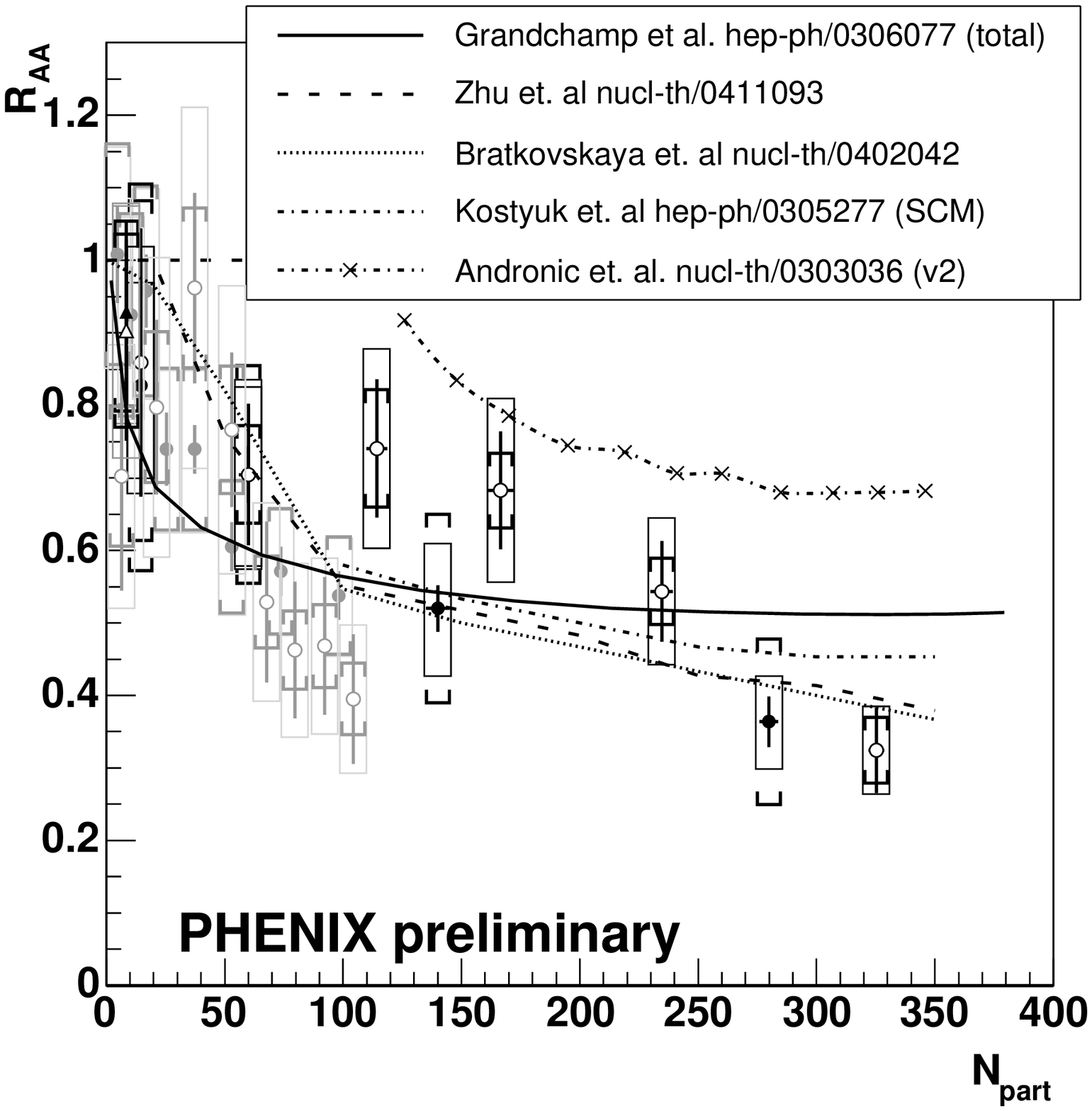}
\end{tabular}
\end{center}\vspace*{-0.25in}
\caption[]{$R_{AA}$ data from Fig.~\ref{fig:NA50PX}b together with predictions from models (a) without, (b) with recombination of $c\, \bar{c}$ quarks. \label{fig:nightmare}}
\end{figure}
This effect is illustrated quantitatively in Fig.~\ref{fig:nightmare}. Models~\cite{CapellaFerreiro05,BKCS04,GrandchampPRL92} which reproduce the SPS $J/\Psi$ suppression with or without a QGP predict a near total absence of $J/\Psi$ at RHIC beyond 150 participants without recombination. With recombination turned on, the RHIC data are reproduced (Fig.~\ref{fig:nightmare}b), with one notable exception~\cite{AndronicPLB517} which predicts a larger $R_{AA}$ for $J/\Psi$ than at CERN. In fact, a $J/\Psi$ enhancement would have been the smoking gun for the QGP. It will be interesting to see whether this occurs at the LHC. 
    
    Does the agreement of the CERN and RHIC data for $R_{AA}$ in $J/\Psi$ production eliminate $J/\Psi$ suppression as a signature of deconfinement? Is it possible that the $J/\Psi$ ($c,\bar{c}$) is no different in its QGP sensitivity than the $\phi$-meson ($s,\bar{s}$)? The recent increases in the predicted dissociation temperatures~\cite{Wong05}, give a possible way out~\cite{Satz05}. Satz has proposed  that the $\chi_c$ and the $\Psi^{'}$ were suppressed both at CERN and at RHIC, but in neither place was the direct $J/\Psi$ suppressed, since a temperature $\sim 2T_c$ was not reached. However temperature sufficient to melt the $J/\Psi$ should be reached at the LHC. Does this mean that we have to wait until 2009 to prove or disprove an idea proposed in 1986? Fortunately there are other tests to be made on the RHIC data. If the $\chi_c$ or $\Psi^{'}$ were to be observed at RHIC, that would settle the issue. For recombination to be true, $J/\Psi$ flow should be observed~\cite{GrandchampPRL92}, and both the rapidity and the $p_T$ distributions of $J/\Psi$ should be much narrower due to recombination than for directly produced $J/\Psi$~\cite{ThewsMangano}. 
    
    One thing is perfectly clear from this discussion: the claim of the QGP discovery from $J/\Psi$ suppression at the SPS was, at best, premature.   

\subsection{Thermal photons,leptons} 

\begin{figure}[!htb]
\begin{center}
\begin{tabular}{cc}
\includegraphics[scale=0.4,angle=0]{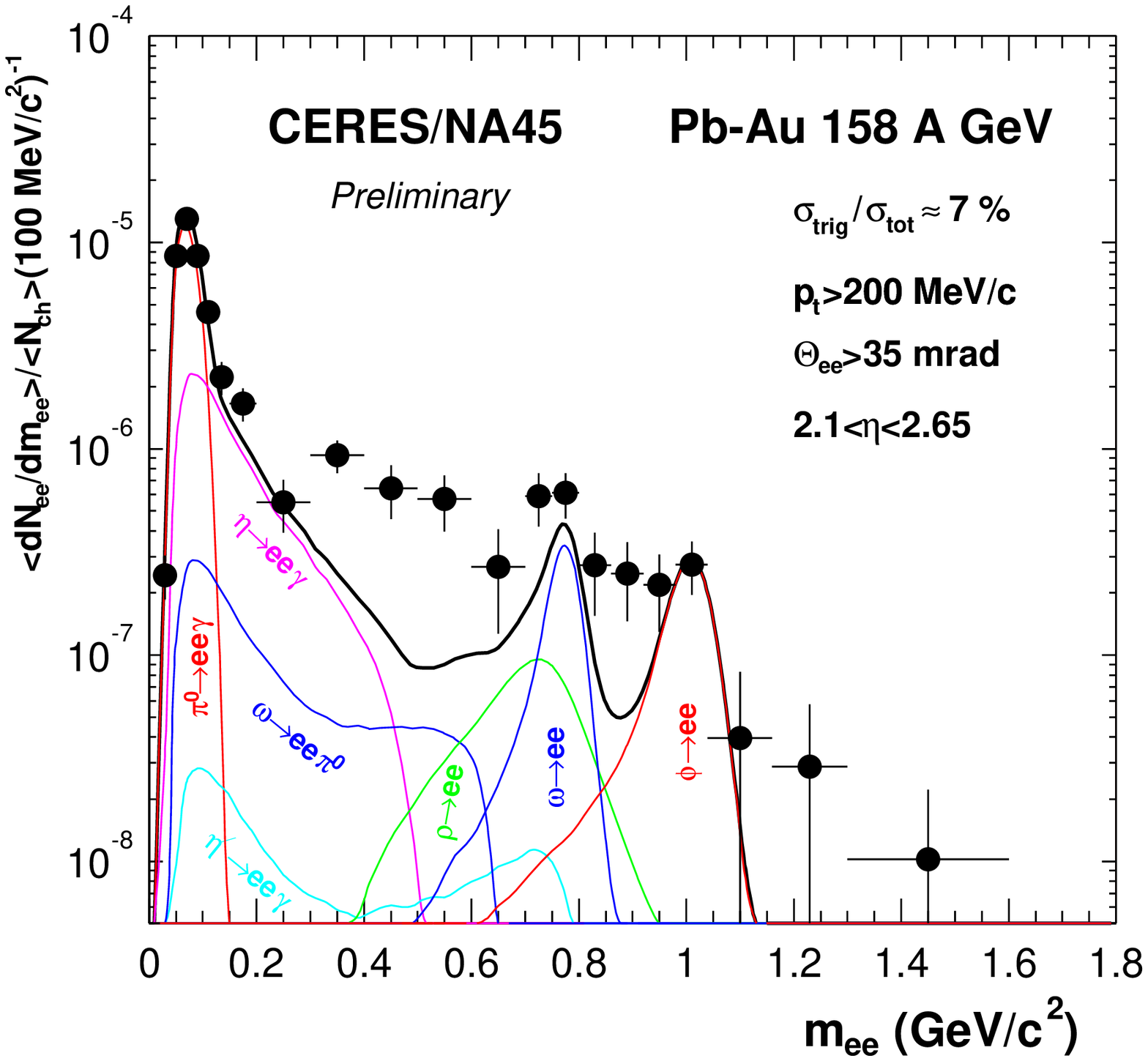}&
\includegraphics[scale=0.3,angle=0]{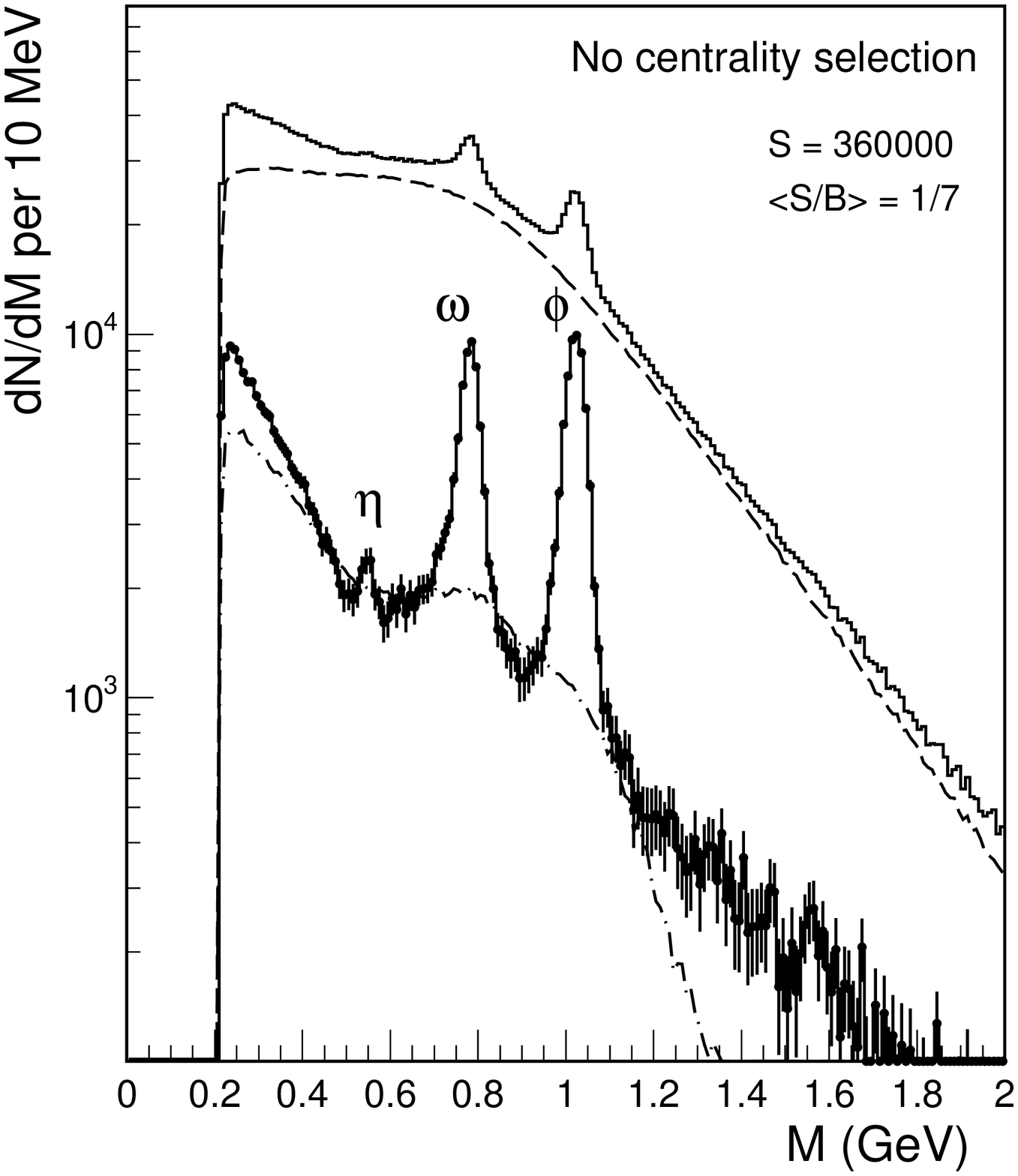}
\end{tabular}
\end{center}\vspace*{-0.25in}
\caption[]{a) Invariant mass spectrum of $e^+ e^-$ pairs in central $158A$ GeV/c Pb-Au collisions, where each $e^+$ or $e^-$ has $p_T> 200$ MeV/c, normalized to the number of charged particles in the detector~\cite{CERESQM05}, compared to a coctail of known hadronic sources. The `enhancement' from $0.2 < m_{ee} < 0.6$  GeV occurs for the pair transverse momentum $p^{ee}_{T} \leq 0.5$ GeV/c~\cite{CERESPLB422}. b) (right) Invariant mass spectrum of $\mu^+ \mu^-$ from minimum bias $158A$ GeV/c In+In collisions~\cite{NA60QM05}. Total data (upper histogram), combinatorial background (dashed), fake matches (dot-dashed), and net spectrum after subtraction. \label{fig:CERESNA60}}
\end{figure}

	The hardest experiments in relativistic heavy ion physics involve production of lepton pairs with low mass and photons at low $p_T$ due to the huge backgrounds. A  measurement of $e^+ e^-$ pairs in $158A$ GeV/c Pb+Au (Fig.~\ref{fig:CERESNA60}a) had created great excitement  because it looked as if the ``extraordinary signal" of a thermal $\rho^0\rightarrow$~dilepton peak~\cite{Pisarski82} had appeared, which stimulated  much model-making. A new high resolution measurement, this year, in $158A$~GeV In+In collisions, with beautiful $\omega\rightarrow \mu^+ \mu^-$ and $\phi\rightarrow \mu^+ \mu^-$ peaks~\cite{NA60QM05} should help clarify the issue (Fig.~\ref{fig:CERESNA60}b).  

\begin{figure}[!htb]
\begin{center}
\begin{tabular}{cc}
\includegraphics[scale=0.36,angle=0]{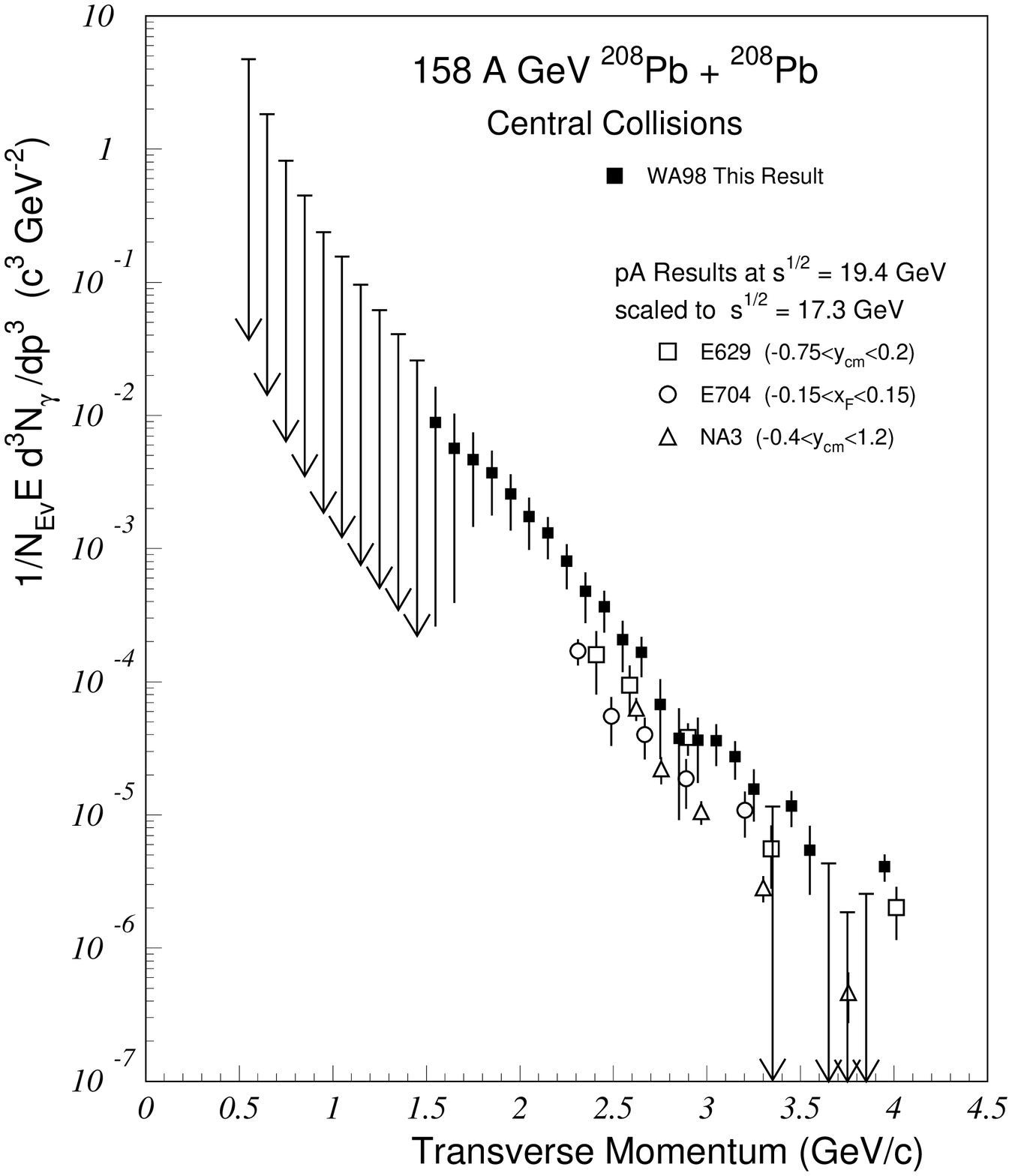}&
\includegraphics[scale=0.36,angle=0]{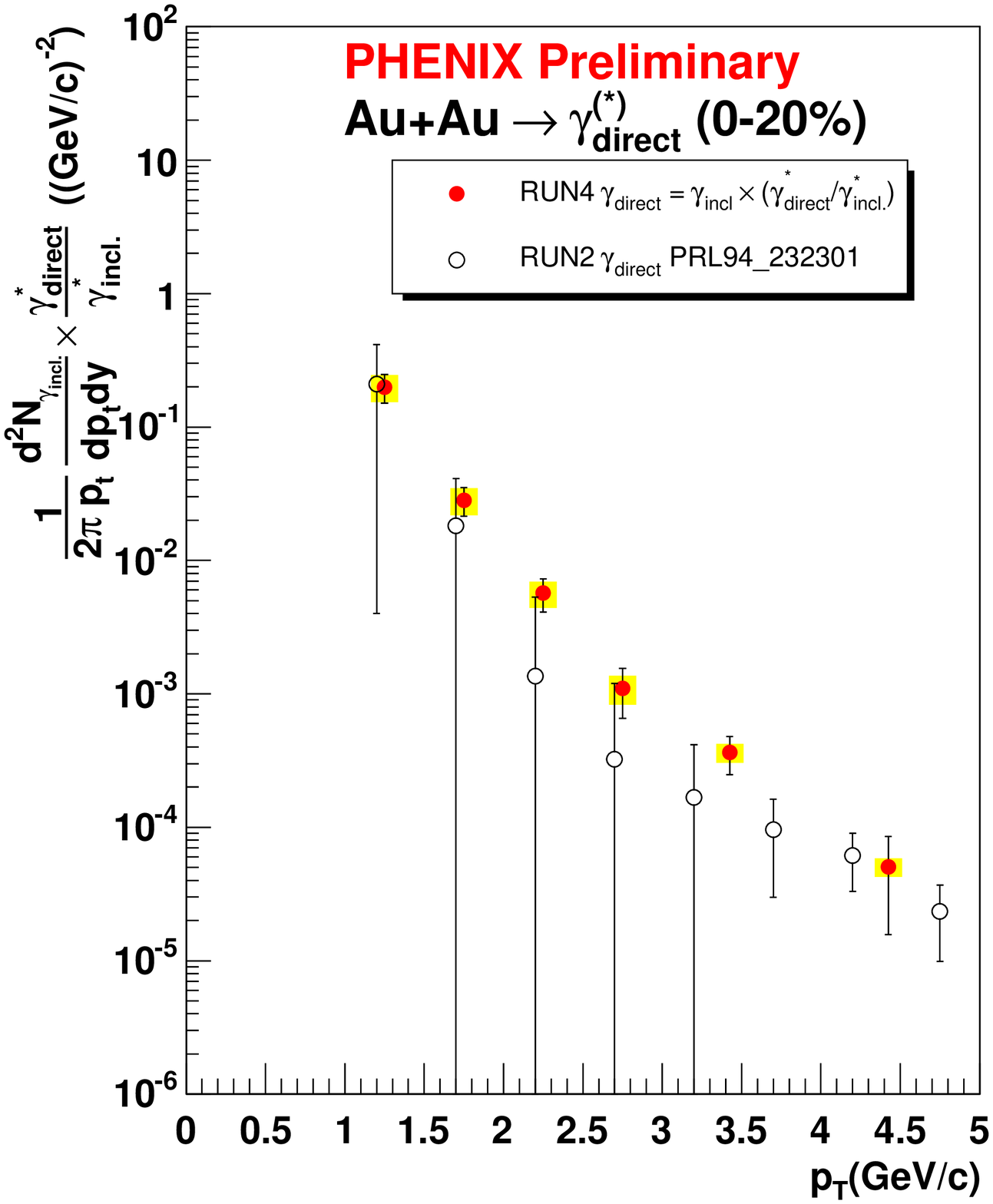}
\end{tabular}
\end{center}\vspace*{-0.25in}
\caption[]{a) (left) Invariant yield of prompt photons from $158A$ GeV Pb+Pb collisions~\cite{WA98photon} together with point-like scaled p+A measurements. Data points with downward arrows indicate unbounded 90\% Confidence Level upper limits. b) Invariant yield of prompt photons as a function of $p_T$ from central $\sqrt{s_{NN}}$ Au+Au collisions~\cite{PXphotonQM05} via photon detection \opencircle, and from measurements \fullcircle of $e^+ e^-$ pairs with $90 \leq m_{ee} \leq 300$ MeV, where only internal conversions from prompt photons (and some $\eta\rightarrow \gamma e^+ e^-$) contribute, compared to $0< m_{ee} < 30$ MeV, where all internal conversions (mostly $\pi^0\rightarrow \gamma e^+ e^-$) contribute.    \label{fig:photons}}
\end{figure}

   It is generally agreed that the observation of thermal photon or lepton pair  production would represent the `black body radiation' of the QGP. An exciting step in this direction was also made this year. Prompt photon production with $p_T > 1.5$ GeV/c had been observed in $158A$~GeV Pb+Pb collisions~\cite{WA98photon} (Fig.~\ref{fig:photons}a) 
but these photons were consistent with being from hard-scattering, with an enhancement compared to point-like scaled p-p collisions. Measurements of prompt photon production by a new method, internal conversion, were presented this year from Au+Au collisions at $\sqrt{s_{NN}}=200$ Gev~\cite{PXphotonQM05}, which showed a clear exponential spectrum in the range $1\leq p_T\leq 3$ GeV/c (Fig.~\ref{fig:photons}b) . Is this the smoking gun of the QGP or do p-p collisions also show a low $p_T$ exponential for direct photons as they do for pions? Incredibly, this is not known from previous measurements and is an example of how unknown aspects of p-p physics are unearthed by A+A measurements. The p-p  measurement is now underway at RHIC. This is an example of what may be faced with any apparent A+A discovery at the LHC, where the p-p physics is totally unknown~\cite{egSeeFrankfurt05}.

\section{Conclusion}
   Was a new state of matter discovered in Pb+Pb collisions at $\sqrt{s_{NN}}\sim 17$ GeV? Possibly, but it surely wasn't the gaseous QGP~\cite{CollinsPerry,Shuryak80,Shuryak05,KMR86,BIR03,Alberico05,Weiner05} envisioned at that time; although the logic---since the data couldn't be described by hadron gas models, it must be something else, and since the only other model (at the time) was the gaseous QGP, that's what we must have discovered---deserves, at least, a smile for its audacity. Nevertheless, this region of A+A collisions is interesting as the transition from ``Baryonic to mesonic freezeout''~\cite{PBMCleymans02}, so it is very possible that a new state of matter does exist in this region, such as an initial state of baryon resonances, raised over the $K^+ \Lambda$ threshold by successive excitation, or even something more interesting~\cite{GazdzickiQM04}. Clearly, this region deserves further study.
   
\begin{figure}[!htb]
\begin{center}
\includegraphics[scale=0.5,angle=0]{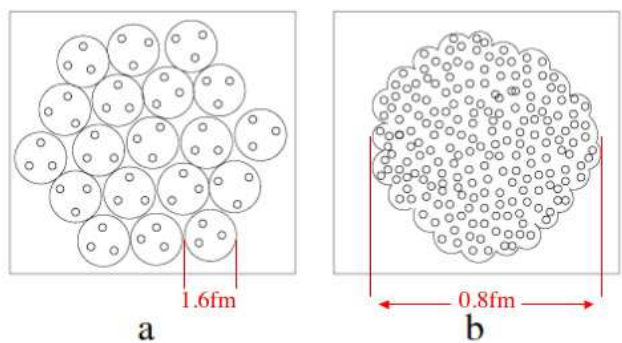}
\end{center}\vspace*{-0.25in}
\caption[]{``Strongly interacting matter as nuclear matter at a density of closely packed nucleons (a) and at a much higher density (b).''~\cite{SatzRPP63} The dimension of (b) is based on RHIC results. \label{fig:grapes}}
\end{figure}

      Was the gaseous QGP discovered at RHIC? Certainly not! The medium discovered at RHIC has the properties of a hot, dense, strongly interacting, (possibly) perfect fluid, the sQGP~\cite{THWPs,Thoma05}. Although the phenomena so far observed at RHIC are far from being understood, the scope of experiments has moved from the `discovery' phase to the `exploration' phase, to characterize in detail the properties of the medium. Has extended deconfinement~\cite{Weiner05} over a large volume been discovered at RHIC? There is no evidence for (or against) extended deconfinement. However the observation of $\sim 10$ times the color charge density of a nucleon in a volume roughly the size of a nucleon, where all the color charge is active on a partonic test probe as it passes through the medium, is indicative of deconfinement on a small scale.  
       This is not unreasonable, since there is nothing yet in the theory of QCD that will {\em reduce} the confinement radius with increasing temperature and density. 
Such a collection of quarks and gluons is not describable in terms of ordinary color-neutral hadrons, because there is no known self-consistent theory of matter composed of ordinary hadrons at the measured densities. The PHENIX experiment~\cite{PXWP} calls this state, ``dense partonic matter."       
The situation is reminiscent of the QGP as described by Satz in a previous review article in this journal~\cite{SatzRPP63} (Fig.~\ref{fig:grapes}), except that the distance scale in Fig.~\ref{fig:grapes}b is comparable to the size of a nucleon rather than the size of the whole nucleus.  

      Is the gaseous QGP ruled out? No. It is possible that, at the LHC, the much larger $\sqrt{s_{NN}}=5500$ GeV will produce a medium with an initial temperature high enough so that $\alpha_s(T)$ will become small enough so that the interaction becomes weak enough that a gaseous QGP is produced. This is also the region, where another possible new state of nuclear matter, the Color Glass Condensate~\cite{GMc}, should become dominant, if it exists. 
      
       Clearly, the phase diagram of hot, compressed nuclear matter is much richer and more complex than originally envisioned, and much work remains to be done at all  temperature and density scales. 
       
       \ack
The author would like to thank his many colleagues for helpful discussions and for allowing the use of their research results; and would like to apologize to those whose work was not able to be mentioned.


\begin{thebibliography}{999}
\bibitem{specificity} It should be noted that the requirement of specificity to A+A collisions immediately rules out the QGP in p-p collisions, which is disputable. See references~\cite{Weiner05,Alexopoulos02}.
\bibitem{Kinematicnote} For equal mass collisions the Laboratory and c.m. systems are simply shifted in rapidity by $Y^{\rm cm}=Y^{\rm beam}/2$.
\bibitem{RHICcent} At RHIC, centrality is usually measured by similar distributions recorded in Beam-Beam counters at relatively forward rapidities, typically $3\leq |\eta|\leq 4$.
\bibitem{CalScales} Treatment of the calorimeter absolute energy scales is different in all experiments. Using measurements of $E_T$ distributions for Au+Au at $\sqrt{s_{NN}}=$19.6, 130 and 200 GeV, PHENIX~\cite{PXppg019} estimates that on a consistent absolute energy scale $dE_T/d\eta$ at 2\% $\sim$ (0-5\%) upper centrality increases by a factor of 2 from AGS to SPS $\sqrt{s_{NN}}=5-17$ GeV and another factor of 2.6 from $\sqrt{s_{NN}}=17-200$ GeV.
\bibitem{butAQM} However, since the constituent quark degrees of freedom in the Additive Quark Model seem to come in to play~\cite{BCMOR,Ochiai,Voloshin,De}  for $\sqrt{s_{NN}}\geq 31$ GeV, this issue merits further investigation.
\bibitem{goodQM99quote} There is also a very interesting quote in Ref.~\cite{HeinzQM99} describing the data in Fig.~\ref{fig:SPSstrange1}b, ``this pattern contradicts expectations from final state hadronic rescattering: since hadrons with more strange quarks are heavier and strangeness must be created in pairs." This is essentially the same argument used by the authors of Ref.~\cite{AbbottPRD45} in explaining why ``a naive picture of successive collisions" could not explain the strangeness enhancement in p+Au collisions (Fig.~\ref{fig:E802K}a) at lower $\sqrt{s_{NN}}$.
\end{thebibliography}
\end{document}